\newcommand{\pa}{\partial}
\newcommand{\be}{\begin{equation}}
\newcommand{\ee}{\end{equation}}
\newcommand{\bea}{\begin{eqnarray}}
\newcommand{\eea}{\end{eqnarray}}

\renewcommand{\a}{\alpha}

\renewcommand{\b}{\beta}

\newcommand{\q}{\theta} 
\newcommand{\bq}{\bar\q}
\newcommand{\bw}{w'}
\newcommand{\ep}{\epsilon}

\newcommand{\bt}[1]{{\bar t}}
\newcommand{\ts}{\textstyle}
\newcommand{\ea}{\!\!\! = \!\!\!}
\newcommand{\half}{{\ts \frac{1}{2}}}

\def \ts{\textstyle}
\def \de{\delta}
\def \De{\Delta}
\def \si{\sigma}

\def \al{\alpha}
\def \pr{\partial}

\def \ep{\epsilon}
\def \vep{\varepsilon}
\def \half{{\textstyle {1 \over 2}}}
\def \thir{{\textstyle {1 \over 3}}}
\def \quar{{\textstyle {1 \over 4}}}
\def \A{{\cal A}}
\def \B{{\cal B}}
\def \C{{\cal C}}
\def \DD{{\cal D}}
\def \E{{\cal E}}
\def \F{{\cal F}}
\def \GG{{\cal G}}
\def \H{{\cal H}}

\def \N{{\cal N}}
\def \O{{\cal O}}

\def \M{{\cal M}}

\def \zz{\chi'}
\def \bal{\al'}
\def \xx{\chi}

\newcommand{\eqn}[2]{\begin{equation}\label{#1}#2\end{equation}}

\newcommand{\eqna}[2]{\begin{eqnarray}\label{#1}#2\end{eqnarray}}
\newcommand{\eqalign}[1]{#1}
\newcommand{\foot}[1]{\footnote{#1}}

\documentclass[11pt]{article}
\input epsf
\usepackage{amsmath,amsfonts,epsfig,color,latexsym}
\usepackage{amssymb}
\def\appendix{{\newpage\section*{Appendices}}\let\appendix\section%
        {\setcounter{section}{0}
        \gdef\thesection{\Alph{section}}}\section} 

\csname @addtoreset\endcsname{equation}{section}
\begin{document}

\begin{titlepage}
{\hbox to\hsize{${~}$\hfill 
{DAMTP-04-45}}}
{\hbox to\hsize{${~}$ \hfill {LAPTH-1047/04}}} \vskip 0.2pt \hfill {\tt 
hep-th/0405180}

\begin{center}
\vglue .3in {\Large\bf On Four-Point Functions of $\half$-BPS Operators\\ in General Dimensions}

\vskip 1cm

{\large{$\rm{F.A. ~Dolan}^{(a)}$, {\large{$\rm{L. ~Gallot}^{(b)}$}}
and  $\rm{E.~Sokatchev^{(b)}}$}}
\\[.3in]
\small

$^{\rm(a)}${\it Department of Applied Mathematics and Theoretical Physics, \\
Wilberforce Road, Cambridge, CB3 0WA, England}
\\
[.03in] $^{\rm(b)}${\it Laboratoire d'Annecy-le-Vieux de Physique 
Th\'{e}orique\footnote[1]{UMR 5108 associ{\'e}e {\`a}
 l'Universit{\'e} de Savoie}
LAPTH, B.P. 110, \\ F-74941 Annecy-le-Vieux et l'Universit\'{e} de Savoie}
\\[.3in]
\normalsize

{\bf ABSTRACT}\\[.0015in]
\end{center}
We study four-point correlation functions of $\half$-BPS operators of arbitrary weight for all dimensions $d=3,4,5,6$ where superconformal theories exist. Using harmonic superspace techniques, we derive the superconformal Ward identities for these correlators and present them in a universal form. We then solve these identities, employing Jack polynomial expansions. We show that the general solution is parameterized by a set of arbitrary two-variable functions, with the exception of the case $d=4$, where in addition functions of a single variable appear. We also discuss the operator product expansion using recent results on conformal partial wave amplitudes in arbitrary dimension. 

\vskip 7pt ${~~~}$ \newline

\end{titlepage}    

\section{Introduction}

The AdS/CFT correspondence has stimulated an enormous activity in the study of superconformal theories in the past few years. The most typical example is the ${\cal N}=4$ super-Yang-Mills theory in four dimensions whose conjectured holographic dual is the type IIB supergravity (string) theory on $AdS_5 \times S^5$ \cite{AdS}. Another case of interest is the superconformal six-dimensional theory of the $(2,0)$ self-dual tensor multiplet. It is believed to describe the world-volume fluctuations of the  M-theory five-branes on $AdS_7 \times S^4$ \cite{d6}. Three-dimensional $\N=8$ superconformal theories have been considered in relation to anti-de Sitter black holes of gauged $SO(8)$ supergravity on $AdS_4 \times S^7$ \cite{d3}.

A considerable part of this recent activity has been concentrated on
the so-called $\half$-BPS states. These are very special
representations of the underlying superconformal symmetry group of the
above theories, which are annihilated by half of the
supercharges. Thus, the superconformal kinematics fixes their
conformal weight (dimension) at its free theory value. Such states (or
the corresponding composite operators in field theory) are called
``short" or ``protected". Translated in terms of correlation
functions, this means that the two-point functions of $\half$-BPS
operators do not receive any quantum corrections. An even stronger
result is that their three-point functions are also protected.
Non-trivial dynamics starts appearing  at the level of the four-point
functions. There one sees the rich spectrum of the OPE of $\half$-BPS
operators, which contains protected as well as unprotected (``long multiplet") states. Superconformal kinematics puts strong restrictions on the general form of such correlators and on their OPE content, which have been successfully tested against numerous results in perturbative field theory and in AdS supergravity. 

In this paper we present a systematic study of the superconformal
properties of four-point functions of $\half$-BPS operators in all
cases of interest. Namely, we derive and solve the constraints (or
superconformal Ward identities) following from superconformal symmetry
in spaces of dimension $d=3,4,5,6$. The corresponding superconformal
algebras are $(P)SU(2,2|\N)$ with $\N=2,4$ in $d=4$, $OSp(8^*|2\N)$ with
$\N=1,2$ in $d=6$, $OSp(8|4,\mathbb{R})$ in $d=3$, as well as the
exceptional algebra $F(4)$ in $d=5$.

An early attempt at such a study was undertaken in \cite{Pickering},
where the simplest case of $\half$-BPS operators of weight two in
$d=4$ $\N=2$ superconformal theory was considered and the
corresponding superconformal Ward identities were derived using
harmonic/analytic superspace techniques \cite{HSS,HaH}. This result
was generalized in \cite{EPSS} to the four-point functions of
stress-tensor multiplets (or $\half$-BPS operators of weight two) in
$d=4$ $\N=4$ super-Yang-Mills theory. There the general solution to
the 
Ward identities was found, making use of the full crossing symmetry of
the amplitude. It was shown that the solution is parameterized by a
function of a single conformally invariant variable and by another,
arbitrary function of two variables. A dynamical argument based on the
field-theoretic insertion procedure \cite{Intri} lead to the
conclusion in \cite{EPSS} that the single-variable function does not
receive quantum corrections (a phenomenon called ``partial
non-renormalization"), in excellent agreement with results from $AdS_5 \times S^5$ supergravity
\cite{AFP} and from perturbative field theory \cite{AEPS}.  Later on,  in \cite{DO} the same Ward identities were
derived, 
in a component field approach, and solved without using
crossing symmetry.  There it was also shown that the single-variable
part of the solution (which, without crossing symmetry, involves two
independent single-variable functions)
admits an OPE (or conformal partial wave)
expansion over a restricted class of superconformal representations,
the so-called short and semishort multiplets. Four-point correlators
of $\half$-BPS operators of weight $k>2$ in the $d=4$ $\N=4$ theory
were studied in \cite{HH,ADOS,APSS}. In particular, in \cite{HH},
using analytic superspace techniques it was shown that for arbitrary
$k$ the amplitude is parameterized by  $\half k(k-1)$ functions of two
variables  and by $k$ independent single-variable functions. The latter were
shown to be expanded over protected superconformal states only.\footnote{See also \cite{NO} where similar results are obtained in a component field approach.} Partial non-renormalization (i.e., the absence of quantum corrections  to the single-variable functions) for $k=3,4$ was confirmed both perturbatively and in AdS supergravity in \cite{ADOS,APSS}. An  alternative explanation of this phenomenon was given in \cite{HH}, where it was linked to the protectedness of the three-point functions of short and semishort multiplets.   

The implications of superconformal symmetry for four-point functions
in six space-time dimensions were studied in
\cite{AS}. There the Ward identities for the four-point function of composite
operators of weight two, made out of $(2,0)$ self-dual tensor
multiplets, were derived and solved, using crossing symmetry. 
In contrast with the four-dimensional case, it was found here
that the solution could be parameterized by a single two-variable 
function. Again, these general predictions were compared with success to explicit results
from $AdS_7 \times S^4$ supergravity. 

In the present paper we are able to extend the result of \cite{AS} to the $d=6$ case\footnote{See also some comments about $d=6$ in \cite{Heslop}.} of arbitrary  weight $k$ and without crossing symmetry, as well as to the cases $d=3,5$. In the process of solving the superconformal Ward identities we obtain a set of arbitrary two-variable functions, as well as another set of restricted functions. Unlike the case $d=4$, the latter can be absorbed into suitable redefinitions of the former. This is possible because in $d\neq 4$ both types of functions are acted upon by a differential operator with a non-trivial kernel. It is precisely this kernel which is responsible for the absorption of the restricted sector of the solution. In the $d=4$ case this operator is the identity and the absorption is impossible.  

The method we use in this paper for deriving the Ward identities is based on
the appropriate versions of harmonic superspace for each type of
superconformal symmetry. The defining property of the $\half$-BPS
operators, realized in $d=3,4,5,6$ superspace, is that they depend
only on half of the Grassmann (odd) variables. Such superfields should
not be confused with the familiar $d=4$ ${\cal N}=1$ chiral
superfields. In the chiral case the half is chosen with regard to the
Lorentz group (left- or right-handed chirality), and this only makes
sense in $d=4$. The $\half$-BPS superfields depend on a different half
of the odd variables obtained by projecting their R symmetry index
rather than the Lorentz one. Thus, $\half$-BPS superfields can be
defined independently of the space-time dimension. Such superfields
are called ``Grassmann analytic" \cite{GA,HSS,HaH}. Grassmann analyticity
can be achieved without loss of  manifest R symmetry if one introduces
internal symmetry (also called ``harmonic") variables. The resulting
Grassmann analytic harmonic superspace (or its version called
``analytic superspace" \cite{HaH,HH}) is the ``native" superspace for
the realization of $\half$-BPS representations of the various superconformal algebras in $d=3,4,5,6$ (see \cite{HH2,FS} for a review on BPS states in harmonic/analytic superspaces). 

Another technique we use in this paper is the expansion of the four-point amplitude in a basis of two-variable  symmetric polynomials, the so-called Jack polynomials \cite{Jack}.  The Ward identities constraining these
four-point functions take a very simple and universal form for all dimensions $d$, upon making a judicious choice of
variables parametrizing the conformal and R symmetry invariants. However,
their solution is far from trivial, because the
functions satisfying the Ward identities are symmetric
separately in the conformal and R symmetric invariant variables.
Without this symmetry the solutions are not difficult to find, but the Jack polynomial expansion is indispensable when it comes to taking the symmetry requirements into account.  Such polynomials have been used before to find,
in general dimensions,  conformal partial wave amplitudes representing
contributions to any scalar four-point correlation function  of an
operator, and its descendants, in the OPE of two of the operators in
the four-point function \cite{fadho}.  Since the conformal group $SO(2,d)$ is non-compact, the conformal partial wave amplitudes are infinite-order polynomials 
in the conformal invariants.  For four dimensions the resulting expansion is directly equivalent to one in terms
of Schur polynomials,  which were used in   \cite{HH} to find $\N=4$ four-point functions compatible with superconformal symmetry. We  use the expansion of \cite{fadho} in terms of Jack polynomials to
perform conformal partial wave analysis of the $k=2$ solution.

Let us now briefly describe the content of the paper.  In Section \ref{csf} we discuss the four-point invariant variables associated with  the $d$-dimensional conformal group $SO(2,d)$, as well as  with the
$SO(n)$ R symmetry group of the corresponding superconformal theories.
These symmetries can be realized on light-like vectors of $SO(2,d)$ (see, e.g., \cite{MackSalam}) or on complex ``light cone" (or ``null") vectors of $SO(n)$.\footnote{The use of such variables probably goes back to the beginning of the 20th century, see \cite{Knapp}. They have been employed, in a very explicit form,  in the studies of conformal field theories in the 1970s (see \cite{TZ}, \cite{DMPPT} and references therein). In the context of four-point functions of $\half$-BPS operators null vectors have been used in \cite{ADOS,APSS}, and recently in \cite{NO}.} They serve as the coordinates
of space-time or of the internal (harmonic) space, respectively. We then introduce
a frame fixing procedure, which consists in using all the available
conformal/internal symmetry in order to reduce the coordinates to
just two independent invariant variables in the space-time sector and
to two (for R symmetry $SO(n)$, $n>3$) or one (for R symmetry $SO(3)$)
variable in the internal sector.\footnote{Such variables have already
been used in \cite{EPSS,DO,HH} for 
solving the superconformal Ward identities. 
Here we give them another, very simple interpretation as, e.g., the light-cone coordinates of two-dimensional
space-time.} We also discuss the general conformal and R symmetry
properties of two-, three- and four-point functions of the primary
scalar states of the  $\half$-BPS multiplets. We write these amplitudes as polynomial expansions 
in the internal invariant variables  with coefficients which are functions of the space-time invariants.

In Section \ref{sshbs} we apply the same frame fixing procedure\footnote{This generalizes the method proposed in \cite{Pickering}. A similar procedure has been used before in \cite{fh2} in the context of $d=4$ $\N=1$ supersymmetry.} to the two-, three- and four-point invariants in the Grassmann analytic superspaces suitable for the description of $\half$-BPS superfields. We choose a superconformal frame involving only the conformal and internal invariant variables, as well as the Grassmann analytic variables at one of the four points. These variables undergo linear residual supersymmetry transformations, which allow us to easily find the completion of the even variables to superconformal invariants. We show that these invariants become singular when one of the internal variables approaches one of the space-time variables. 

In Section \ref{SCWIs} we use these results to supersymmetrize the four-point amplitudes. The crucial point here is the requirement that the amplitude be free from singularities. This is precisely the origin of the superconformal Ward identities. The frame fixing procedure guarantees that the constraints we find are necessary and sufficient conditions for superconformal invariance.  We present the Ward identities in a universal form suitable for any space-time dimension and any R symmetry group. In this section we also solve the Ward identities in the $d=4$ case, thus reproducing the known results. 

In Section \ref{Wisrs} we discuss the Ward identities for theories in
any dimension $d$ and with R symmetry group $SO(3)$ (these include the
cases $d=6$ $\N=(1,0)$, $d=5$ and $d=4$ $\N=2$). We need to solve two
partial differential equations involving the $k+1$ coefficient functions of the amplitude of weight $k$. We start by finding the general solution in the simplest case of weight
$k=1$ in terms of Bessel functions of the first and second kind, depending on whether the dimension  
of the spacetime is odd or even, respectively.  Subsequently we show that the $k=2$ case can be reduced to the $k=1$ one. This is done by a suitable redefinition of the coefficients of the amplitude, for which we employ an operator identity.  This identity is also used to show that we may solve for the general $k$ case in a similar
fashion.  We find that the solution is parameterized by $k-1$ unrestricted two-variable functions along with the
restricted solution from the $k=1$ case. An important point is that the restricted part of the solution for
each $k\geq 2$, coming from the $k=1$ reduction, may be completely absorbed into the two-variable part via a certain redefinition of the latter, but only for $d\neq 4$. This is shown here for $d=6$ but holds more generally, as we see in Section \ref{CRSTJP} where Jack polynomials are employed to find solutions.

In Section \ref{CRSTJP} we solve the Ward identities  for theories
with R symmetry $SO(n)$, $n \geq 3$ (these include the cases $d=3$
$\N=8$, $d=4$ $\N=4$ and $d=6$ $\N=(2,0)$). We use expansions in terms
of Jack polynomials to solve for the simplest $k=1$ case, showing how
solutions for the $k=1$ case for $SO(3)$ may be unified in terms of these. We also prove and generalize certain claims made in Section \ref{Wisrs} pertaining to
arbitrary dimension.  Finally, we solve for the $k=2$ and general
$k$ cases using these and other results based on Jack polynomial
expansion.  The solution is parameterized by $\half k (k-1)$
unrestricted two-variable functions and $k$ restricted ones which,
for $d\neq 4$, may be absorbed into a redefinition of the two-variable
functions.

In Section \ref{CPWA} we use results from \cite{fadho} 
to expand the $k=2$ solutions to the Ward identities for $SO(n)$ 
in terms of conformal partial wave amplitudes. This is nontrivial for the unrestricted two-variable part of the solution. Still, we are able to conclude that the conformal partial wave expansion of this part of the solution corresponds to long multiplets in the OPE, whose primary state is a scalar. It is also shown here how one of the restricted solutions, even
though it can be absorbed into a redefinition of a two-variable function,
corresponds to shortened multiplets in the OPE, whose primary state is again 
a scalar and all operators appearing have twist $d-2$. Also in this section, the corresponding results of \cite{DO} for $d=4$ $\N=4$ are recovered.

Generalizing this approach to conformal partial wave
expansion for other $k$ appears very non-trivial - at least
when using the method described here.  We hope that
the basis independent form of the solutions to
the Ward identities for $d\neq 4$, might aid such
expansion more generally.  We feel that the
issue is related to making an appropriate choice
of basis with which to expand the solution, so
as to extract harmonic polynomials in the
R symmetry invariants and conformal partial waves. 

\bigskip {\bf Note added.} After the first version of this paper was submitted to the e-archive, the paper \cite{Heslop6d} appeared in which the case $d=6$, $k=2$ is treated by a different method. In Section \ref{k2case} we show that our results for this case and those of \cite{Heslop6d} are equivalent.

\section{Conformal and R symmetries of the four-point amplitudes}\label{csf}

The subject of our study are the four-point correlation functions of the so-called $\half$-BPS operators. The $\half$-BPS states, also called ``short supermultiplets", are special representations of the  superconformal algebras in $d=3,4,5,6$. The symmetry group of the even sector of these subalgebras is the product of the $d$-dimensional conformal group $SO(2,d)$ and the R symmetry group. In all the cases of interest the R symmetry group is locally isomorphic to an orthogonal group $SO(n+2)$ with $n=1,3,4$ or $6$. Postponing the discussion of the superconformal properties to Section \ref{sshbs}, here we concentrate on the bosonic symmetries of these amplitudes. 

\subsection{Conformal cross-ratios and related variables}\label{cf} 

Dirac proposed a natural realization of the representations of the conformal group, the so-called ``light ray" realization (see, e.g., \cite{MackSalam}). The idea is to introduce real light-like (or null) vector variables $X^m$, $m=0,1,\ldots,d+1$,
\begin{equation}\label{light}
  X\cdot X = (X^0)^2 - (X^1)^2 - \ldots - (X^{d})^2 + (X^{d+1})^2 = 0\,,
\end{equation}
on which the conformal group $SO(2,d)$ acts linearly. The representations of the conformal group can be realized on fields $\phi(X)$ on the light cone, homogeneous of degree $\ell$, $\phi(\rho X) = \rho^\ell\phi(X)$. The algebraic condition (\ref{light}) and the homogeneity condition effectively eliminate two degree of freedom, so we are left with $d$ independent variables. A convenient choice for the $d$ coordinates of space-time is
\begin{equation}\label{convch}
  X^m = \left(x^\mu, \frac{1}{2}(1+x^2),\frac{1}{2}(1-x^2)\right)\,, \qquad \mu=0,1,\ldots,d-1\,.
\end{equation}
Here $x^2=(x^0)^2-(x^1)^2-\ldots-(x^{d-1})^2$ and $x^\mu$ is a vector of the Lorentz group $SO(1,d-1)$ on which the conformal group $SO(2,d)$ acts {\it non-linearly}. 

One of our main goals is to construct invariants. In the case of the conformal group this is quite obvious. Since the vectors $X^m$ are light-like, we cannot form an $SO(2,d)$ invariant with a single vector. With two such vectors we can obtain a covariant by taking their scalar product $X_1\cdot X_2$. Here covariance means that the homogeneity condition requires that each vector transform with its own scaling factor, $X_1\cdot X_2 \ \to \rho_1\rho_2X_1\cdot X_2$. Then it is clear that invariants can be constructed by using four null vectors and forming the so-called ``conformal cross-ratios" 
\begin{equation}\label{-1}
  u=\frac{X_1\cdot X_2\,X_3\cdot X_4}{X_1\cdot X_3\,X_2\cdot X_4}\,, \qquad v=\frac{X_1\cdot X_4\,X_2\cdot X_3}{X_1\cdot X_3\,X_2\cdot X_4}\,.
\end{equation}  
In terms of the unconstrained  $d$-vectors $x^\mu_{1,2,3,4}$ these cross-ratios read
\begin{equation}\label{1}
  u=\frac{x^2_{12}x^2_{34}}{x^2_{13}x^2_{24}}\,, \qquad v=\frac{x^2_{14}x^2_{23}}{x^2_{13}x^2_{24}}\,,
\end{equation}  
where $x^2_{ij} = (x_i-x_j)^2$.

In what follows we shall work in a special conformal frame, commonly used in studies of conformal theories. With the help of translations, conformal boosts, dilatations and part of the Lorentz
group $SO(1,d-1)$, we can fix the frame\footnote{In this paper, in order to avoid space-time singularities we always assume that ${x_i}\neq {x_j}$.}
\begin{equation}\label{1'}
   x^\mu_2 = (1,0,\ldots,0)\,, \quad x^\mu_3 = (x^\mu_4)^{-1}  ={0}\,,
 \end{equation}
where $(x^\mu_4)^{-1} ={x^\mu_4}/{x^2_4}$.
This choice breaks $SO(1,d-1)$ down to $SO(d-1)$. We can specify the frame even further by using the residual rotational symmetry  $SO(d-1)$, so that the vector $x_1$ has only two independent components:
\begin{equation}\label{68}
  x^\mu_1 = \left(1+\frac{1}{{2}}(z+z'),\frac{1}{{2}}(z-z'),0,\ldots,0\right)\,, 
\end{equation}  
or, equivalently,
\begin{equation}\label{6812}
  x^\mu_{12} =x^\mu_1-x^\mu_2 = \left(\frac{1}{{2}}(z+z'),\frac{1}{{2}}(z-z'),0,\ldots,0\right)\,. 
\end{equation}
Thus, in the frame  (\ref{1'}), (\ref{6812}) the cross-ratios  (\ref{1}) become 
\begin{equation}\label{02}
  u = \frac{zz'}{(1+z)(1+z')}\,, \qquad v = \frac{1}{(1+z)(1+z')}\,.
\end{equation}

In fact, such variables are most natural and are widely used in two-dimensional conformal field theory.\footnote{E.S. is grateful to I. Todorov for this remark.} Indeed, let us replace the vector coordinates $x^\mu$ of the two-dimensional  space by the light-cone coordinates 
\begin{equation}\label{3}
  \zeta^\pm = x^0 \pm x^1 \ \Leftrightarrow \ x^\mu = \left(\frac{1}{{2}}(\zeta^+ + \zeta^-),\frac{1}{{2}}(\zeta^+ - \zeta^-)\right)\,, \qquad \zeta^+\zeta^- = x^2\,.
\end{equation}
The two-dimensional Lorentz group $SO(1,1)$ acts by rescaling of $\zeta^\pm$, so we need not square the coordinate differences $\zeta^\pm_{ij}$ to obtain Lorentz invariant cross-ratios. Thus, we can form the cross-ratios 
\begin{eqnarray}
   z = \frac{\zeta^+_{12}\zeta^+_{34}}{\zeta^+_{14}\zeta^+_{23}}\,,&& \qquad\ \ z' = \frac{\zeta^-_{12}\zeta^-_{34}}{\zeta^-_{14}\zeta^-_{23}} \,; \nonumber\\
   1+z = \frac{\zeta^+_{13}\zeta^+_{24}}{\zeta^+_{14}\zeta^+_{23}}\,,&& \quad 1+z' = \frac{\zeta^-_{13}\zeta^-_{24}}{\zeta^-_{14}\zeta^-_{23}}\,. \label{4}
\end{eqnarray}
Substituting (\ref{4}) in the expressions for $u,v$ (\ref{1}), we obtain precisely (\ref{02}). It is then natural to extend the definition (\ref{02}) to any space-time dimension.

It should be stressed that the change of variables (\ref{02}) is not bijective. Indeed, solving eqs. (\ref{02}) for $z, z'$ we find
\begin{equation}\label{0202}
  z = \frac{1}{2v}\left(1-u-v \pm \sqrt \Delta  \right)\,, \quad z' = \frac{1}{2v}\left(1-u-v \mp \sqrt \Delta  \right)\,, 
\end{equation}
where 
\begin{equation}\label{0203}
  \Delta = (1-u-v)^2 -4uv = -4 \frac{\ep{\mathbf x}^2}{((x^0)^2 +\ep {\mathbf x}^2)^2}\,.
\end{equation}
To obtain (\ref{0203}) we have used  the conformal frame (\ref{1'}) with $x^\mu_1 = (x^0,{\mathbf x})$, and with $\ep=-1$ in the Minkowski case and $\ep=+1$ in the Euclidean case. This explains why the variables $z, z'$ are real in the former case and complex in the latter. The choice of sign in (\ref{0202}) is conventional and should not affect any physical quantity. This follows from the definition of $z,z'$ (\ref{02}), which is symmetric under the exchange $z\leftrightarrow z'$. Therefore, all the results we are going to obtain should have this symmetry.

We note that the equation $z=z'$ (i.e., $\Delta=0$) defines the ``border line" between the Euclidean and Minkowskian domains in the conformal cross-ratios plane. The physical correlation functions should be well defined in the entire Minkowskian (Euclidean) domain, including the border. We shall come back to this point later on.

The relevance of such variables to the superconformal Ward identities was first pointed out in \cite{EPSS}. The variables $\xi,\eta$ used there are related to $z,z'$ as follows:
\begin{equation}\label{newtoold}
  z\rightarrow\xi\,, \qquad z' \rightarrow \frac{1}{\eta}\,.
\end{equation} 
They are the standard variables for solving the integrability condition for the superconformal Ward identities. They are also commonly used in the literature on multiloop integrals (see, e.g., \cite{UD}). Later on in Ref.  \cite{DO} it was shown that a more convenient choice of variables allows one to directly solve the Ward identities without considering their integrability condition. The variables $z',x$ of  \cite{DO} are related to our $z,z'$ as follows: 
\begin{equation}\label{reldo}
  z \rightarrow \frac{z'}{1-z'}\,, \qquad z' \rightarrow \frac{x}{1-x}\,.
\end{equation}
Finally, in Ref. \cite{HH} it was pointed out that such variables arise in yet another way, by diagonalizing the coordinate matrix (see Section \ref{fpsc}).       

\subsection{R symmetry invariant variables}\label{rsiv}

Besides the conformal group, the amplitudes we are discussing in this paper have another symmetry, the compact R symmetry group which is locally isomorphic to an orthogonal group $SO(n+2)$ with $n=1,3,4$ or $6$. Its representations can be realized in terms of internal variables similar to the space-time variables of the conformal group.

\subsubsection{The cases $SO(5)$, $SO(6)$ and $SO(8)$}

In these cases the $\half$-BPS multiplets have primary states in symmetric traceless tensor representations of $SO(n)$. For example, the scalars of the $d=4$ ${\cal N}=4$ SYM multiplet form an $SO(6)$ vector, while the primary state of the stress-tensor multiplet is in the {\bf 20'} of $SU(4)\sim SO(6)$ (a rank two tensor of $SO(6)$).  It is natural to associate such representations to {\it complex null vectors} of $SO(n+2)$. The idea is very much like that of Dirac's light ray realization of the conformal group. One introduces null vector variables $Y^{\mathbf{i}}$, $\mathbf{i}=1,\ldots,n+2$,
\begin{equation}\label{lightR}
  Y\cdot Y = (Y^1)^2 + \ldots + (Y^{n+2})^2 = 0\,,
\end{equation}
on which the group $SO(n+2)$ acts linearly. This condition, together with a homogeneity requirement similar to that in the conformal case, reduces the $SO(n+2)$ vector $Y$ to an independent $SO(n)$ vector, e.g.,
\begin{equation}\label{convchR}
  Y^{\mathbf{i}} = \left(y^i, \frac{1}{2}(1-\vec{y}^2),\frac{i}{2}(1+\vec{y}^2)\right)\,, \qquad i=1,\ldots,n\,,
\end{equation}
on which $SO(n+2)$ acts {\it non-linearly}.  Note that the $SO(n+2)$ null vector $Y^{\mathbf{i}}$ must be complex, and so is the unconstrained $SO(n)$ vector $y^i$.

Given four sets of such internal coordinates we can construct R symmetry invariants in close analogy with the conformal case: 
\begin{equation}\label{-11}
  U=\frac{Y_1\cdot Y_2\,Y_3\cdot Y_4}{Y_1\cdot Y_3\,Y_2\cdot Y_4}\,, \qquad V=\frac{Y_1\cdot Y_4\,Y_2\cdot Y_3}{Y_1\cdot Y_3\,Y_2\cdot Y_4}\,.
\end{equation}  
In terms of the independent $n$-vectors $y^i_{1,2,3,4}$ these internal cross-ratios read
\begin{equation}\label{-111}
  U=\frac{y^2_{12}y^2_{34}}{y^2_{13}y^2_{24}}\,, \qquad V=\frac{y^2_{14}y^2_{23}}{y^2_{13}y^2_{24}}\,.
\end{equation}  
Using part of the $SO(n+2)$ symmetry\footnote{Strictly speaking, since
the vectors $\vec{y}$ are complex, we should use the complexification
of $SO(n)$. However, in what follows we shall not make this
distinction.} we can fix the frame (cf. (\ref{1'}))
\begin{equation}\label{68'12}
  y^i_2 = (1,0,\ldots,0)\,, \quad y^i_3 =  (y^i_4)^{-1}={0}\,. 
\end{equation}
This breaks $SO(n+2)$ down to $SO(n-1)$. Then, using this residual
rotational symmetry, we can further fix the frame
\begin{equation}\label{68'12'}
  \vec{y}_{12}  = \left(\frac{1}{{2}}(w+\bw),\frac{1}{{2}}(w-\bw),0,\ldots,0\right)\,, 
\end{equation} 
in which there remain only two independent complex variables $w,\bw$. In terms of these new variables the invariant cross-ratios read (cf. (\ref{02}))
\begin{equation}\label{02int}
  U = \frac{w\bw}{(1+w)(1+\bw)}\,, \qquad V = \frac{1}{(1+w)(1+\bw)}\,.
\end{equation}

\subsubsection{The case  $SO(3)$}\label{tcso3}

This case is special since the elementary $\half$-BPS multiplets are related to the fundamental (spinor) rather than the vector representation of $SU(2)\sim SO(3)$. For example, the scalars of the $d=4$ ${\cal N}=2$ matter multiplet (hypermultiplet) belong to an $SU(2)$ doublet. Then the internal variables should form a spinor $Y^{\mathbf{a}}$, ${\mathbf{a}}=1,2$, of $SU(2)$  (or of its complexification $SL(2,\mathbb{C})$). Once again, we consider functions of $Y$ homogeneous of degree $\ell$, $\phi(\rho Y) = \rho^\ell\phi(Y)$, where $\rho$ is complex.\footnote{The internal variables we use here are similar, although not identical, to the ``harmonic" variables of \cite{HSS}. The latter form an $SU(2)$ matrix and are homogeneous with a $U(1)$ factor.} This allows us to choose a frame in which $Y^{\mathbf{a}}$ has only one independent component,
\begin{equation}\label{spinharm}
  Y^{\mathbf{a}} = (y,1)\,.
\end{equation}
Further, given two points we can form the $SL(2,\mathbb{C})$ invariant contraction 
\begin{equation}\label{su2inv}
  Y_1\cdot Y_2 \equiv Y^{\mathbf{a}}_1\epsilon_{\mathbf{ab}}Y^{\mathbf{b}}_2 = Y^{\mathbf{1}}_1Y^{\mathbf{2}}_2 -  Y^{\mathbf{2}}_1Y^{\mathbf{1}}_2 = y_1-y_2 \equiv y_{12}\,.
\end{equation}
Given four points and taking into account the cyclic identity
\begin{equation}\label{38''}
  y_{12}y_{34} - y_{13}y_{24} + y_{14}y_{23} = 0\,, 
\end{equation}
we can construct a single independent invariant variable, e.g.,
\begin{equation}\label{crrsu2}
  w = \frac{Y_1\cdot Y_2\,Y_3\cdot Y_4}{Y_1\cdot Y_4\,Y_2\cdot Y_3} = \frac{y_{12}y_{34}}{y_{14}y_{23}}\,.
\end{equation}
We may say that this case is in a sense the analog of the $d=2$ conformal one.

Finally, we can use the $SL(2,\mathbb{C})$ symmetry to fix the frame
\begin{equation}\label{frsu2}
  y_1=1+w\,, \quad  y_2=1\,, \quad y_3= y_4^{-1}=0 \,.
\end{equation}
Note that this choice completely breaks down the R symmetry, the remaining independent variable $w$ being invariant.

\subsection{Two-, three- and four-point functions}\label{fpa}

The most important property of the $\half$-BPS operators is that their two-, three- and four-point functions are completely determined by the lowest (primary) component in their superfield expansion (see Section \ref{sshbs}). This lowest component is in turn strongly restricted by conformal and R symmetry covariance.

\subsubsection{Two- and three-point functions}\label{tpf}

The best-known example of $\half$-BPS states are the composite operators made out of the $d=4$ ${\cal N}=4$ super-Yang-Mills on-shell field-strength multiplet. The lowest component of this multiplet is a set of six real scalars $\phi^{\mathbf{i}}(x)$, $\mathbf{i}=1,\ldots,6$ in the vector representation of $SO(6)$. Introducing the $SO(6)$ null vector (\ref{convchR}) we can define the scalar field
\begin{equation}\label{016}
  \phi(x,y) = i\sqrt{2}\phi^{\mathbf{i}}(x) Y^{\mathbf{i}} =  \frac{i}{\sqrt{2}}\left[(\phi^5+i\phi^6) + 2\phi^i y^i  - (\phi_5-i\phi_6) \vec{y}^2\right]\,.
\end{equation} 
This field, regarded as a function of the null vector $Y^{\mathbf{i}}$, transforms with unit weight under the projective action of the R symmetry group, 
\begin{equation}\label{weightf}
  \phi(x,\rho Y) = \rho \phi(x,Y)\,. 
\end{equation}
Notice that the dependence on the internal variables $y^i$ is polynomial.\footnote{The null vector $Y$ is complex, therefore the projected field $\phi(x,y)$ cannot be real. Nevertheless, in the harmonic \cite{HSS} (or analytic \cite{HH}) superspace approach one can define a combination of complex conjugation and of a particular reflection on the internal manifold, under which $\phi(x,y)$ goes into itself and thus the coefficients in (\ref{016}) can be made real.} 

In this notation the two-point function of the free scalars
\begin{equation}\label{017}
  \langle \phi^{\mathbf{i}}(x_1)\phi^{\mathbf{j}}(x_2) \rangle = \frac{\delta^{\mathbf{ij}}}{x^2_{12}}
\end{equation}
becomes
\begin{equation}\label{018}
  \langle \phi(x_1,y_1)\phi(x_2,y_2) \rangle =\frac{Y_1\cdot Y_2}{x^2_{12}} =  \frac{y^2_{12}}{x^2_{12}}\,.
\end{equation}  

The ${\cal N}=4$ $\half$-BPS primary states of weight $k$ are realized as gauge-invariant composite operators
\begin{equation}\label{defcompop}
  {\cal O}^k(x,y) = {\rm Tr}\; (\phi(x,y))^k = {\rm Tr}\; (\phi^{\mathbf{i}_1} \cdots \phi^{\mathbf{i}_k}) Y^{\mathbf{i}_1} \cdots Y^{\mathbf{i}_k}\,, 
\end{equation}
transforming in the symmetric traceless representation of $SO(6)$ (or the $[0,k,0]$ of $SU(4)$),
\begin{equation}\label{trwk}
  {\cal O}^k(x,\rho Y) = \rho^k {\cal O}^k(x,Y)\,.
\end{equation}
These operators have conformal weight equal to their R weight $k$, in the free as well as in the interacting field theory. The most remarkable feature of the BPS operators is that their conformal dimension is ``protected", i.e. does not receive any quantum corrections. The two-point function of such operators is just the $k$-th power of (\ref{018}),
\begin{equation}\label{019}
  \langle {\cal O}^k(x_1,y_1){\cal O}^k(x_2,y_2) \rangle = \left( \frac{Y_1\cdot Y_2}{x^2_{12}} \right)^k = \left(\frac{y^2_{12}}{x^2_{12}}\right)^k\,.
\end{equation} 

Exactly as in ordinary conformal theory, where the three-point
functions of scalars are obtained by multiplying two-point functions,
we can construct the three-point function of $\half$-BPS operators of
weights $k_1,k_2,k_3$ as follows:
\begin{eqnarray}
  && \!\!\!\!\!\!\! \langle {\cal O}^{k_1}(x_1,y_1){\cal O}^{k_2}(x_2,y_2) {\cal
 O}^{k_3}(x_3,y_3)\rangle \nonumber\\
  &&  \qquad\qquad \qquad = C(g)\left(\frac{y^2_{12}}{x^2_{12}} \right)^{\frac{k_1+k_2-k_3}{2}} \left(\frac{y^2_{13}}{x^2_{13}} \right)^{\frac{k_1+k_3-k_2}{2}} \left(\frac{y^2_{23}}{x^2_{23}} \right)^{\frac{k_2+k_3-k_1}{2}}\,,\label{3ptn4}
\end{eqnarray}
where $C(g)$ is a normalization constant, in general depending on the gauge coupling $g$. Remarkably, for $\half$-BPS operators not only their weights $k$, but also the normalization  $C(g)$ is ``protected" \cite{3ptprotected}. 

Another, even simpler example of a $\half$-BPS multiplet is the elementary $d=4$ ${\cal N}=2$ matter multiplet (hypermultiplet). It can be obtained as a submultiplet of the ${\cal N}=4$ SYM multiplet under the reduction ${\cal N}=4\ \to \  {\cal N}=2$.   The lowest (primary) component of the hypermultiplet is an $SU(2)$ doublet of complex scalars $\phi_{\mathbf{a}}(x)$ whose two-point function is 
\begin{equation}\label{14}
  \langle\phi_{\mathbf{a}}(x_1)\bar\phi^{\mathbf{b}}(x_2)\rangle = \frac{\delta_{\mathbf{a}}{}^{\mathbf{b}}}{x^2_{12}}\,.
\end{equation}                                               
Introducing the internal variables from Section \ref{tcso3}, we can write down the doublet as a polynomial of degree one, $\phi(x,y) = \phi_{\mathbf{a}}(x) Y^{\mathbf{a}} = \phi_2(x) + y \phi_1(x)$. Analogously, the conjugate fields become $\widetilde{\phi}(x,y) = \bar\phi^{\mathbf{a}} \epsilon_{\mathbf{ab}} Y^{\mathbf{b}} = \bar\phi^1(x) - y\bar\phi^2(x)$, where we have used the same variables $Y^{\mathbf{a}}$ and not their conjugates. In this notation the two-point function (\ref{14}) takes the form
\begin{equation}\label{15}
  \langle{\phi}(x_1,y_1)\widetilde\phi(x_2,y_2)\rangle = \frac{Y_1\cdot Y_2}{x^2_{12}} =  \frac{y_{12}}{x^2_{12}}\,.
\end{equation} 
We can also define ${\cal N}=2$ $\half$-BPS composite operators with
primary states ${\cal O}^k = {\rm Tr}(\phi(x,y))^k$ of weight $k$ (or
isospin $2k$) whose two-point function is the $k$-th power of
(\ref{15}):
\begin{equation}\label{15'}
  \langle {\cal O}^k(x_1,y_1)\widetilde{{\cal O}}^k(x_2,y_2)\rangle = \frac{y_{12}^k}{{x}^{2k}_{12}}\,.
\end{equation}
The internal variables in (\ref{15'}) form a polynomial of degree $k$.  Three-point functions are defined by analogy with (\ref{3ptn4}).

The $d=4$ cases discussed above can immediately be generalized to $\half$-BPS states in $d=3,5,6$. The $d$-dimensional free scalar fields have canonical conformal dimension
\begin{equation}\label{varep}
  \varepsilon = \frac{d}{2}-1\,.
\end{equation}
Accordingly, the two-point function of the scalars from the elementary on-shell supermultiplet becomes
\begin{equation}\label{01711}
  \langle \phi^{\mathbf{i}}(x_1)\phi^{\mathbf{j}}(x_2) \rangle = \frac{\delta^{\mathbf{ij}}}{(x^2_{12})^{\varepsilon}}\,,
\end{equation} 
where $\mathbf{i,j}$ are indices of the corresponding R symmetry group. Multiplying by the appropriate vector (or spinor) internal variables $Y$ we obtain the two-point functions of elementary and composite states:
\begin{equation}\label{01800}
  \langle \phi(x_1,y_1)\phi(x_2,y_2) \rangle =\frac{Y_1\cdot Y_2}{(x^2_{12})^{\vep}} =  \frac{y^2_{12}}{(x^2_{12})^{\varepsilon}}\,.
\end{equation}

\subsubsection{Four-point functions}\label{fpf}

Let us first discuss the case $d=4$ ${\cal N}=4$. The four-point
correlator of the primary states of $\half$-BPS operators of weight
$k$ is obtained by connecting each point to the other three points
with $k$ elementary two-point functions (``propagators") in all
possible ways. This gives rise to $\frac{1}{2}(k+1)(k+2)$ propagator
structures, each having the required conformal and R symmetry properties. The
general amplitude is then given by a linear combination of these
structures with invariant coefficients depending on the conformal
cross-ratios $u,v$:
\begin{eqnarray}
  &&  \langle {\cal O}^k(x_1,y_1){\cal O}^k(x_2,y_2){\cal
  O}^k(x_3,y_3){\cal O}^k(x_4,y_4)\rangle^{{\cal N}=4} \nonumber\\
  && = \sum_{0\leq m+n \leq k} a_{mn}(u,v) \left(\frac{y^2_{12}y^2_{34}}{x^2_{12}x^2_{34}} \right)^{k-m-n} \left(\frac{y^2_{13}y^2_{24}}{x^2_{13}x^2_{24}} \right)^{m} \left(\frac{y^2_{14}y^2_{23}}{x^2_{14}x^2_{23}} \right)^{n}\,. \label{71}
\end{eqnarray}
The dependence on the internal variables $y$ is polynomial, which
reflects the fact that we are dealing with finite-dimensional
representations of $SU(4)\sim SO(6)$. The number of terms
$\frac{1}{2}(k+1)(k+2)$ in this sum is also the number of $SU(4)$
representations occurring in the tensor product of two representations
$[0,k,0]$. The polynomial of degree $k$ of two variables in (\ref{71})
provides a ``propagator basis" for such correlators. It is related,
although in a non-trivial way, to the OPE basis where the expansion
goes over $SU(4)$ representations - see Section \ref{CPWA}.

The weights of the four operators in (\ref{71}) do not have to be the
same. Considering a set of four different weights $k_1 \geq k_2 \geq
k_3 \geq k_4$, we can generalize  (\ref{71}) to the following
factorized form: 
\begin{eqnarray}
  && \langle {\cal O}^{k_1}(x_1,y_1){\cal O}^{k_2}(x_1,y_1){\cal
 O}^{k_3}(x_3,y_3){\cal O}^{k_4}(x_4,y_4)\rangle \nonumber\\
  &&{}\qquad\qquad\qquad =\langle {\cal
 O}^{k'_1}(x_1,y_1){\cal O}^{k'_2}(x_2,y_2) {\cal
 O}^{k'_3}(x_3,y_3)\rangle \nonumber\\
 && {}\qquad\qquad\qquad\quad \times  \langle{\cal O}^{k_4}(x_1,y_1){\cal O}^{k_4}(x_2,y_2){\cal
  O}^{k_4}(x_3,y_3){\cal O}^{k_4}(x_4,y_4)\rangle\,, \label{7171}
\end{eqnarray}
where $k'_{i} = k_{i}-k_4$. As we shall see later on, this generalization has no effect on the superconformal Ward identities, the latter concern only the coefficients of the correlator of weight $k_4$ in (\ref{7171}).

We remark that if the four operators ${\cal O}^k$ are considered identical, the amplitude is invariant under point permutations (``crossing symmetry"). This symmetry imposes certain relations among the coefficient functions $a_{mn}$. In the present paper we do not consider the implications of crossing symmetry.

It is convenient to rewrite the amplitude (\ref{71}) by pulling out a
single propagator factor of weight $k$,
\eqna{71prop}{
&& \langle {\cal O}^k(x_1,y_1){\cal O}^k(x_2,y_2){\cal
  O}^k(x_3,y_3){\cal O}^k(x_4,y_4)\rangle^{{\cal N}=4} \cr
&& \qquad\qquad\qquad\qquad\qquad = \left(\frac{y^2_{12}y^2_{34}}{x^2_{12}x^2_{34}}  \right)^{k}
  \sum_{0\leq m+n \leq k} a_{mn}(u,v)  \left(\frac{V}{v} \right)^{n}
  \left(\frac{u}{U} \right)^{m+n} \,.
}
This factor gives the amplitude the required conformal and R weight, while the sum is an invariant function of the cross-ratios (\ref{-1}), (\ref{-111}).

Further, we can go to the conformal and R symmetry frames (\ref{1'}),
(\ref{6812}), (\ref{68'12}), (\ref{68'12'}) and rewrite the amplitude
as a function of the invariant variables $z,z'$ and $w,\bw$:
\begin{equation}\label{73}
  \lim_{x_4,y_4\to\infty} \left(\frac{x^{2}_{4}}{y^2_4}\right)^k \langle {\cal O}^k(x_1,y_1){\cal O}^k(x_2,y_2){\cal
  O}^k(x_3,y_3){\cal O}^k(x_4,y_4)\rangle^{{\cal N}=4} = \left(\frac{y^2_{12}}{x^2_{12}} \right)^{k} G_k(z,z';w,\bw)\,,
\end{equation}
where 
\begin{equation}\label{7333}
  G_k(z,z';w,\bw) = \sum_{0\leq m+n \leq k} a_{mn}(z,z')\, \left(\frac{(1+w)(1+\bw)}{(1+z)(1+z')} \right)^m \left(\frac{zz'}{w\bw}\right)^{m+n}\,.
\end{equation}
The coefficient functions $a_{mn}(z,z')$ should satisfy the additional condition
\begin{equation}\label{symmetr}
  a_{mn}(z,z') = a_{mn}(z',z)\,.
\end{equation}
Indeed, initially these coefficients were functions of the cross-ratios, $a_{mn}(u,v)$. As pointed out earlier, in the change of variables (\ref{0202}) from $u,v$ to $z,z'$ there is a sign ambiguity. The choice of this sign is conventional and should not affect any physical quantity.  The same applies to the internal variables. So, we define
\begin{equation}\label{exchsym}
  G_k(z,z';w,\bw) = G_k(z',z;w,\bw) = G_k(z,z';\bw,w) \,.
\end{equation}
This requirement amounts to extending the superconformal transformations by global exchange symmetries \cite{HH}. 

The four-point amplitude is even simpler in the case $d=4$ ${\cal
N}=2$. Instead of (\ref{71}) we have
\begin{equation}\label{38'}
 \langle {\cal O}^k(x_1,y_1){\cal O}^k(x_2,y_2){\cal
  O}^k(x_3,y_3){\cal O}^k(x_4,y_4)\rangle^{{\cal N}=2}   = \sum^k_{n=0} a_{n}(z,z') \left(\frac{y_{12}y_{34}}{x^2_{12}x^2_{34}} \right)^{k-n}  \left(\frac{y_{14}y_{23}}{x^2_{14}x^2_{23}} \right)^{n}\,.
\end{equation}
The appearance of only two distinct propagator structures in
(\ref{38'}) follows from the cyclic identity (\ref{38''}). The sum has
$k+1$ terms which is also the number of $SU(2)$ representations
occurring in the tensor product of two representations of isospin
$k/2$ each. Further, we can pull out a propagator factor, e.g.,
$y_{12}^ky_{34}^k/x^{2k}_{12}x^{2k}_{34}$, which gives the correlator
the required conformal and $SU(2)$ weight. The rest of it then becomes
an invariant polynomial of degree $k$ in the single variable
\begin{equation}\label{012}
  \frac{y_{14}y_{23}}{y_{12}y_{34}}\ \frac{x^2_{12}x^2_{34}}{x^2_{14}x^2_{23}} \ \rightarrow \ \frac{zz'}{w}
\end{equation}
in the frame (\ref{1'}), (\ref{6812}), (\ref{frsu2}). So, in  this
frame we have
\begin{equation}\label{38}
  \lim_{x_4, y_4\to\infty} \left(\frac{x^{2}_{4}}{y_4}\right)^k \langle {\cal O}^k(x_1,y_1){\cal O}^k(x_2,y_2){\cal
  O}^k(x_3,y_3){\cal O}^k(x_4,y_4)\rangle^{{\cal N}=2} = \left(\frac{y_{12}}{x^2_{12}} \right)^k G_k(z,z';w)\,,
\end{equation}
where
\begin{equation}\label{73333}
  G_k(z,z';w)  = \sum^k_{n=0} a_n(z,z') \left(\frac{zz'}{w}\right)^n\,.
\end{equation}                                                  

The generalization of the above to other dimensions is obvious as we
just need to use the corresponding propagators. Thus, the four-point
amplitude (\ref{73}) becomes
\begin{equation}\label{73alld}
  \lim_{x_4,y_4\to\infty} \left(\frac{x^{2\varepsilon}_{4}}{y^2_4}\right)^k\langle {\cal O}^k(x_1,y_1){\cal O}^k(x_2,y_2){\cal
  O}^k(x_3,y_3){\cal O}^k(x_4,y_4)\rangle^{(\vep)} = \left(\frac{y^2_{12}}{x^{2\varepsilon}_{12}} \right)^{k} G^{(\varepsilon)}_k(z,z';w,\bw)\,,
\end{equation}
where 
\begin{equation}\label{7333alld}
  G^{(\varepsilon)}_k(z,z';w,\bw)  = \sum_{0\leq m+n \leq k} a_{mn}(z,z')\, \left(\frac{(1+w)(1+\bw)}{(1+z)^\varepsilon(1+z')^\varepsilon} \right)^m \left(\frac{z^\varepsilon {z'}^\varepsilon}{w\bw}\right)^{m+n}\,,
\end{equation}
and similarly for (\ref{73333}),
\begin{equation}\label{73333alld}
  G^{(\varepsilon)}_k(z,z';w)  = \sum^k_{m=0} a_m(z,z') \left(\frac{z^\varepsilon {z'}^\varepsilon}{w}\right)^n\,.
\end{equation}  
Note that (\ref{73333alld}) can be obtained from (\ref{7333alld}) by taking the limit $w'\to\infty$ and by identifying
\begin{equation}\label{identifying}
  a_m(z,z') = \sum^k_{n=m} C^m_n \frac{a_{n0}(z,z')}{(1+z)^{\varepsilon n}(1+z')^{\varepsilon n}}\,,
\end{equation}
where $C^m_n$ are the binomial coefficients, $(1+1)^n=\sum_m C^m_n$.

For the purposes of solving the superconformal Ward identities
(Sections \ref{Wisrs}, \ref{CRSTJP}) and for application to conformal
partial wave
expansion (Section \ref{CPWA}) it is convenient to make the following change of variables:
\begin{equation}\label{chvar}
  \chi = \frac{z}{1+z}\,, \qquad \chi' = \frac{z'}{1+z'}\,, \qquad \alpha = \frac{1+w}{w}\,, \qquad \alpha' = \frac{1+w'}{w'}\,.
\end{equation}                                                                                        
In terms of these we have
\begin{equation}\label{7333alld'}
  G^{(\varepsilon)}_k(\chi,\chi';\alpha,\alpha') = \sum_{0\leq m+n \leq k} a_{mn}(\chi,\chi')\, 
  \frac{(\chi\chi')^{\varepsilon(m+n)}}{[(1-\chi)(1-\chi')]^{\varepsilon n}}\, (\alpha\alpha')^m [(1-\alpha) (1-\alpha')]^n \end{equation}   
and, by setting $\alpha'=1$,
\begin{equation}\label{73333alld'}
  G^{(\varepsilon)}_k(\chi,\chi';\alpha)  = \sum^k_{m=0} a_m(\chi,\chi')\, (\chi\chi')^{\varepsilon m}\, \alpha^m \,,
\end{equation}  
where now $a_m(\chi,\chi') = a_{m0}(\chi,\chi')$.


\section{Superconformal symmetry and $\half$-BPS states}\label{sshbs}

\subsection{Superconformal covariants and invariants of the $\half$-BPS type}

The four-point amplitudes considered in Section \ref{fpf} transform as covariants of the conformal and R symmetry group of weight $k$ at each point. Our main task in this section is to complete them to full superconformal covariants. 

The crucial property of the two-, three- and four-point superconformal (co)invariants of the $\half$-BPS type is that their dependence on the Grassmann variables is uniquely fixed by superconformal symmetry, given the lowest (i.e., $\q=0$) component in their expansion. This is quite obvious from the following counting argument. A four-point Grassmann-analytic (co)invariant depends on four sets of odd variables $\q_{1,2,3,4}$, each having half of the components of a Lorentz and R symmetry spinor. On the other hand, the superconformal algebra has two sets of odd parameters, one for Q-supersymmetry and another for S-supersymmetry, each being a full spinor. Disregarding any possible singularities, we conclude that these parameters can be used to choose a frame in which all $\q_{1,2,3,4} = 0$. Inversely, ``undoing" this frame we obtain the unique superconformal completion of a conformal and R symmetry four-point (co)invariant of this type. Another way of putting this is to say that there exist no nilpotent four-point superconformal invariants of the Grassmann analytic type. Obviously, the same applies to (co)invariants with two or three points.

The simple counting argument above seems to imply that the four-point correlators of  $\half$-BPS operators are entirely determined by the properties of their lowest component, i.e., just by conformal invariance and R symmetry, and that superconformal symmetry is implemented by straightforward completion of the conformal (co)invariants to superconformal ones. This is not exactly true, and the reason is the singular nature of the superconformal transformations needed to remove all the $\q$s. As to the potential space-time singularities, we may postulate that we keep all four points $x_{1,2,3,4}$ apart, so that no such singularities occur. However, we are not allowed to treat the harmonic (R symmetry) singularities in the same way. The point is that the harmonic functions are {\it globally defined analytic functions} on the harmonic space. The latter is a complex compact coset of the R symmetry group. This property of {\it harmonic analyticity} guarantees that the harmonic dependence is just polynomial. In other words, we want do deal with finite-dimensional unitary representations of the compact R symmetry group which can be written down in the form of harmonic polynomials.

\subsection{Two- and three-point superconformal covariants}\label{tpsc}

The analytic (polynomial) dependence of the harmonic variables may
indeed be lost when we try to remove all the $\q$s. A simple example
illustrates this phenomenon very well - the two-point function of the elementary $d=4$ ${\cal N}=2$ on-shell superfield, the hypermultiplet. In Section \ref{tpf} we have found its form (\ref{15}) as a function of the space-time variables $x_{12}$ and as a {\it polynomial} in the harmonic variables $y_{12}$. The question now is to find its completion to a full superconformal covariant and to make sure that the polynomial dependence on the harmonic variables persists. 

The hypermultiplet can be described \cite{HSS} by a $\half$-BPS (or Grassmann analytic) superfield which, as its name suggests, depends only on half of the Grassmann variables $\q^\a{}_\mathbf{a}$ and $\bq^{\a'}{}_\mathbf{a}$. Here $\a,\a'$ are Lorentz spinor indices and $\mathbf{a}$ is a spinor index of the R symmetry group $SU(2)$. The half suitable for Grassmann analyticity is obtained by a harmonic projection of the type described in Section \ref{tcso3},
\begin{equation}\label{harmproq}
  \q^\a_+ = \q^\a{}_\mathbf{a} Y^\mathbf{a}\,, \qquad \bq^{\a'}_+ = \bq^{\a'}{}_\mathbf{a} Y^\mathbf{a}\,, 
\end{equation}
so that they only carry a $U(1)$ (or $GL(1,\mathbb{C})$ after complexification) weight. 

We do not need to study the full superconformal group for the supersymmetrization of the two-point function (\ref{15}). It is well known that the conformal two-point functions are determined essentially by Poincar\'e invariance together with dilatations (no need to use proper conformal invariance).   Similarly, their Grassmann-analytic superconformal extensions are completely fixed by Poincar\'e (or Q-) supersymmetry, R symmetry and dilatations (no need to use conformal, or S-supersymmetry). Adapting the counting argument from the beginning of this section to Q-supersymmetry alone, we see that a two-point function of the Grassmann analytic type depends on as many odd variables $(\q^\a_+)_{1,2}$, $(\bq^\a_+)_{1,2}$ as the number of Q-supersymmetry parameters, so it is indeed completely determined by Q-supersymmetry. 

Let us consider the left-handed part of Q-supersymmetry with parameters $\ep^\a{}_\mathbf{a}$. The transformations of the coordinates of the Grassmann analytic superspace have the following form: 
\begin{equation}\label{qt4}
  \delta x^{\a\a'} = \ep^\a_- \bq^{\a'}_+\,, \qquad \delta\q^{\a}_+ =  \ep^\a{}_\mathbf{a} Y^\mathbf{a} = \ep^{\a}_+ + y\ep^\a_-\,, \qquad \delta\bq^{\a'}_+ = \delta y = 0\,,
\end{equation}   
where $x^{\a\a'} = x^\mu (\sigma_\mu)^{\a\a'}$ and the Q-supersymmetry parameters $\ep^\a$ are projected with the internal variable $Y$, in accordance with (\ref{harmproq}). These transformations allow us to fix a special frame similar to the conformal and R symmetry frames discussed in Section \ref{csf},
\begin{equation}\label{19}
  \mbox{left-handed Q frame:}\qquad \q^{\a}_{2+} = 0\,.
\end{equation} 
In this frame we still have the residual Q-supersymmetry with parameters
\begin{equation}\label{19'}
 \ep^{\a}_+ = -y_2\ep^\a_- \,,
\end{equation}
under which the remaining $\q_{1+}$ undergoes the transformation
\begin{equation}\label{20}
  \delta_Q\q^{\a}_{1+} =  y_{12}\ep^\a_-\,. 
\end{equation}
It is then clear that if we wish to go a step further and eliminate
$\q_{1+}$ as well, we would create a harmonic singularity of the type
$y^{-1}_{12}$, and this would not be compatible with the requirement
of harmonic polynomial dependence (i.e., with harmonic analyticity). So, we must keep $\q_{1+}$ in order to have control on the harmonic singularities. 

The extension of $x_{12}$ to a Q-supersymmetry invariant is very easy to find\footnote{A similar construction for two- and three-point functions in $d=4$ $\N=1$ superspace has been considered in \cite{Osborn:cr}, and in \cite{Park} for $d=3,6$.} in the left-handed Q frame (\ref{19}), together with the analogous right-handed frame $\bq^{\a'}_{2+}=0$:
\begin{equation}\label{21'}
    \check{x}^{\a\a'}_{12} = x^{\a\a'}_{12} - {y^{-1}_{12}} \q^{\a}_{1+} \bq^{\a'}_{1+}\,.
\end{equation}  
When the two harmonic points coincide, $y_{12} = 0$, the supersymmetric extension (\ref{21'}) is clearly singular. What matters however is the singularity of the complete supersymmetrized two-point function (\ref{15}), \begin{equation}\label{22}
  \langle \phi(x_1,\q_1,\bq_1,y_1)\widetilde\phi(x_2,0,0,y_2) \rangle =  \frac{y_{12}}{\check{x}^2_{12}}\,.
\end{equation}
Expanding in $\q\bq$, we see that the harmonic numerator suppresses
the singularity of the denominator at the level $\q\bq$. At the top
level $(\q \bq )^2$ the space-time factor is $\square_1 x^{-2}_{12} =
0$ (recall that we neglect space-time singularities). Thus, the
supersymmetric two-point function (\ref{22}) is {\it free from
harmonic singularities}.\footnote{In fact, the top level in the
expansion of (\ref{22}) is a contact term, $y^{-1}_{12} \delta(x_{12})
(\q_1\bq_1)^2$. This is needed for the two-point function to become
the solution of the Green's function equation for the hypermultiplet
\cite{HSS}.}  

The full dependence on both $\q_1$ and $\q_2$ in (\ref{22}) can be restored by ``undoing" the frame fixing, i.e. by performing the inverse Q-supersymmetry transformation which lead to (\ref{19}) and to its right-handed analog. This does not change the behaviour of the two-point function at the coincident point $y_{12}=0$.

The supersymmetrization of the two-point functions (\ref{15'}) of
$\half$-BPS operators of weight $k$ is achieved in exactly the same
way, by just replacing $x$ by $\check{x}$:
\begin{equation}\label{2200}
  \langle {\cal O}^k(x_1,\q_1,\bq_1,y_1)\widetilde{{\cal O}}^k(x_2,0,0,y_2)\rangle = \frac{y_{12}^k}{\check{x}^{2k}_{12}}\,.
\end{equation} 

We have obtained the two-point functions (\ref{22}), (\ref{2200}) using  Q-supersymmetry alone. However, from the counting argument above we know that Q-supersymmetry is sufficient to guarantee the uniqueness of the two-point functions of the $\half$-BPS type. This implies that (\ref{22}), (\ref{2200}) automatically have the required transformation properties under the full superconformal symmetry.

The other elementary $\half$-BPS multiplet in  $d=4$, the on-shell ${\cal N}=4$ SYM multiplet, is treated similarly \cite{HoweWest}. Here the harmonic variables form an $SO(4)\sim SU(2)\times SU(2)$ vector $y^{aa'}$ and the odd variables of the Grassmann analytic superspace $\q^{\a a'}$, $\bq^{a\a'}$ carry Lorentz and $SU(2)$ spinor indices of both types. Now (\ref{21'}) becomes
\begin{equation}\label{21}
   \check{x}^{\a\a'}_{12} = x^{\a\a'}_{12} -   \q^{\a a'}_1 (y_{12})^{-1}_{a'a}\bq^{a\a'}_1 
\end{equation}
and the supersymmetrization of the two-point function (\ref{018}) is
\begin{equation}\label{22'}
  \langle \phi(x_1,\q_1,\bq_1,y_1)\phi(x_2,0,0,y_2) \rangle =  \frac{y^2_{12}}{\check{x}^2_{12}}\,.
\end{equation} 
Expanding in $\q_1\bq_1$, we see that the harmonic numerator suppresses the singularity of the denominator up to the level $(\q_1\bq_1)^2$. At the next levels the space-time factor contains $\square_1 x^{-2}_{12} = 0$. Thus, the supersymmetric two-point function (\ref{22'}) is free from harmonic singularities.

Three-point superconformal correlators of $\half$-BPS states are obtained by simply replacing $x^2_{ij} \to \check{x}^2_{ij}$ in (\ref{3ptn4}). Once more, counting the number of $\q$s versus the number of Q- and S-supersymmetry parameters, we see that this supersymmetric extension is unique. Moreover, it is also free from harmonic singularities, since each propagator factor in it has this property.

Finally, the generalization of the above construction to $d=3,5,6$ is straightforward and is left to the reader.\footnote{We just mention that the harmonic superspace formulation of the $d=6$ $(2,0)$ self-dual tensor multiplet was given in \cite{Howe6d}. The properties of the two- and three-point functions of $\half$-BPS operators for general dimensions have been discussed in \cite{EFS,fersok}, also in a harmonic superspace framework.}

\subsection{Four-point superconformal invariants}\label{fpsc}

The four-point amplitude (\ref{71prop}) is a conformal and R symmetry covariant of weight $k$ at each point. Actually, the necessary weight is carried by the propagator factor, the coefficient functions $a_{mn}(u,v)$ are invariant. The supersymmetrization of this amplitudes goes in two steps. Firstly, we replace $x^2_{ij} \to \check{x}^2_{ij}$ in the propagator factor, as explained in Section \ref{tpsc}. This step does not create any harmonic singularities since the propagators are free from them. Secondly, we have to complete the space-time $u,v$ and harmonic $U,V$ cross-ratios to full superconformal invariants. 

Instead of the cross-ratios, it is easier to supersymmetrize the alternative variables $z,z'$ and $w,\bw$. A systematic procedure for constructing such four-point invariants has been proposed in \cite{HH}.\footnote{In Appendix \ref{AppA} we apply a version of this procedure to one particular case, $d=4$ ${\cal N}=4$, in order to answer a question raised in Section \ref{SCWIs}.} However, as explained in Section \ref{SCWIs}, for the purpose of deriving the superconformal Ward identities we need only the linearized (lowest order in the $\q$ expansion) invariants. This is easily done by examining the linearized supersymmetry transformations and finding the appropriate $\q\q$ terms which compensate the variations of $z,z'$ and $w,\bw$. We find it very useful to generalize the conformal and R symmetry frames of Section \ref{csf} to the superconformal case.  

Let us now look into the details, case by case.

\subsubsection{$\mathbf{d=6}$, $\mathbf{{\cal N}=(2,0)}$ and $\mathbf{{\cal N}=(1,0)}$}

Here we are considering the superalgebra $OSp(8^*|2\N)$ with
$\N=1,2$, the supersymmetric extension of the $d=6$ conformal group is $SO(2,6)$. The corresponding space-time coordinates form an $SO(1,5)$ vector $x^\mu$. It is convenient to rewrite $x^\mu$ as a bispinor (matrix) by contracting it with gamma matrices,
\begin{equation}\label{xd6}
  x^{\a\b} = -x^{\b\a} = x^\mu (\Gamma_\mu)^{\a\b}\,.
\end{equation}
We recall that $d=6$ supersymmetry is chiral, so only  gamma  matrices of one chirality are used.\footnote{The $d=6$ case is a good illustration of how inadequate the term ``Chiral Primary Operators",  commonly used to denote $\half$-BPS operators, is. In $d=6$ every supermultiplet is chiral. The characteristic feature of the $\half$-BPS superfields is that they depend on half of the odd coordinates of this chiral superspace.} The spinor indices $\a,\b$ belong to the fundamental representation of $SU^*(4)\sim SO(1,5)$. The R symmetry group in the case ${\cal N}=(2,0)$ is $USp(4) \sim SO(5)$, so according to Section \ref{rsiv} we need an $SO(3)$ vector coordinate $y^i$. The latter can be cast in matrix form,
\begin{equation}\label{yd6}
  y^{ab}= y^{ba} = y^i (\gamma^i)^{ab}
\end{equation}
where the spinor indices $a,b$ belong to the fundamental representation of $SU(2)$ and $\gamma^i$ are the $SO(3)$ gamma matrices, i.e., the three symmetric Pauli matrices. The odd coordinates of the Grassmann analytic superspace $\q^{\a a}$, corresponding to $\half$-BPS multiplets, carry both space-time ($SU^*(4)$) and internal ($SU(2)$) spinor indices. Note that the full $d=6$  ${\cal N}=(2,0)$ superspace contains twice this number of odd variables, forming a spinor of $USp(4)$.  

In this paper we work in a frame where $x_2^\mu=(1,0,\ldots,0)$ and  $y^i_2=(1,0,0)$. The matrix form of these fixed vectors is obtained by making a convenient choice of the relevant gamma matrices in the space-time and internal sectors:
\begin{equation}\label{Gammad6}
  \Gamma_0 = 
  \begin{pmatrix}
    0 & -\sigma_0 \\
    \sigma_0 & 0 
  \end{pmatrix}\,,  \qquad \gamma_1=\sigma_0=\delta^{ab}\,.
\end{equation}
In this frame the residual Lorentz symmetry is $SO(5)$ and the R symmetry is further broken from $SO(3)$ down to $SO(2)$, so the position of the indices $a,b$ is now irrelevant. 

In what follows we need the  Q- and S-supersymmetry transformations of the superspace coordinates. In Section  \ref{csf} we fixed conformal and R symmetry four-point frames in which the independent even variables are the coordinates at point 1. In the supersymmetric version of these frames we can eliminate all the odd variables but $\q_1$:
\begin{equation}\label{oddframed6}
  \q^{\a a}_2=\q^{\a a}_3=(x^{-1}_4)_{\a'\a}\q^{\a a}_4=0\,.
\end{equation}
This is done by using 3/4 of the Q- and S-supersymmetry transformations. The remaining 1/4 supersymmetry should be compatible with this choice. It is not hard to see that the residual supersymmetry transformations of the independent coordinates $x=x_1-x_2$, $y=y_1-y_2$ and $\q=\q_1$ should have the form\footnote{The complete superconformal transformations can be derived in the supermatrix approach of \cite{HH}. After fixing the corresponding frame they reduce to (\ref{trcood6}). The point we want to make here is that working directly in this frame allows us to obtain all that we need in just a few elementary steps.}
\begin{eqnarray}
  \delta x^{\a\b} &=& -\ep^{\a b}\q^{\b b} - \q^{\a b}\ep^{\b b} \nonumber\\
  \delta y^{ab }&=& \ep^{\a a} (\tilde\Gamma_0)_{\a\b} \q^{\b b} + \q^{\a a} (\tilde\Gamma_0)_{\a\b} \ep^{\b b} \label{trcood6}\\
  \delta \q^{\a a} &=& y^{ab}\ep^{\a b} - x^{\a\b}(\tilde\Gamma_0)_{\b\gamma}\ep^{\gamma a}\,, \nonumber
\end{eqnarray}  
where $(\tilde\Gamma_0)_{\a\b} = -\Gamma_0^{\a\b}$, $\Gamma_0\tilde\Gamma_0 = \mathbb{I}$. Indeed, from (\ref{Gammad6}) it is clear that $\delta \q_2=\sigma_0\ep - \Gamma_0\tilde\Gamma_0\ep = 0$. In addition, since by fixing the conformal and R symmetry frames we have already used most of the bosonic symmetry (translations, conformal boosts, dilatations, part of $SO(1,5)$ and $SO(5)$), only the residual $SO(5)\times SO(2)$ should appear in the commutator of two supersymmetry transformations (\ref{trcood6}). This explains the linear form of the transformations (\ref{trcood6}) and even fixes the coefficients, up to a simultaneous rescaling of $\q$ and $\ep$.

We now recall that the space-time and internal space frames can be further specified as shown in (\ref{6812}), (\ref{68'12}).  The choice 
\begin{equation}\label{Gammad6'}
  \Gamma_1 = 
  \begin{pmatrix}
    0 & -\sigma_3 \\
    \sigma_3 & 0 
  \end{pmatrix}\,, \qquad \gamma_2 = \sigma_3 \,,
\end{equation}      
makes the matrices $x$ and $y$ diagonal:
\begin{equation}\label{diagformd6}
  x^\a{}_\b \equiv x^{\a\gamma}(\tilde\Gamma_0)_{\gamma\b} = \left(
  \begin{array}{rr}
    {\begin{array}{rr} z  &    \\   
                          &  z'\end{array}} 
                                        & {\makebox[1cm][c]{\Huge$0$}} \\
     {\makebox[1cm][c]{\Huge $0$}}\quad & \!\!\!\!\!\!\!\!\!\! \!\!\!\!\!\!
                                          {\begin{array}{cc} z  &    \\   
                                                                &  z'\end{array}}
  \end{array}
     \right)\,, \qquad y^{ab} = 
  \begin{pmatrix}
    w & 0 \\
    0 & \bw 
  \end{pmatrix}
     \,.
\end{equation}

This choice casts the supersymmetry transformation of $\q$ into a very suggestive form:
\begin{eqnarray}
  \delta\q^1 &=& \left(
 \begin{array}{rr}
    {\begin{array}{rr} w-z  &    \\   
                          &  w-z'\end{array}} 
                                        & {\makebox[1cm][l]{\Huge$0$}}\quad \\
     {\makebox[1cm][l]{\Huge $0$}}\quad\quad & \!\! \!\!\!\!\!\!
                                          {\begin{array}{cc} w-z  &    \\   
                                                                &  w-z'\end{array}}
  \end{array} 
     \right) \ep^1 + (\ep\q\q)\,;\nonumber\\
  \delta\q^2 &=& \left(
 \begin{array}{rr}
    {\begin{array}{rr} \bw-z  &    \\   
                          &  \bw-z'\end{array}} 
                                        & {\makebox[1cm][l]{\Huge$0$}}\quad\,\,\,\, \\
     {\makebox[1cm][l]{\Huge $0$}}\quad\quad\,\, & \!\! \!\!\!\!\!\!
                                          {\begin{array}{cc} \bw-z  &    \\   
                                                                &  \bw-z'\end{array}}
  \end{array}  
     \right) \ep^2  + (\ep\q\q) \,, \label{castsusyd6}
\end{eqnarray}
where $\q^{1,2}$ are the two projections of the R symmetry doublet and $(\ep\q\q)$ denotes non-linear terms to be discussed below. We see that each component of $\q^\a$ is shifted by the corresponding parameter $\ep^\a$ multiplied by the difference of an internal ($w,\bw$) and a space-time ($z,z'$) invariant variables. Were it not for these differences which might vanish, we could proceed to a further frame fixing where all the $\q$s are eliminated (recall the counting argument from the beginning of Section \ref{sshbs}). However, the corresponding supersymmetry transformations are clearly singular when, e.g., $w\to z$, etc. The requirement of absence of such singularities is the origin of the superconformal Ward identities, as explained in Section \ref{SCWIs}.

Next, we give the transformations of the even variables:
\begin{eqnarray}
  \delta z &=& \ep^{1 a} \q^{3a} + \q^{1 a}\ep^{3a} \nonumber\\
  \delta z' &=& \ep^{2a} \q^{4a} + \q^{2 a}\ep^{4a} \label{evvartr}\\
  \delta w &=& -2\ep^{\a 1} (\tilde\Gamma_0)_{\a\b}\q^{\b 1} \nonumber\\
  \delta \bw &=& -2\ep^{\a 2} (\tilde\Gamma_0)_{\a\b}\q^{\b 2}\,.  \nonumber
\end{eqnarray}

We stress that in (\ref{castsusyd6}) we only see the linearized part of the supersymmetry transformations of the $\q$s. Indeed, the particular choice (\ref{diagformd6}) is not invariant under the transformations (\ref{trcood6}). In order to restore the diagonal form (\ref{diagformd6}) we need to make additional, compensating $SO(5)\times SO(2)$ transformations.  For instance, in the space-time sector the $SO(5)$ transformation is given by the matrix $A^\a{}_\b$, $A^\a{}_\a=0$ with entries, e.g., 
\begin{equation}\label{compenslor}
  A^1{}_2  = {\frac{1}{z' - z}}(\ep^{1a} \q^{4a} + \q^{1 a}\ep^{4a})\,,\quad {\rm etc.}
\end{equation}
This does not affect the transformation of $z,z'$ in (\ref{evvartr}), but creates new, non-linear terms of the type $A^\a{}_\b\q^{\b a}$ in $\delta\q^{\a a}$ (\ref{castsusyd6}). The compensating transformation in the internal sector has a similar effect. It is important that these non-linear terms do not create new singularities of the $1/(w-z)$, etc. type (notice, however, the singularity $1/(z' - z)$).

Now we are ready to construct the superconformal completions of the variables $z,z',w,\bw$. Inspecting the linearized transformations (\ref{castsusyd6}), (\ref{evvartr}), we see that the following expressions:
\begin{eqnarray}
  \hat z &=& z - \frac{\q^{11}\q^{31}}{w-z} - \frac{\q^{12}\q^{32}}{\bw-z} + O((\q)^4)\nonumber\\
  \hat{z}' &=& z' - \frac{\q^{21}\q^{41}}{w-z'} - \frac{\q^{22}\q^{42}}{\bw-z'} + O((\q)^4) \label{hatd6}\\
  \hat w &=& w - 2\frac{\q^{11}\q^{31}}{w-z} - 2\frac{\q^{21}\q^{41}}{w-z'}  + O((\q)^4)\nonumber\\
  \hat{w}' &=& \bw - 2\frac{\q^{12}\q^{32}}{\bw-z} - 2\frac{\q^{22}\q^{42}}{\bw-z'} + O((\q)^4) \nonumber
\end{eqnarray}
are invariant to lowest order in $\q$,
\begin{equation}\label{invlow}
  \delta\hat z = \delta\hat{z}' =\delta\hat w = \delta\hat{w}' = O(\ep (\q)^3)\,.
\end{equation}     
To obtain the full non-linear invariants we ought to take into account the compensating transformations (\ref{compenslor}), as well as to transform the denominators in (\ref{hatd6}), which leads to higher-order terms. This process terminates after a few steps, but is rather cumbersome. There exists a more efficient method for constructing the full invariants (see \cite{HH} and Appendix \ref{AppA}). However, as explained in Section \ref{SCWIs}, for the purposes of deriving the superconformal Ward identities it is sufficient to know the linearized form (\ref{hatd6}).

The most important feature of the supersymmetric extensions (\ref{hatd6}) is their singular behavior when, e.g., $w\to z$, etc. Our main task in Section \ref{SCWIs} will be to find the conditions for suppressing these singularities in the expansion of the four-point amplitude. In fact, we already know some particular combinations of the variables (\ref{hatd6}) which should be free form such singularities. Indeed, as shown in Section \ref{tpsc}, the supersymmetrized propagators are not singular. Combining four six-dimensional propagators, we can construct conformal and R symmetry invariants of the cross-ratio type, e.g., 
\begin{equation}\label{explstr}
  \frac{y^2_{12}y^2_{34}}{y^2_{14}y^2_{23}} \frac{x^4_{14}x^4_{23}}{x^4_{12}x^4_{34}} \ \rightarrow\ \frac{y^2_{12}}{x^4_{12}} = \frac{w\bw}{(zz')^2}\,, 
\end{equation}
where we have used the conformal and R symmetry frames (\ref{1'}), (\ref{68'12}). We have already seen such invariants in the expansion of the amplitude $G^{(\varepsilon)}_k$ (\ref{7333alld}) (here $\varepsilon=2$). The supersymmetrization of (\ref{explstr}) can be achieved by replacing $x_{12}\to \check{x}_{12}$, as explained in Section \ref{tpsc}. Alternatively, since the four-point superconformal invariants of the $\half$-BPS type are unique, this should be equivalent to supersymmetrizing the variables $w,\bw,z,z'$ in (\ref{explstr}) according to (\ref{hatd6}):
\begin{equation}\label{equivhats}
  \frac{y^2_{12}}{\check{x}^4_{12}} = \frac{\hat{w}\hat{w}'}{(\hat{z}\hat{z}')^2}\,.
\end{equation}
Now, we know from Section \ref{tpsc} that the left-hand side of (\ref{equivhats}) is free from harmonic singularities, i.e., it is well defined for any value of the ``running" variable $y_1$ ($y_2$ is fixed in the frame  (\ref{68'12})). Consequently, the right-hand side of (\ref{equivhats}) should also be well defined for any values of $w,w'$. In particular, the singularity, e.g.,  $(w-z)^{-1}$ should be absent. To check this we pick out the singular terms
\begin{eqnarray}
  \Delta z &=& \hat z - z = - \frac{\q^{11}\q^{31}}{w-z} + \mbox{reg. terms}\,, \nonumber\\
  \Delta w &=& \hat w - w = - 2\frac{\q^{11}\q^{31}}{w-z} + \mbox{reg. terms}\,, \label{singterm}
\end{eqnarray}
i.e., $\Delta w_{\scriptsize\rm sing} = 2 \Delta z_{\scriptsize\rm sing}$ in this particular regime. Then the potentially singular factors in the right-hand side of (\ref{equivhats}) combine into the regular term
\begin{equation}\label{regdelta}
  (\Delta z_{\scriptsize\rm sing} \pa_z + \Delta w_{\scriptsize\rm sing} \pa_w) \left(\frac{w}{z^2}\right) = \frac{2(z-w)}{z^3}\Delta z_{\scriptsize\rm sing} = \frac{2}{z^3}\q^{11}\q^{31}  
\end{equation}
(recall that here we only construct linearized superconformal invariants). The same applies to the other singularities present in (\ref{hatd6}).

Anticipating the discussion of the other dimensions below, we can state that (\ref{regdelta}) generalizes to  \begin{equation}\label{regdeltagen}
  \Delta w_{\scriptsize\rm sing} = \varepsilon \Delta z_{\scriptsize\rm sing} \sim (w-z)^{-1}\,, \qquad (\Delta z_{\scriptsize\rm sing} \pa_z + \Delta w_{\scriptsize\rm sing} \pa_w) \left(\frac{w}{z^\varepsilon}\right) = \mbox{regular} 
\end{equation} 
in the singular regime $w\to z$, and similarly for the other singular limits.

Finally, the case ${\cal N}=(1,0)$ is easily obtained by reduction. There the R symmetry group is $USp(1)\sim SU(2)$, so we have a single complex variable $w=y_1-y_2$ in the internal sector. The reduction is done by suppressing the R symmetry indices $a,b$ and by setting $\bw=\infty$ in the equations above. In particular, we find
\begin{eqnarray}
  \hat z &=& z - \frac{\q^{1}\q^{3}}{w-z}  + O((\q)^4)\nonumber\\
  \hat{z}' &=& z' - \frac{\q^{2}\q^{4}}{w-z'}  + O((\q)^4) \label{hatd610}\\
  \hat w &=& w - 2\frac{\q^{1}\q^{3}}{w-z} - 2\frac{\q^{2}\q^{4}}{w-z'}  + O((\q)^4)\,.\nonumber
 \end{eqnarray}

\subsubsection{$\mathbf{d=5}$}

Here we are considering the exceptional superalgebra $F(4)$, the unique supersymmetric extension of the $d=5$ conformal group is $SO(2,5)$, with R symmetry $USp(1)\sim SU(2)$.
This case can be treated in close analogy with the case $d=6$ ${\cal N}=(1,0)$ above, we just need to adapt the supersymmetry transformations (\ref{trcood6}). The reduction from the $d=6$ Lorentz group $SO(1,5)$ to the $d=5$ one $SO(1,4)$ is done using the $SO(1,4)$ invariant symplectic matrix
\begin{equation}\label{sympm}
  \Omega^{\a\b} = -\Omega_{\a\b} =
  \begin{pmatrix}
    0 & \sigma_2 \\
    \sigma_2 & 0 
  \end{pmatrix} =  \Gamma_5
  \,, \qquad \Omega_{\a\b}\Omega^{\b\gamma} = \delta_\a{}^\gamma\,,
\end{equation} 
so that the $d=5$ matrix $x^{\a\b}$ is symplectic-traceless, $x^{\a\b}\Omega_{\b\a}=0$. Then (\ref{trcood6}) becomes
\begin{eqnarray}
  \delta x^{\a\b} &=& -\ep^\a\q^{\b} - \q^{\a}\ep^\b - \frac{1}{2}\Omega^{\a\b}\;\ep^\gamma \Omega_{\gamma\delta} \q^\delta\nonumber\\
\delta y &=& \frac{3}{2} \ep^{\a} (\tilde\Gamma_0)_{\a\b} \q^{\b}\label{trcood5}\\
  \delta \q^\a &=& y\ep^\a - x^{\a\b}(\tilde\Gamma_0)_{\b\gamma}\ep^\gamma\,, \nonumber
\end{eqnarray} 
so that $\delta x^{\a\b}\Omega_{\b\a}=0$ and the frame  $\q_2=0$, $x_2 = \Gamma_0$, $y_2=1$ is preserved. The coefficient $3/2$ in $\delta y$ is fixed from the requirement that the commutator of two supersymmetry transformations produce an $SO(4)$ rotation of $x$ and $\q$ (this commutator vanishes on $y$ itself, since we have broken down the R symmetry completely by setting $y_2=1$).

After this the case $d=5$ becomes very similar to the case $d=6$ ${\cal N}=(1,0)$. In particular, we find
\begin{equation}\label{trthetad5}
  \delta\q = \left(
  \begin{array}{rr}
    {\begin{array}{rr} w-z  &    \\   
                          &  w-z'\end{array}} 
                                        & {\makebox[1cm][l]{\Huge$0$}}\quad \\
     {\makebox[1cm][l]{\Huge $0$}}\quad\quad & \!\! \!\!\!\!\!\!
                                          {\begin{array}{cc} w-z  &    \\   
                                                                &  w-z'\end{array}}
  \end{array} 
     \right) \ep  + (\ep\q\q) \,;
\end{equation}             
\begin{eqnarray}
  \delta z &=& \ep^1 \q^{3} + \q^{1}\ep^3\,,\nonumber\\
  \delta z' &=& \ep^2 \q^{4} + \q^{2}\ep^4\,,\label{evvartrd5}
\end{eqnarray}   
so the superconformal completion of the variables $z,z'$ is
\begin{eqnarray}
  \hat z &=&  z - \frac{\q^{1}\q^{3}}{w-z} + O((\q)^4)\nonumber\\
  \hat{z}' &=&  z' - \frac{\q^{2}\q^{4}}{w-z'} + O((\q)^4)\label{hatd5}\\
  \hat w &=& w - \frac{3}{2}\frac{\q^{1}\q^{3}}{w-z} - \frac{3}{2}\frac{\q^{2}\q^{4}}{w-z'} + O((\q)^4)\,.\nonumber
\end{eqnarray}
As a consistency check we may consider the invariant (cf. (\ref{012}) for $d=4$)
\begin{equation}\label{regcr}
 \frac{y_{14}y_{23}}{y_{12}y_{34}}\ \left(\frac{x^2_{12}x^2_{34}}{x^2_{14}x^2_{23}}\right)^{3/2} \ \rightarrow \ {w}^{-1}{(zz')^{3/2}}\,.
\end{equation}
It is easy to see that this combination remains regular if we use the variables (\ref{hatd5}), according to the general statement (\ref{regdeltagen}). Notice that the only difference between (\ref{hatd610}) for the case $d=6$ $\N=(1,0)$ and (\ref{hatd5}) for the case $d=5$ is in the coefficient in $\hat w$. 

\subsubsection{$\mathbf{d=4}$, $\mathbf{{\cal N} = 4}$ and $\mathbf{{\cal N} = 2}$}

Here we are considering the superalgebras  $PSU(2,2|4)$ and  $SU(2,2|2)$, the supersymmetric extensions of the $d=4$ conformal group $SO(2,4)$, with R symmetry $SU(4)\sim SO(6)$ and $SU(2) \sim SO(3)$, respectively. 

The case $d=4$ ${\cal N}=4$ is similar to the case $d=6$ ${\cal N}=(2,0)$. We choose
\begin{equation}\label{chogamd4}
  \Gamma_0 = \sigma_0 = \delta^{\a\a'} \,, \quad \Gamma_1 = \sigma_3\,; \qquad  \gamma_1 = \sigma_0 = \delta^{aa'} \,, \quad \gamma_2 = \sigma_3\,,
\end{equation}
so primed and unprimed spinor indices (space-time and internal) become equivalent and their position (up or down) is irrelevant. The even coordinate matrices become diagonal\footnote{The fact that the variables $z,z'$ and $w,w'$ appear as the eigenvalues of the matrices $x^{\a\a'}$ and $y^{ab}$ was first pointed out in \cite{HH}.}
\begin{equation}\label{xind4}
  x^{\a\b} \equiv x^{\a\b}_{12} = 
  \begin{pmatrix}
    z & 0 \\
    0 & z' 
  \end{pmatrix}\,, \qquad  y^{ab} \equiv y^{ab}_{12}= 
  \begin{pmatrix}
    w & 0 \\
    0 & \bw 
  \end{pmatrix}\,. \end{equation}
The supersymmetry transformations involve two parameters, a left- and a right-handed one:
\begin{eqnarray}
  \delta x^{\a\b} &=& \ep^\a{}_b\bq^{b\b} + \q^{\a b}\bar\ep^{b}{}_\b \nonumber\\
  \delta y^{ab} &=& -\bar\ep^a{}_\a \q^{\a b}  + \bq^{a\b}\ep^\b{}_b  \label{trcood4}\\
  \delta \q^{\a b} &=& \ep^\a{}_a y^{ab} - x^{\a\b}\ep^\b{}_b\,, \nonumber \\
  \delta \bq^{a \b} &=& -\bar\ep^a{}_\a x^{\a\b} + y^{ab}\bar\ep^b{}_\b \,.  \nonumber
\end{eqnarray}
Once again, the form of these transformations follows from the preservation of the frame $\q_2=\bq_2=0$, $x_2=y_2=\sigma_0$ and from the requirement that their commutator reduce to $SO(3)\times SO(3)$ transformations of the coordinates.  In the fixed frame (\ref{xind4}) we obtain the transformations
\begin{eqnarray}
   \delta \left(
  \begin{array}{ll}
      \q^{11}  &  \q^{12} \\
      \q^{21}  &  \q^{22}   
  \end{array}
     \right) &=& \left(
  \begin{array}{rr}
      (w-z)\ep^1{}_1  &  (\bw-z)\ep^1{}_2 \\
      (w-z')\ep^2{}_1  &  (\bw-z')\ep^2{}_2   
  \end{array}
     \right)  + (\ep\q\q) \nonumber\\
  \delta \left(
  \begin{array}{ll}
      \bq^{11}  &  \bq^{12} \\
      \bq^{21}  &  \bq^{22}   
  \end{array}
     \right)  &=& \left(
  \begin{array}{rr}
      (w-z)\bar\ep^1{}_1  &  (w-z')\bar\ep^1{}_2 \\
      (\bw-z)\bar\ep^2{}_1  &  (\bw-z')\bar\ep^2{}_2   
  \end{array}
     \right)  + (\ep\q\q) \label{castsusyd4}
\end{eqnarray}   
along with  
\begin{eqnarray}
  \delta z &=& \ep^1{}_a \bq^{a1} + \q^{1 a}\bar\ep^a{}_{1}\nonumber\\
  \delta z' &=& \ep^2{}_a \bq^{a2} + \q^{2 a}\bar\ep^a{}_2\label{evvartrd4}\\
  \delta w &=& -\bar\ep^1{}_\a\q^{\a1} + \bq^{1\a}\ep^\b{}_1 \nonumber\\
  \delta \bw &=& -\bar\ep^2{}_\a\q^{\a2} + \bq^{2\a}\ep^\b{}_2 \,.\nonumber
\end{eqnarray}            

This yields the linearized superconformal completions
\begin{eqnarray}
  \hat z &=& z - \frac{\q^{11}\bq^{11}}{w-z} - \frac{\q^{12}\bq^{21}}{\bw-z} + O((\q)^4)\nonumber\\
  \hat{z}' &=& z' - \frac{\q^{21}\bq^{12}}{w-z'} - \frac{\q^{22}\bq^{22}}{\bw-z'} + O((\q)^4) \label{hatd4}\\
  \hat w &=& w - \frac{\q^{11}\bq^{11}}{w-z} - \frac{\q^{21}\bq^{12}}{w-z'} + O((\q)^4)\nonumber\\
  \hat{w}' &=& \bw  - \frac{\q^{12}\bq^{21}}{\bw-z}- \frac{\q^{22}\bq^{22}}{\bw-z'} + O((\q)^4)\,.\nonumber
\end{eqnarray} 
We note that the general relation (\ref{regdeltagen}) holds in the singular regime $w\to z$.

Finally, the reduction from ${\cal N}=4$ to ${\cal N}=2$ is achieved by suppressing the R symmetry indices $a,b$ and by setting $\bw=\infty$. In this way we find
\begin{eqnarray}
  && \hat z = z - \frac{\q^{1}\bq^{1}}{{w}-z} + O((\q)^4)  \nonumber\\
  && \hat{z}' = z' - \frac{\q^{2}\bq^{2}}{{w}-z'} + O((\q)^4) \label{hatd4n2}\\
  && \hat w = w - \frac{\q^{1}\bq^{1}}{{w}-z}  - \frac{\q^{2}\bq^{2}}{{w}-z'} + O((\q)^4)  \,.   \nonumber
\end{eqnarray}

\subsubsection{$\mathbf{d=3}$, $\mathbf{{\cal N}=8}$}

Here we are considering the superalgebra   $OSp(8|4,\mathbb{R})$, the supersymmetric extensions of the $d=3$ conformal group $SO(2,3)$, with R symmetry $SO(8)$.     
This case is the ``mirror" of the case $d=6$ ${\cal N}=(2,0)$. Indeed, now the space-time variables form an $SO(1,2)$ vector, similar to the $SO(3)$ internal vector in $d=6$. The internal variables form an $SO(6)$ vector which is the analog of the $SO(1,5)$ space-time vector in $d=6$. The odd variables are the same, with the internal and space-time indices interchanged.

\section{Superconformal Ward identities}\label{SCWIs}

The supersymmetrization of the four-point correlators (\ref{73alld}), (\ref{7333alld}), (\ref{73333alld}) goes in two steps. Firstly, we replace $x^2_{ij}\to \check{x}^2_{ij}$ in the propagator factors, as explained in Section \ref{tpsc}. This step does not create any harmonic singularities since the propagators are free from them. The non-trivial step is replacing the arguments $z,z',w,\bw$ of the invariant amplitude $G^{(\varepsilon)}_k$ by the (linearized) superconformal invariants found in Section \ref{fpsc}. The potential problem are the singularities present in (\ref{hatd6}), (\ref{hatd610}), (\ref{hatd5}), (\ref{hatd4}), (\ref{hatd4n2}). They occur when the internal variables $w,\bw$ approach the space-time variables $z,z'$. It is clear that such singularities must not take place. Indeed, the amplitude (\ref{7333alld}) is well defined for any value of $w,\bw$ (in fact, at any point in the space of the auxiliary internal variables $y$) and this must be so for its supersymmetric partners. Thus, we have to make sure that these singularities are suppressed. 

In what follows we restrict ourselves to four-point functions of $\half$-BPS operators of equal weights $k$. The generalization to four different weights is trivial. According to (\ref{7171}), it consists in multiplying the four-point amplitude with equal weights by a three-point function. The supersymmetrization of the latter is straightforward and does not create or remove any singularities. 

\subsection{Ward identities and their solution in the case $d=4$ ${\cal N}=2$}\label{win2}

The superconformal completions of the invariant variables (\ref{hatd4n2}) become singular either when ${w}\to z$ or when ${w}\to z'$. Let us assume for the time being that $z\neq z'$. This allows us to treat the two singularities independently. The linearized ${\cal N}=2$ transformations of the $\q$s are obtained from (\ref{castsusyd4}):
\begin{equation}\label{013}
  \delta\q^1 = \left({w}-z \right)\ep^1 + (\ep\q\q)\,, \qquad  \delta\q^2 = \left({w}-z'\right)\ep^2 + (\ep\q\q) \,.
\end{equation}
We see that when, for instance, ${w}\to z' \neq z$ we are allowed to make an additional superconformal transformation which sets $\q^1=0$.\footnote{Strictly speaking, we need to know whether the non-liner terms in (\ref{013}) bring in new singularities of the same type. Adapting the $d=6$ discussion around eq. (\ref{compenslor}), we can see that the singularity in the non-liner terms is of a different type, $(z-z')^{-1}$, which we do not consider for the moment.} In this frame $\Delta z= \hat z - z =0$ and the only source of potential singularities are the nilpotent terms 
\begin{equation}\label{014}
  \Delta z' = \Delta w = -\frac{\q^2\bq^2}{{w}-z'}\,, \qquad (\Delta z')^2=(\Delta w)^2=0\,.
\end{equation}
It is important to realize that in the new frame the linearized superconformal invariants (\ref{hatd4n2}) become exact, since there can be no non-linear terms in the single surviving nilpotent pair $\q^2\bq^2$. The possibility to fix such a frame while staying in the vicinity of one particular singular point, is crucial for our argument. This explains why we need not know the full non-linear version of (\ref{hatd4n2}) (see, however, Appendix \ref{AppA} for a discussion of the singularity $(z-z')^{-1}$).

Thus, the supersymmetrized invariant amplitude (\ref{73333})  gives rise to the following singular term:
\begin{equation}\label{39}
  G^{{\cal N}=2}_k(z,z' + \Delta z';{w}+\Delta w) = G^{{\cal N}=2}_k(z,z';{w}) -\frac{\q^2\bq^2}{{w}-z'} (\pa_{z'}+\pa_w) G^{{\cal N}=2}_k(z,z';{w}) \,.
\end{equation}
It is then clear that the condition for suppressing the singularity is
\begin{equation}\label{40}
  \left. (\pa_{z'}+\pa_w) G^{{\cal N}=2}_k(z,z';{w})\right\vert_{{w}\to z'} = 0\,.
\end{equation}
This constraint is what we call a {\it superconformal Ward identity} for $d=4$ ${\cal N}=2$.

Now, recall that $G^{{\cal N}=2}_k$ in (\ref{73333}) is a polynomial
in the single variable ${zz'}/{w}$, made out of non-singular propagators, which has the property
\begin{equation}\label{homogvar}
  (\pa_{z'}+\pa_w)\left(\frac{zz'}{w}\right) = 0\,.
\end{equation}
So, condition (\ref{40}) only affects the coefficient functions in (\ref{73333}):
\begin{equation}\label{coefffu}
  \left. \sum^k_{n=0} \pa_{z'}a_n(z,z') \left(\frac{zz'}{w}\right)^n \right\vert_{{w}\to z'} = \pa_{z'}\sum^k_{n=0}a_n(z,z') z^n = 0\,.
\end{equation}
In other words, the function
\begin{equation}\label{41}
  G^{{\cal N}=2}_k(z,z';z') = \sum^k_{n=0} a_n(z,z') z^n \equiv h(z)
\end{equation}
must depend on the single variable $z$ only. This condition is another form of the superconformal Ward identity (\ref{40}).

Similarly, the examination of the singularity near ${w}\to z \neq z'$
 leads to the superconformal Ward identity 
\begin{equation}\label{4000}
  \left. (\pa_{z}+\pa_w) G^{{\cal N}=2}_k(z,z';{w})\right\vert_{{w}\to z} = 0\,,
\end{equation}
i.e., to the single variable function
\begin{equation}\label{4100}
  G^{{\cal N}=2}_k(z,z';z) = \sum^k_{n=0} a_n(z,z') z'^n \equiv h(z')\,.
\end{equation} 
The fact that $h(z')$ is the same function as $h(z)$ follows from the exchange symmetry $a_n(z,z') = a_n(z',z)$ of the coefficients of the amplitude.

The contribution of the function $h(z)$ to the amplitude can be obtained by setting, e.g., $a_n=0$ for $n=2,\ldots,k$. This is equivalent to considering the case $k=1$. The remaining two coefficients $a_0,a_1$ can be solved for from the linear equations (\ref{41}), (\ref{4100}) in terms of $h(z)$ and $h(z')$, which gives
\begin{equation}\label{41'}
 a_0 + a_1 \frac{zz'}{{w}} \ \rightarrow \  \frac{z h(z') - z' h(z)}{z-z'} + \frac{ h(z) - h(z')}{z-z'}\,\frac{zz'}{{w}} \,.
\end{equation}
We remark that although we have derived the Ward identity (\ref{40}) under the assumption $z\neq z'$, the expression (\ref{41'}) remains  well defined when $z\to z'$.  

In (\ref{41'}) we only see the first two coefficients of the amplitude. Since the single variable contribution was obtained by studying the limit ${w}\to z'$ (or ${w}\to z$), it is clear that the remaining $k-1$ coefficients should form a contribution which vanishes in this limit. It can be put in the form of a {\it factorized polynomial}:
\begin{equation}\label{42}
  ({w}-z)({w}-z')\sum^{k-2}_{n=0} A_n(z,z') \left(\frac{zz'}{{w}}\right)^{n+2} 
\end{equation}
(we recall that this is a polynomial in $1/w$ of degree $k$).
It automatically solves the Ward identities (\ref{40}) and  (\ref{4000}).
Indeed, expanding (\ref{42}) in the powers of $1/w$ and expressing $a_n$, $n=0,\ldots, k$ from (\ref{38}) in terms of $A_n$, $n=0,\ldots,k-2$, it is easy to see that the  single variable function (\ref{41}) identically vanishes. Another way to show that (\ref{42}) is a solution to the Ward identities is to realize that the prefactor in (\ref{42}) suppresses both singularities coming from the expansions of $\hat z$ and $\hat{z}'$ in $A_n(z,z')$. At the same time, the prefactor itself is regular, as follows from (\ref{014}).

Combining eqs. (\ref{41'}) and (\ref{42}), we can write down the complete solution  to the superconformal Ward identities, i.e. the most general four-point amplitude of $\half$-BPS $d=4$ ${\cal N}=2$ operators of weight $k$ in the following form:
\begin{eqnarray}
  G^{{\cal N}=2}_k(z,z';{w}) &=& \frac{z h(z') - z' h(z)}{z-z'} + \frac{ h(z) - h(z')}{z-z'}\,\frac{zz'}{{w}} \nonumber\\
  && + ({w}-z)({w}-z')\sum^{k-2}_{n=0} A_n(z,z') \left(\frac{zz'}{{w}}\right)^{n+2}\,, \label{1042}
\end{eqnarray}
where $h(z)$ is an arbitrary single variable function and $A_n(z,z')=A_n(z',z)$ is a set of $k-1$ arbitrary symmetric functions. Setting ${w}=z'$ or ${w}=z$ in (\ref{1042}) reproduces the definitions of the  single variable functions (\ref{41}), (\ref{4100}).

Finally, let us come back to the issue of singularities in the regime $z\to z'$. In this case both linearized supersymmetry transformations (\ref{013}) may become singular simultaneously, therefore we are not allowed to shift away any of the $\q$s. Moreover, the non-linear terms in $\delta\q$ are singular when $z\to z'$ (recall (\ref{compenslor}) and the analogous terms for any $d$). In order to make sure that we have not missed any further constraints we ought to find the full non-linear version of (\ref{014}) and then expand the amplitude (\ref{38}) taking into account the Ward identity (\ref{40}). A thorough analysis of this issue in the case $d=4$ ${\cal N}=4$ is presented in Appendix \ref{AppA}. Here we just give an example, restricting ourselves to the simplest case $d=4$ ${\cal N}=2$, weight $k=2$, and considering only the factorized form (\ref{42}) which represents the non-trivial (quantum) part of the amplitude:
\begin{eqnarray}
  (\hat {w}-\hat z)(\hat {w}-\hat{z}') A(\hat z,\hat{z}')&=&({w}-z)({w}-z')A \nonumber\\
  && -\q^1\bq^1(A+({w}-z')A_z) -\q^2\bq^2(A+({w}-z)A_{z'}) \nonumber\\
  && + \q^1\bq^1\q^2\bq^2 \left(A_{zz'} - \frac{A_{z'}-A_{z}}{z'-z}  \right) \label{015}
\end{eqnarray}
(we have dropped the non-singular propagator factor). It is then clear that a sufficient condition which removes the singularity $(z'-z)^{-1}$ is that the coefficient function $A(z,z')$ admit an expansion in the even positive powers of $z'-z$. To put it differently, the function $A(z,z')$ should be well defined on the line $z=z'$ and in addition it should be symmetric in $z,z'$, as postulated earlier. The examination of the singularity $(z'-z)^{-1}$ in the full amplitude for arbitrary $k$, including the  single variable part $h(z)$, does not lead to any new conditions. 

\subsection{Ward identities and their solution in the case $d=4$ ${\cal N}=4$}\label{tcn4}

As in the case ${\cal N}=2$, we first examine the singular behaviour under the assumption that $z\neq z'$. From the linearized transformations (\ref{castsusyd4}) of the $\q$s it is clear that treating the harmonic singularities as isolated, for instance, taking $\bw\to z'\neq z$ but keeping all other factors in (\ref{castsusyd4}) non-vanishing, a further frame fixing is possible in which the only non-vanishing odd variables are $\q^{22}$ and $\bq^{22}$. This reduces  (\ref{hatd4}) to
\begin{equation}\label{2400}
  \Delta z' = \Delta\bw = -\frac{\q^{22}\bar{\q}^{22}}{\bw-z'}\,, \qquad (\Delta z')^2=(\Delta\bw)^2=0\,.
\end{equation}
Expanding the amplitude in $\Delta z', \Delta\bw$ creates a simple pole at the coincident point $\bw=z'$:
\begin{equation}\label{3900}
  G^{{\cal N}=4}_k(z,z' + \Delta z';w,\bw + \Delta\bw) = G^{{\cal N}=4}_k(z,z';w,\bw) -\frac{\q^{22}\bar{\q}^{22}}{\bw-z'} (\pa_{z'} + \pa_{\bw}) G^{{\cal N}=4}_k(z,z';w,\bw) \,.
\end{equation}
It is then clear that the condition for suppressing the singularity is
\begin{equation}\label{condd4n4}
  \left. (\pa_{z'}+\pa_{\bw}) G^{{\cal N}=4}_k(z,z';w,\bw)\right\vert_{{\bw}\to z'} = 0\,.
\end{equation}
The amplitude $G^{{\cal N}=4}_k$ in (\ref{7333}) is a polynomial in two non-singular propagator-type variables having the properties
\begin{equation}\label{propvarn4}
  \left. (\pa_{z'}+\pa_{\bw})\left(\frac{(1+w)(1+\bw)}{(1+z)(1+z')} \right)\right\vert_{\bw\to z'} = 0\,, \qquad  (\pa_{z'}+\pa_{\bw})\left(\frac{zz'}{w\bw}\right) = 0\,.
\end{equation}
Thus, only the coefficient functions in (\ref{7333}) are affected by the constraint (\ref{condd4n4}):
\begin{equation}\label{affectcoef}
  \left. \sum_{0\leq m+n \leq k} \pa_{z'} a_{mn}(z,z')\, \left(\frac{(1+w)(1+\bw)}{(1+z)(1+z')} \right)^m \left(\frac{zz'}{w\bw}\right)^{m+n}\right\vert_{{\bw}\to z'} = 0\,.
\end{equation}
In other words, the function
\begin{equation}\label{41000}
  G^{{\cal N}=4}_k(z,z';w,z') = \sum_{0\leq m+n \leq k} a_{mn}(z,z') \frac{z^{m+n}}{(1+z)^m}\ \frac{(1+w)^m}{w^{m+n}} \equiv H(z,w)
\end{equation}
must depend only on $z$ (it is manifestly holomorphic in $w$ as well). The dependence on $w^{-1}$ in $H(z,w)$ is polynomial,
\begin{equation}\label{polyh}
  H(z,w) = \sum_{n=0}^k h_n(z)w^{-n}\,,
\end{equation}
giving rise to $k+1$  single variable functions 
\begin{equation}\label{75}
  h_n(z)= \sum_{p=0}^{n}\sum_{q=n}^{k}a_{q-p,p}(z,z')C_{q-p}^{n-p}\frac{z^{q}}{(1+z)^{q-p}}\,, \qquad n=0,\ldots,k\,.
\end{equation}

In fact, these functions are not all independent. Indeed, examining the singularity $w\to z$ and taking account of the exchange symmetry of the coefficient functions $a_{mn}(z,z')$ leads to the equivalent set of $k+1$ functions $H(z',\bw) = \sum_{n=0}^k h_n(z')\bw^{-n}$. Then we find 
\begin{equation}\label{76}
H(z,z) = \sum_{n=0}^k h_n(z)z^{-n} = \sum_{0 \leq m+n \leq k} a_{mn}(z,z') =  H(z',z') = \sum_{n=0}^k h_n(z')z'^{-n} = {\cal C}\,,
\end{equation} 
where ${\cal C}$ is a constant. 

The part of the correlator which is not affected by the Ward identity (\ref{75}) (and its analogs obtained by exchange of variables) should vanish in all four singular limits. This suggests to cast it in the form of a {\it factorized polynomial}:
\begin{eqnarray}\label{77}
{\cal F}^{{\cal N}=4}_k(z,z';w,\bw)&\ea & (w-z)(\bw-z)(w-z')(\bw-z')\nonumber\\
&& \times\!\!\!\!\!\! \sum_{0 \leq m+n \leq k-2}\!\!\!\!\!\! A_{mn}(z,z') \left(\frac{(1+w)(1+\bw)}{(1+z)(1+z')} \right)^m \left(\frac{zz'}{w\bw}\right)^{m+n+2} \,.
\end{eqnarray}   
Indeed, the prefactor in (\ref{77}) suppresses all singularities coming from the expansions of $\hat z$ and $\hat{z}'$ in $A_{mn}(z,z')$.  At the same time, the prefactor itself is regular, as follows from (\ref{hatd4}).  In ref. \cite{EPSS} it has been shown that the quantum corrections to the correlator with $k=2$, generated through the field-theoretic insertion procedure, take the factorized form (\ref{77}).\footnote{For the generalization to higher weights see \cite{ADOS,APSS}. The factorized form (\ref{77}) in terms of variables of this type appeared in \cite{HH}. }

So, we see that the $\frac{1}{2}(k+1)(k+2)$ coefficients $a_{mn}(z,z')$ of the amplitude (\ref{71}) split into two subsets. One is given by the $\frac{1}{2}k(k-1)$ terms of the polynomial of degree $k-2$ in (\ref{77}), and they are not restricted in any way by the superconformal Ward identities. The remaining $2k+1$ coefficients are expressed in terms of the $k+1$ single variable functions (\ref{75}) subject to the condition (\ref{76}), which makes $k$ independent  single variable functions. The complete amplitude can be written down as follows:
\begin{equation}\label{complam}
  G^{{\cal N}=4}_k(z,z';w,\bw) = {\cal H}^{{\cal N}=4}_k(z,z';w,\bw) + {\cal F}^{{\cal N}=4}_k(z,z';w,\bw)\,,
\end{equation}
where ${\cal H}^{{\cal N}=4}_k$ is the  single variable contribution. As shown in \cite{NO}, the latter can be cast in a form generalizing (\ref{41'}):
\begin{eqnarray}\label{holpart}
&&{\cal H}^{{\cal N}=4}_k(z,z';w,\bw) = \\
&&- {\cal C} +{(w-z)({\bw}-{z'})[H(z,{\bw})+H({z'},w)]-(w-{z'})({\bw}-z)[H(z,w)+H({z'},{\bw})] \over (z-{z'})(w-{\bw})} \,, \nonumber
\end{eqnarray} 
where the constant ${\cal C}$ has been defined in (\ref{76}). Indeed, it clearly has the required exchange symmetry properties. Further, setting $\bw=z'$ we immediately reproduce the definition (\ref{41000}) of $H(z,w)$ (or of $H({z'},{\bw})$ by setting $w=z$).

Finally, let us mention the issue of the singularity when $z\to z'$. In this regime it is not possible to treat the various singularities as independent. Then we need to expand the amplitude in the full non-linear $\Delta z, \Delta z'$ and to study the occurrences of poles in $z-z'$. To simplify the calculation, we have done this in the simplest case $k=2$ and we have taken the amplitude in the factorized form (\ref{77}). The details are given in Appendix \ref{AppA}. The conclusion is that the natural requirement of symmetry under the exchange of $z$ and $z'$ is sufficient to suppress all such singularities.

\subsection{Ward identities in the cases $d=3,5,6$}
 
The treatment of the $d=4$ singularities presented above can easily be generalized to all other dimensions. The case $d=6$ ${\cal N}=(2,0)$ is similar to $d=4$ ${\cal N}=4$. Inspecting the linearized supersymmetry transformations (\ref{castsusyd6}), we see that choosing a particular singular limit, e.g., $\bw\to z'$, affects only $\q^{22}$ and $\q^{42}$. So, in this limit we can set to zero all $\q$s but this pair. Consequently, the singular terms in (\ref{hatd6}) appear only in
\begin{equation}\label{2400d6}
  \Delta\bw = 2\Delta z' = -2\frac{\q^{22}{\q}^{42}}{\bw-z'}\,, \qquad (\Delta z')^2=(\Delta\bw)^2=0\,.
\end{equation}
Thus, the linearized expressions (\ref{regdeltagen}) now become exact. Then the expansion of the general amplitude (\ref{7333alld}) terminates at level one and gives rise to a simple pole $(\bw-z')^{-1}$. The condition for suppressing this pole is
\begin{equation}\label{condalld}
  \left. (\pa_{z'}+\varepsilon\pa_{\bw}) G^{(\varepsilon)}_k(z,z';w,\bw)\right\vert_{{\bw}\to z'} = 0\,,
\end{equation} 
where $\varepsilon =d/2-1 =2$. This is also the form of the constraint
(\ref{condd4n4}) for the case $d=4$ ${\cal N}=4$, with $\varepsilon =
1$. As mentioned earlier, the case $d=3$ ${\cal N}=8$ is the ``mirror"
of the case $d=6$ ${\cal N}=(2,0)$ obtained by exchanging the
space-time and internal sectors, which again leads to (\ref{condalld})
with $\varepsilon = {1\over 2}$. Thus, eq. (\ref{condalld}), together with its $z\leftrightarrow z'$, $w\leftrightarrow \bw$ counterparts, are the superconformal Ward identities for any $d$ and for R symmetry $SO(n)$, $n>3$.

In terms of the coefficient functions the constraint (\ref{condalld}) becomes
\begin{equation}\label{condalldcoef}
  \sum_{0\leq m+n \leq k}z'^{(\varepsilon-1)(m+n)}\, \pa_{z'} a_{mn}(z,z') \frac{z^{\varepsilon(m+n)}}{(1+z)^{\varepsilon m}}\ \frac{(1+w)^m}{w^{m+n}} = 0\,.
\end{equation}
The crucial difference from the case $d=4$ $(\varepsilon=1)$ is in the presence of the $z'$-dependent term. For $\varepsilon \neq 1$ it does not allow us to pull out the derivative, thus obtaining a set of single variable functions. The best we can do now is to expand in the powers of $w^{-1}$ and derive a set of $k+1$ differential constraints (Ward identities). The other singular limits imply analogous conditions which are obtained by exchanging the variables.

The cases $d=6$ ${\cal N}=(1,0)$ and $d=5$ are similar to $d=4$ ${\cal N}=2$. For instance, in $d=5$, examining (\ref{trthetad5}), (\ref{hatd5}) we see that $\Delta w = \frac{3}{2}\Delta z'$ in the singular limit $w\to z'$. This leads to the general superconformal Ward identities for the cases with R symmetry $SO(3)$:
\begin{equation}\label{40alld}
  \left. (\pa_{z'}+\varepsilon\pa_w) G^{(\varepsilon)}_k(z,z';{w})\right\vert_{{w}\to z'} = 0\,,
\end{equation}
and analogously with $z\leftrightarrow z'$. In terms of the coefficient functions (\ref{40alld}) becomes
\begin{equation}\label{WId5}
  \sum_{n=0}^k z^{\varepsilon n} z'^{(\varepsilon-1) n} \pa_{z'} a_n(z,z')  = 0\,.
\end{equation}

For $d\neq4$ solving the above constraints is a rather non-trivial
task. In the special case $d=6$ ${\cal N}=(2,0)$ with $k=2$ this has
been done in \cite{AS}, making use of the crossing symmetry of the
amplitude. In Sections \ref{Wisrs}, \ref{CRSTJP} we give the general
solution, for any $k$ and any $d$, without crossing
symmetry. As mentioned earlier, this is most conveniently done in terms of the variables (\ref{chvar}), whereby the Ward identity (\ref{condalld}) reads 
\begin{equation}\label{condchvar}
  \left. (\chi\pa_{\chi} - \varepsilon\alpha\pa_{\alpha}) G^{(\varepsilon)}_k(\chi,\chi';\alpha,\alpha')\right\vert_{\alpha\to 1/\chi} = 0\,,
\end{equation}   
and similarly for (\ref{40alld}),
\begin{equation}\label{40alldchvar}
  \left. (\chi\pa_{\chi} - \varepsilon\alpha\pa_{\alpha}) G^{(\varepsilon)}_k(z,z';\alpha)\right\vert_{\alpha\to 1/\chi} = 0\,.
\end{equation}

\section{Ward identity solutions for  R  symmetry
$SO(3)$}\label{Wisrs}

In this section we consider the Ward identities for four-point
functions involving $\half$-BPS operators with R symmetry
$SO(3)$. This corresponds to the cases $d=6$ $\N=(1,0)$, $d=5$ and
$d=4$ $\N=2$ discussed above.  We want to solve the Ward identity (\ref{40alldchvar}) along with its partner equation under $\chi\leftrightarrow\chi'$, for general $\varepsilon$. In Section
\ref{win2} we did this only for $\vep=1$. We saw that the Ward
identities effectively concern only the first two coefficients
$a_0,a_1$ in the amplitude (see (\ref{41'})), the rest remain
unconstrained (see (\ref{42})). In fact, keeping only $a_0,a_1$
amounts to considering the case $k=1$. This case is of no physical
interest, since one cannot have a gauge invariant $\half$-BPS operator
of weight 1, which would be the elementary on-shell
multiplet. Nevertheless, this case is very useful as a starting point
in solving the Ward identities. It allowed us to introduce the single
variable function $h$ of Section \ref{win2}. Now we are going to the
same, but this time for general $\varepsilon$. The next step will be
to introduce the analog of (\ref{42}), i.e. the arbitrary part of the
amplitude which automatically solves the Ward identities. The crucial
difference from the case $\vep=1$ will be that for $\vep\neq 1$ we
will be able to absorb the single variable functions into the
unconstrained part of the amplitude.   

In summary, the main result of this section is that
\eqn{mainsol1}{
G^{(\vep)}_k(\xx,\zz;\al)|_{\vep\neq 1}=\sum_{n=0}^{k-2}(\xx\zz)^{(n+2)\vep}\,\De_\vep\,
(\xx\al-1)(\zz\al-1)b_n(\xx,\zz)\al^{n}\,,
}
in terms of a symmetric differential operator $\De_\vep$, defined later in Section \ref{genksolo}.
Equation (\ref{mainsol1})  may be rewritten in a basis independent way as in
(\ref{solgennetyeah}).

\subsection{Solution for the case $k=1$}\label{solk1}

The constraint equation (\ref{40alldchvar}) and its partner under $\chi\leftrightarrow\chi'$ become, for $k=1$, 
\eqn{keonet}{
{\pr\over \pr \xx}a_0+{u^{\vep}\over \xx}{\pr\over \pr \xx}a_1=0\,,\qquad 
{\pr\over \pr \zz}a_0+{u^{\vep}\over \zz}{\pr\over \pr \zz}a_1=0\,,
}
where $u=\chi\chi'$ (recall (\ref{-1})). Clearly, the case $\vep=1$ is exceptional, since, e.g.,  $u/\xx = \zz$ and we can pull out the derivative $\pa_\xx$. This immediately leads to the solution in the form of a single variable function \cite{DO}. However, the situation is not as simple as this for $\vep\neq 1$. Now we are going to follow the method first used in \cite{EPSS} for $\vep=1$ and then in \cite{AS} for $\vep=2$. It consists of regarding (\ref{keonet}) as a pair of coupled partial differential equations, whose integrability conditions can be written down in the form
\eqn{integneto}{ 
D_{\vep}\,\tilde{a}_0=0\,,\qquad\qquad  D_\vep \,a_1=0\,,
}
where $a_0=u^\vep\,\tilde{a}_0$ and the symmetric differential
operator
is given by
\eqn{opanyd}{
D_{\vep}={\pr^2\over \pr \xx\,\pr \zz}-\vep\,{1\over \xx-\zz}\Big({\pr\over
\pr \xx}-{\pr\over \pr \zz}\Big)\,. 
}  
It satisfies the obvious identities
\eqn{identitiesone}{
{1\over \xx}\,\pr_{\zz}\,u^{\vep}\pr_\xx-\xx\leftrightarrow
{\zz}{}=-(\xx-\zz)u^{\vep-1}D_\vep\,,\quad  
{\xx}\,\pr_{\zz}\,u^{-\vep}\pr_\xx u^\vep-\xx\leftrightarrow
{\zz}=-(\xx-\zz)D_\vep\,,
}
with the help of which the consistency conditions for (\ref{keonet}) can be written in the form (\ref{integneto}).

To solve either of (\ref{integneto}) we make the change of variables,
\eqn{changnet}{
p=\xx+\zz\,,\qquad\qquad q=\xx-\zz\,,
}
whereby we may easily find
\eqn{dchange}{
D_\vep={\pr^2\over \pr p^2}-{\pr^2\over \pr q^2}-2\vep{1\over
q}{\pr\over \pr q}\,.
}
This form of the operator allows for solutions by separation of variables to
(\ref{integneto}).
Writing either of $\tilde{a}_0$ or $a_1$ as $f(p)g(q)$ then
(\ref{integneto}) implies that
\eqn{soleqs}{
{f''(p)\over f(p)}={1\over g(q)}
\Big(g''(q)+2 \vep\,{1\over q}\,g'(q)\Big)=-c^2\,,
} 
where $c^2$ is a real constant and where, due to the symmetry condition 
$a_j(\xx,\zz)=a_j(\zz,\xx)$, the function $g(q)$ must be even, $g(q)=g(-q)$.  From here we may easily find for $f(p)$ that
\eqn{solfynet}{
f(p)=A(c)\cos c p+B(c)\sin c p\,,
}
while for $g(q)$ we have that
\eqn{solgznet}{
g(q)=C(c)q^{{1\over 2}-\vep}\,J_{{1\over 2}-\vep}(c q)+D(c)
q^{{1\over 2}-\vep}\,Y_{{1\over 2}-\vep}(c q)\,,
}
involving Bessel functions of the first and second kind, respectively.  Taking into account the evenness condition $g(q)=g(-q)$ and the properties of the Bessel functions, we conclude that $C(c)=0$ for
$\vep=1,2,\dots$ (corresponding to even dimensions) and $D(c)=0$ for
$\vep=\half, {3\over 2},\dots$ (corresponding to odd dimensions).  
For $\tilde{a}_0=h_0^{(\vep)},\,{a}_1=h_1^{(\vep)}$, we
may write the full solution to (\ref{integneto}) as a superposition of (\ref{solfynet}) and (\ref{solgznet}),
\eqna{fullsolnet}{\eqalign{
h^{(\vep)}_j(\xx,\zz)=(\xx-\zz)^{{1\over 2}-\vep}\,
\int\,{\rm d} c\,\big(A_j^{(\vep)}(c)\cos(c \xx+c\zz)+
B_j^{(\vep)}(c)\sin(c \xx+c \zz)\big)\cr
\times \left\{\begin{array}{r}   Y_{{1\over 2}-\vep}(c \xx-c \zz)  
\,\,\,\,\,\,{\rm for}\,\,\,\,\vep=1,2,\dots \cr
J_{{1\over
2}-\vep}(c \xx-c \zz)\,\,\,\,\,\,{\rm for}\,\,\,\,\vep=\half,{\ts{3\over 2}},
\dots\end{array}\right.
\,,}
}
with arbitrary coefficients $A_j^{(\vep)}(c),\,B_j^{(\vep)}(c)$.

This solution can be expressed in terms of two basic solutions, the one for $\vep=1$ and the other for $\vep=\half$:
\eqna{identuseful}{\eqalign{
h^{(\vep)}_j(\xx,\zz) & \ea {}&(D_\vep)^{\vep-1}h^{(1)}_j(\xx,\zz)\,,\quad
\vep=1,2,\dots\,,\cr
h^{(\vep)}_j(\xx,\zz) & \ea {}&(D_\vep)^{\vep-{1\over 2}}
h^{({1\over 2})}_j(\xx,\zz)\,,\quad
\vep=\half,{\ts{3\over 2}},\dots\,.}
}     
To prove this we may use the recurrence relations
\eqn{usualrecurb}{
J_{n+1}(q)+J_{n-1}(q)={2\,n\over q}J_n(q)\,,\qquad 
Y_{n+1}(q)+Y_{n-1}(q)={2\,n\over q}Y_n(q)\,,\nonumber
}
to show that
\eqn{relationinop}{
D_\vep {\rm K}_{n}(p,q)=-\big(2(\vep+n)-1\big)c^2\,{\rm K}_{n-1}(p,q)
\,,\nonumber
}
where ${\rm K}_{n}(p,q)=q^n(A(c)\cos c p+
B(c)\sin c p )\,\big(J_{n}(c q),\,Y_n(c q)\big)$.
It is easy to derive from this that 
$$(A_j^{(\vep)}(c),\,B_j^{(\vep)}(c))=(\vep-1)!(-2 c^2)^{\vep-1}
(A_j^{(1)}(c),\,B_j^{(1)}(c))$$ for integer $\vep$, and 
$$(A_j^{(\vep)}(c),\,B_j^{(\vep)}(c))=(\vep-{\ts{1\over 2}})!(-2 c^2)^{\vep-{1\over 2}}
(A_j^{({1\over 2})}(c),\,B_j^{({1\over 2})}(c))$$ for half-integer $\vep$.

Let us come back for a moment to the particular case $\vep=1$. We already know the solution in this case (recall (\ref{41'})). Here we can reproduce it by substituting $h^{(1)}_j(\xx,\zz)\ \to\ \varphi(\xx,\zz)/(\xx-\zz)$ in (\ref{integneto}), which immediately leads to 
\eqn{fournet}{
h^{(1)}_j(\xx,\zz)={h_j(\xx)-h_j(\zz)\over \xx-\zz}\,,
}
for arbitrary $h_j$. It is not hard to see that the general solution
(\ref{fullsolnet}) reduces to just (\ref{fournet}) if
$\vep=1$. Indeed, the Bessel function $Y_{-1/2}(t) \sim
t^{-1/2} \sin t$, so (\ref{fullsolnet}) becomes the Fourier integral representation of (\ref{fournet}).

In Section \ref{JP} we find a similar result as in
(\ref{identuseful}) for the $k=1$ case with $SO(n)$ R symmetry, but this time in the form of expansions in terms of Jack polynomials whose properties are reviewed in Appendix \ref{AppB}. There also
an alternative and convenient expansion is given for $h^{({1\over 2})}_j(\xx,\zz)$ in terms of Legendre polynomials.
 
Focusing now on the integer $\vep$ case\footnote{The technically more involved half-integer case is dealt with in Section \ref{CRSTJP} where the analysis simplifies in terms of Jack polynomials.}, further conditions on $h_0(\xx),\,h_1(\xx)$ of  (\ref{fournet}) are implied by (\ref{keonet}). For $\vep=1$ it is easy to see that (cf. (\ref{41'}))
\eqn{vepfokonetxx}{
h_0(\xx)= C-{1\over \xx}h_1(\xx)\quad \Rightarrow \quad G_{1}^{(1)}(\xx,\zz;\al)=
u\,{(\xx\al-1)h(\xx)-(\zz\al-1)h(\zz)\over \xx-\zz}\,,
}
where $C$ is a constant and $h(\xx)=h_1(\xx)/\xx$.
We may find similar conditions for other
integer $\vep$ so that,
\eqn{vepfokonet}{
\quad G_{1}^{(\vep)}(\xx,\zz;\al)\big|_{\vep=1,2,\dots}=
u^{\vep}(D_\vep)^{\vep-1}\,{(\xx\al-1)h(\xx)-(\zz\al-1)h(\zz)\over \xx-\zz}\,,
}
for arbitrary $h(\xx)$.

Using Jack polynomials we may unify even and odd dimensional results
for the $k=1$ case, as shown in Section \ref{CRSTJP}.  Here we quote
the result which is
\eqn{vepfokonetop}{
\quad G_{1}^{(\vep)}(\xx,\zz;\al)=
u^{\vep}\H^{(\vep)}(\xx,\zz;\al)\,,
}
where
\eqn{frumpet}{
\H^{(\vep)}(\xx,\zz;\al)=
\sum_{\lambda}a_\lambda\,\big(\tilde{P}_{\lambda-1}^{(\vep)}(\xx,\zz)-
\al\,\tilde{P}^{(\vep)}_\lambda(\xx,\zz)\big)\,,} 
with arbitrary $a_\lambda$ and where we define
\eqn{tildepdef}{
\tilde{P}_{\lambda}^{(\vep)}(\xx,\zz)={(2\vep)_\lambda\over
\Gamma(\lambda+1)}
\,P^{(\vep)}_{\lambda\,0}(\xx,\zz)\,,
}
where the Jack polynomials
$P^{(\vep)}_{\lambda_1\,\lambda_2}(\xx,\zz)$ are defined in Appendix \ref{AppB}.
Here also we are using the Pochhammer
symbol, $(a)_b=\Gamma(a+b)/\Gamma(b)$.
We may reproduce (\ref{vepfokonet}) via (\ref{frumpet}) through, 
\eqn{soline}{
{\tilde{P}}_{\lambda}^{(\vep)}(\xx,\zz)=(-1)^{\vep+1}{1\over
(\vep-1)!^2}\,(D_{\vep})^{\vep-1}\Big({\xx^{\lambda+2 \vep-1}
-{\zz}^{\lambda+2\vep-1}\over \xx - \zz}\Big)\,,
}
which follows from results in Appendix
\ref{AppB}.

\subsection{Solution for the general $k$ case.}\label{genksolo}

The first non-trivial case of a correlator corresponds to $k=2$,
i.e., 
to bilinear gauge invariant composite $\half$-BPS operators. 
The Ward identities we need to solve are given by
\eqn{keqtwonet}{
{\pr_\xx}a_0+{u^{\vep}\over \xx}{\pr_\xx}a_1+{u^{2\vep}\over \xx^2}\pr_\xx
a_2=0\,, \qquad {\pa_{\zz}}a_0+{u^{\vep}\over \zz}{\pa_{\zz}}a_1+{u^{2\vep}\over \zz^2}\pa_{\zz}
a_2=0\,.
}

Let us recall once more the situation in four dimensions. There (\ref{keqtwonet}) becomes
\eqn{keqtwonet4d}{
{\pr_\xx}a_0+\zz {\pr_\xx}a_1+\zz^2  \pr_\xx
a_2=0\,, \qquad {\pa_{\zz}}a_0+\xx {\pa_{\zz}}a_1+\xx^2\pa_{\zz}
a_2=0\,.
}   
We have two partial differential equations for three unknown
functions, so one of the functions must remain arbitrary.  Indeed, after the substitutions \cite{Pickering}
\begin{equation}\label{substk2}
  \tilde a_0 = a_0 - u\, a_2\,, \qquad \tilde a_1 = a_1 + (\xx+\zz)a_2
\end{equation}
the equations in (\ref{keqtwonet4d}) are reduced to those from
the $k=1$ case (\ref{keonet}). The possibility to eliminate the coefficient $a_2$ from the constraints follows from the form (\ref{42}) of the unconstrained part of the amplitude, which, for $k=2$, is simply $(\xx\al-1)(\zz\al-1)u^2 b(\xx,\zz)$. Given an amplitude which satisfies the Ward identities, we can always obtain another one by adding such a polynomial. We can use this freedom by choosing
\begin{equation}\label{subb}
  a_2(\xx,\zz) = u\,b(\xx,\zz)\,,
\end{equation}
and subtracting $(\xx\al-1)(\zz\al-1)\, u a_2$ from the original $k=2$
amplitude. Thus, we obtain a new amplitude with $\tilde a_2 = 0$ (i.e., of the  $k=1$
type) and with $\tilde{a}_0,\,\tilde{a}_{1}$ as given in (\ref{substk2}).

We can apply the same strategy to (\ref{keqtwonet}) with general $\vep$. The analog of (\ref{subb}) now is
\eqn{atwonet}{
a_{2}(\xx,\zz)=\De_\vep\,u\,b(\xx,\zz)\,,
}
where $b(\xx,\zz)=b(\zz,\xx)$ is arbitrary. The operator 
\eqn{delt}{
\De_f=(D_\vep)^{f-1}\,u^{f-1}\,,} 
has Jack polynomials as eigenfunctions and is relevant to the subsequent expansions in terms of such polynomials. 

It is important to realize that the substitution (\ref{atwonet}) is in general not invertible, except for $\vep=1$, where it takes the form (\ref{subb}). For $\vep\neq1$, the operator $\De_\vep$ has a non-trivial kernel, which means that the function $b(\xx,\zz)$ is determined up to some freedom. We shall come back to this point shortly.

We may then rewrite (\ref{keqtwonet}) as
\eqn{keqtwoneto}{
\pr_\xx(a_0-u^{2\vep}\De_\vep b)+{u^{\vep}\over
\xx}\,\pr_\xx(a_1+u^\vep\,\De_\vep\,
(\xx+\zz)b)=0\,,
}
which reduces to the $k=1$ case (\ref{keonet}) again. In deriving (\ref{keqtwoneto}) we have used the operator
identity
\eqn{opidnet}{
{u^{j\vep}\over \xx^j}\,\,\pr_\xx \,(D_\vep)^{\vep-1}\,u = {u^{(j-1)\vep}\over \xx^{j-1}}\,
\,\pr_\xx\,u^{\vep}\,(D_\vep)^{\vep-1}(\xx+\zz)-
{u^{(j-2)\vep}\over \xx^{j-2}}\,\, \pr_\xx\,u^{2\vep}
\,(D_\vep)^{\vep-1}\,,
}
which may be extended to cases of half-integer $\vep$ through use of Jack polynomial expansion (see Section  \ref{CRSTJP}). 

The solution to (\ref{keqtwoneto}) is already known from Section \ref{solk1}. We may write it as
\eqn{solstwonet}{
a_0=u^{2\vep}\,\De_\vep\,b+u^\vep\,h_{0}^{(\vep)}\,,\quad a_1=-
u^{\vep}\,\De_\vep\,(\xx+\zz)b+h_{1}^{(\vep)}\,,
}
where $h_j^{(\vep)}$ are given by (\ref{fullsolnet}).

With (\ref{atwonet}) and (\ref{solstwonet}) we may
write the complete solution to the $k=2$ Ward identities as
\eqn{fullsolltwonet}{
G^{(\vep)}_2(\xx,\zz;\al)=u^{2\vep}\De_\vep(\xx\al-1)(\zz\al-1)b(\xx,\zz)+
u^{\vep}\H^{(\vep)}(\xx,\zz;\al)\,,
}
with $\H^{(\vep)}(\xx,\zz;\al)$ given by (\ref{frumpet}). This form of
the solution resembles the four-dimensional case (\ref{1042}) with
$k=2$. Indeed, there we also have two terms, one involving a
restricted (single variable) function (related to $\H^{(1)}$
in (\ref{fullsolltwonet})) and another involving an arbitrary function of two variables ($b(\xx,\zz)$ in (\ref{fullsolltwonet})). However, there is a crucial difference between the case $\vep=1$ and the general case $\vep\neq1$. It has to do with the freedom in  $b(\xx,\zz)$ mentioned above. It allows us to absorb the term $\H^{(\vep)}$ into a redefinition of  $b(\xx,\zz)$. Let us first illustrate with the case $\vep=2$. 

For $\vep=2$, upon making the following redefinition in (\ref{atwonet}),
\eqn{goronet}{
b\rightarrow b-b'\,,\qquad b'(\xx,\zz)=-{1\over u^2}D_2\Big({g(\xx)-g(\zz)\over
\xx-\zz}\Big)\,,
}
where $\De_\vep \,u b'=0$, so that the modification $b'$ has no effect
on $a_2$,
then we may find that,
\eqn{helooop}{
u^{4}\De_2
(\xx\al-1)(\zz\al-1)b'=u^2 D_2\Big({(\xx
\al-1)h(\xx)-(\zz\al-1)h(\zz)\over \xx-\zz}\Big)\,,
}
where 
\eqn{whereo}{
-\xx g'(\xx)+2 g(\xx)=h(\xx)\,.
}
determing $g$ in terms of $h$.
Of course, the right hand side of (\ref{helooop}) is precisely of the
form which cancels $\H^{(2)}$ in (\ref{solstwonet})
as given in (\ref{vepfokonetop}) with
(\ref{vepfokonet}).  This may be generalized most easily to other
$\vep$ in terms of Jack polynomial expansions - we show this
in Section \ref{CRSTJP}.

Now we are ready to move to the general $k$ case. The idea is the
same: we redefine the last $k-1$
coefficients of the amplitude $a_2,\ldots,a_k$ in such a way that they
all drop out from the constraints (\ref{40alldchvar}), thus again reducing
the problem to the $k=1$ case. Using (\ref{opidnet}) and setting 
\eqna{gensolsnet}{\eqalign{
a_k  & ={}& \De_\vep\,u\,b_{k{-2}}\,,\cr
a_{k{-1}}&={}& \De_\vep\,u\,b_{k{-3}}-u^\vep\,\De_\vep\,(\xx+\zz)\,b_{k{-2}}\,,\cr
a_{k{-2}}&={}& \De_\vep\,u\,b_{k{-4}}-u^\vep\,\De_\vep\,(\xx+\zz)\,b_{k{-3}}+
u^{2\vep}\,\De_\vep\,b_{k{-2}}\,,\cr
&\vdots{}&\cr
a_2 & ={}&\De_\vep\,u\,b_{0}-u^\vep\,\De_\vep\,(\xx+\zz)\,b_{1}+
u^{2\vep}\,\De_\vep\,b_{2}\,,
}}
where $b_i(\xx,\zz)=b_i(\zz,\xx)$ are arbitrary,
we may reduce (\ref{40alldchvar}) to the $k=1$ case with
\eqn{gensolonet}{
\pr_\xx\big(a_0-u^{2\vep}\De_\vep\,b_0\big)+ 
{1\over \xx}\,u^\vep\,\pr_\xx\big(a_1+u^\vep\,\De_\vep(\xx+\zz)b_0
-u^{2\vep}\De_\vep b_1\big)=0\,,
}
so that
\eqna{moreeqnet}{\eqalign{
a_1 & ={}&-u^{\vep}\,\De_\vep\,(\xx+\zz)b_0+u^{2\vep}\De_\vep\,b_1
+h_1^{(\vep)}(\xx,\zz)\,,\cr
a_0& ={}& u^{2\vep}\,\De_\vep\,b_0+h_0^{(\vep)}(\xx,\zz)\,.
}}
With (\ref{gensolsnet}) and (\ref{moreeqnet}) we have that,
\eqn{solgennet}{
G^{(\vep)}_k(\xx,\zz;\al)
=\sum_{i=0}^{k-2}u^{(i+2)\vep}\al^{i}\,\De_\vep\,
(\xx\al-1)(\zz\al-1)b_i(\xx,\zz)+u^{\vep}\H^{(\vep)}(\xx,\zz;\al)\,.
}
For $\vep=1$ and with (\ref{vepfokonet}) this reproduces the general 
solution (\ref{1042}). We stress once more the fact that in four dimensions the single variable part of 
the solution (\ref{1042}) cannot be absorbed into the unconstrained part, unlike all other dimensions. {}From 
here (\ref{mainsol1}) follows.

We may in fact write (\ref{mainsol1}) in a basis independent way
by defining the operator,
\eqn{newopyeah}{
\DD_\vep=\left(D_\vep+\vep^2{1\over u}\,{\pr^2\over \pr \al^2}\,\al^2
-\vep{1\over u\al}\,D_1^+\,{\pr\over \pr \al}\,\al^2\right)^{\vep-1}\,,
}
where $D_1^+$ is the Euler operator defined in Appendix \ref{AppB}, whereby
we may find that,
\eqn{solgennetyeah}{
G_k^{(\vep)}(\xx,\zz;\al)|_{\vep\neq
 1}=(\al^2\DD_\vep\,u-\al\DD_\vep\,(\xx+\zz)+\DD_\vep)
 \,\F_{k-2}^{(\vep)}(\xx,\zz;\al)|_{\vep\neq 1}\,,
}
for arbitrary $\F_{k-2}^{(\vep)}(\xx,\zz;\al)$ of degree $k-2$ in
 $\al$,
where we have used the fact that $\H^{(\vep)}(\xx,\zz;\al)$ may be
 absorbed by $b_0$ in (\ref{solgennet}) for $\vep\neq 1$.
(\ref{newopyeah}) reduces to the unrestricted two-variable part of the
 four-dimensional
solution when $\DD_\vep\to1$. 

\section{Ward identity solutions for  R  symmetry $SO(n)$}\label{CRSTJP}

For general $k$, the main result of this section is that
\eqna{mainsol2}{
&& G_k^{(\vep)}(\xx,\zz;\al,\bal)|_{\vep\neq 1}=\!\!\!\!\!\!\!\!\!\sum_{{0\leq n+m\leq k-2}}\!\!\!\!\!\!\!\!\!
\,u^{(m+n+2)\vep}\,v^{-m\vep}\,(\al\bal)^n\,\big((\al-1)(\bal-1))^{m}\cr
&&\qquad \qquad\qquad\qquad \qquad \times \De_\vep (\xx\al-1)(\zz \al-1 ) 
(\xx\bal-1)(\zz\bal-1)\F_{nm}(\xx,\zz)\cr
&&\qquad\qquad\qquad\qquad \qquad+u^\vep\,\H_1^{(\vep)}(\xx,\zz;\al,\bal)\,,
}
where $\F_{nm}$ are unrestricted two-variable functions and
$\H_1^{(\vep)}$ is a restricted solution to the Ward identities (which
in even
dimensions may be expressed in terms of single variable functions).
The unrestricted part of (\ref{mainsol2}) 
 may be rewritten in a basis independent way as in (\ref{basisind}).

\subsection{Solutions for Ward identities in terms of Jack polynomials}\label{JP} 

Jack polynomials \cite{Jack} provide a useful basis for the symmetric
functions of $\xx,\,\zz$
of the four-point function with which to find solutions to the Ward
identities, particularly
in odd dimensions.
Expansions in terms of Schur polynomials (the four dimensional
reduction of Jack polynomials) have been used before in \cite{HH}
enabling  $d=4$, $\N=4$ superconformal four-point functions to be found.
In this section we use Jack polynomials to find the $k=1$ solutions
for $SO(n)$ R symmetry.  This demonstrates their convenience
in unifying solutions in different dimensions.  An important
consequence of the analysis is that such solutions involve expansion
in terms of a single set of Jack polynomials,
$\tilde{P}^{(\vep)}_\lambda(\xx,\zz)$ in (\ref{tildepdef}). This
is a feature of the general $k$ case also whereby, upon using a
technique similar to the $SO(3)$ case to reduce the equations to
a simpler set, they solve reduced sets of equations.

 Also in this
section we demonstrate the claim made about absorption of
$\H^{(\vep)}$
in (\ref{solgennet}) and prove (\ref{opidnet}) (which is 
used further in Section \ref{solgn4} and Appendix \ref{AppC}), for general $\vep$,
 using such expansions. Both rely on an important property of the
 operator
$\De_\vep$ in (\ref{delt}) - it has Jack polynomials as eigenfunctions.
 The proof of (\ref{opidnet}) also relies upon
the solution to an important recurrence relation, (\ref{determb}), in
Section \ref{subne1}.

\subsubsection{The $k=1$ solutions in terms of Jack polynomials}\label{subne1}
For $k=1$, we may find
four independent equations coming from the Ward identities (\ref{condchvar})
constraining $a_{00},\, a_{10},\,a_{01}$
given by,
\eqna{fset}{\eqalign{
&& {\pr_\xx}a_{00}
+{\xx-1\over \xx}\Big({u\over v}\Big)^{\vep}\,{\pr_\xx}a_{01}=0\,,\qquad 
{\pr_{\zz}}a_{00}+{{\zz}-1\over {\zz}}\Big({u\over v}\Big)^{\vep}\,
{\pr_{\zz}}a_{01}=0\,,\cr
&& {\pr_\xx}a_{01}+{1\over 1-\xx}
\,v^\vep\,{\pr_\xx}a_{10}=0\,,\qquad \quad \,\,{\pr_{\zz}}a_{01}
+{1\over 1-{\zz}}
\,v^\vep\,{\pr_{\zz}}a_{10}=0\,,}
}
which we may re-write more conveniently by defining,
\eqn{defining}{
a_{00}=u^\vep \,\hat{a}_{00}\,,\qquad a_{01}=v^\vep\,\hat{a}_{01}\,,
}
so that from the first set of equations in (\ref{fset}) we have that
\eqna{ffset}
{\eqalign{
&& \big(\vep+\xx
\pr_\xx\big)\hat{a}_{00}+\big(\vep+(\xx-1)\pr_\xx\big)\hat{a}_{01}
=0\,,\cr
&& \big(\vep+{\zz}
\pr_{\zz}\big)\hat{a}_{00}+\big(\vep+({\zz}-1)
\pr_{\zz}\big)\hat{a}_{01}=0\,,}
}
while from the second set we have that
\eqna{fgset}{
&& \pr_\xx {a}_{10}-\big(\vep+(\xx-1)\pr_\xx\big)\hat{a}_{01}=0\,,
\cr
&& \pr_{\zz} {a}_{10}-\big(\vep+({\zz}-1)\pr_{\zz}\big)
\hat{a}_{01}=0\,. 
}

Considering integrability constraints on (\ref{fset}) we may find that
these imply that
\eqn{ghope}{
D_\vep\,\hat{a}_{00}=0\,,\qquad D_\vep\,\hat{a}_{01}=0\,,\qquad
D_\vep\,a_{10}=0\,,
}
for which we know the solutions from Section \ref{solk1}.  Further
constraints on these solutions are implied by (\ref{ffset}) and (\ref{fgset}) 
however it proves easiest to satisfy these constraints through
expansions
as below.
We may solve for these constraints quite simply for $\vep=2$ (corresponding
to
six dimensions) as the solution to (\ref{ghope}) in this case is given
by,
\eqn{solsvep2}{
\hat{a}_{00}=D_2\Big({g_1(\xx)-g_1(\zz)\over \xx-\zz}\Big)\,,\quad 
\hat{a}_{01}=D_2\Big({g_2(\xx)-g_2(\zz)\over \xx-\zz}\Big)\,,\quad
{a}_{10}=D_2\Big({g_3(\xx)-g_3(\zz)\over \xx-\zz}\Big)\,,
}
for arbitrary $g_i(\xx)$ and to satisfy (\ref{ffset}) we find that,
\eqn{solsvep2o}{
g_1(\xx)=-{\xx-1\over
\xx}g_2(\xx)+A_0{1\over \xx}+A_1+A_2\,\xx+A_3\,\xx^2\,,}
for $A_i$ being constants, while to satisfy (\ref{fgset}) we find that,
\eqn{solsvep2tw}{
g_3(\xx)=(\xx-1)g_2(\xx)+B_1+B_2\, \xx+B_3\,\xx^2+B_4\,\xx^3\,,
}
for $B_i$ being constants.{\foot{One way of seeing (\ref{solsvep2o}),
(\ref{solsvep2tw}) is to expand (\ref{ffset}), (\ref{fgset}) in terms
of
either $\xx$ or $\zz$ and $\xx-\zz$ whereby, on Taylor expanding in
$\xx-\zz$, the  coefficients of
$(\xx-\zz)^{n-4},\,{n\geq 4}$, are proportional, respectively, to
\eqna{extrastice}{
n\big(g_1^{(n-1)}(\xx)+g_2^{(n-1)}(\xx)\big)+\xx
g_1^{(n)}(\xx)+(\xx-1)g_2^{(n)}(\xx)=0\,,\cr
n\,g_2^{(n-1)}(\xx)+(\xx-1)g_2^{(n)}(\xx)-g_3^{(n)}(\xx)=0\,.\nonumber}
These must clearly vanish for all $n\geq 4$ and
solving these equations leads respectively to (\ref{solsvep2o}),
(\ref{solsvep2tw}).}
}
Inserting these expression back into
(\ref{solsvep2}) we find that the $A_i,\,B_i,\,i=1,2,3$,
contributions vanish irrespective of the choice of these constants.
With (\ref{solsvep2}), (\ref{solsvep2o}) and (\ref{solsvep2tw})
we may find that
\eqn{whatdowefind}{
G_1^{(2)}(\xx,\zz;\al,\bal)=u^2\,D_2\Big({(\xx\al-1)(\xx\bal-1)h(\xx)-(\zz\al-1)(\zz\bal-1)h(\zz)\over
\xx-\zz}\Big)\,,
}
where{\foot{Notice that the $A_0$ term contributes to $G^{(2)}_1$ as
$G_1^{(2)}\sim -A_0$ which is the trivial constant solution to the
Ward identities.}}
\eqn{fg1interms}{
h(\xx)={1\over \xx}g_2(\xx)+A_0{1\over \xx}+B_4\xx\,.
}

For general $\vep$ we now expand,
\eqn{expansionJ}{
\hat{a}_{00}=\sum_{\lambda_1,\lambda_2}\,a_{\lambda_1\,\lambda_2}
\,P^{(\vep)}_{\lambda_1\,\lambda_2}\,,\quad
\hat{a}_{01}=\sum_{\lambda_1,\lambda_2}\,b_{\lambda_1\,\lambda_2}
\,P^{(\vep)}_{\lambda_1\,\lambda_2}\,,\quad
{a}_{10}=\sum_{\lambda_1,\lambda_2}\,c_{\lambda_1\,\lambda_2}
\,P^{(\vep)}_{\lambda_1\,\lambda_2}\,,
}
where
$a_{\lambda_1\,\lambda_2},\,b_{\lambda_1\,\lambda_2},\,c_{\lambda_1\,\lambda_2},\,$
are arbitrary, 
so that we may easily obtain from (\ref{ffset}) 
(using relations from Appendix \ref{AppB}),
\eqna{reao}{\eqalign{
&&{}(\lambda_1+\lambda_2+\vep)(a_{\lambda_1\,\lambda_2}
+b_{\lambda_1\,\lambda_2})\cr
&&{}\qquad -{(\lambda_1+\vep+1)(\lambda_-+1)\over
\lambda_-+\vep+1}\,b_{\lambda_1+1\,\lambda_2}
-{(\lambda_2+1)(\lambda_-+2\vep-1)\over
\lambda_-+\vep-1}\,b_{\lambda_1\,\lambda_2+1}=0\,,}
}
and,{\foot{We may obtain no more independent constraints
 due to the recurrence relations
mentioned in Appendix \ref{AppB} for the operators $D^{\pm}_n$.}}
\eqna{reaot}{\eqalign{
&&{}(\lambda_1+\lambda_2)b_{\lambda_1\,\lambda_2}
-{(\lambda_1+\vep-1)(\lambda_-+2\vep-1)\over
\lambda_-+\vep-1}(a_{\lambda_1-1\,\lambda_2}
+b_{\lambda_1-1\,\lambda_2})\cr
&&{}\qquad\qquad 
-{(\lambda_2-1)(\lambda_-+1)\over \lambda_-+\vep+1}
(a_{\lambda_1\,\lambda_2-1}+b_{\lambda_1\,\lambda_2-1})=0\,.}
}
Similarly, from (\ref{fgset}) we may obtain,
\eqna{reaoth}{\eqalign{
{}&&(\lambda_1+\lambda_2+2\vep)a_{\lambda_1\,\lambda_2}\cr
{}&&\qquad\qquad+{(\lambda_1+\vep+1)(\lambda_-+1)\over
\lambda_-+\vep+1}c_{\lambda_1+1\,\lambda_2}+{(\lambda_2+1)(\lambda_-+2
\vep-1)\over \lambda_-+\vep-1}c_{\lambda_1\,\lambda_2+1}=0\,,}
}
and
\eqn{reaofo}{\eqalign{
(\lambda_1+\lambda_2)c_{\lambda_1\,\lambda_2}
+{(\lambda_1+\vep-1)(\lambda_-+2\vep-1)\over
\lambda_-+\vep-1}a_{\lambda_1-1\,\lambda_2}
+{(\lambda_2-1)(\lambda_-+1)\over
\lambda_-+\vep+1}a_{\lambda_1\,\lambda_2-1}=0\,.}
}
To solve these recurrence relations it proves easiest to solve for
$b_{\lambda_1\,\lambda_2}$ using (\ref{reao}) with (\ref{reaot}), whence we may
solve for $a_{\lambda_1\,\lambda_2}$ using (\ref{reao}) and
$c_{\lambda_1\,\lambda_2}$ using (\ref{reaofo}).  (\ref{reaoth}) demonstrates
consistency.
In this way the equation we get for $b_{\lambda_1\,\lambda_2}$ is
given by,
\eqna{determb}{\eqalign{
&&{(\lambda_-\!+2\,\vep{-2})(\lambda_-\!+2\,\vep{-1})\over
(\lambda_-{+\vep-2})(\lambda_-{+\vep-1})}\,
(\lambda_1+\vep-1)\,(\lambda_2+1)\,{b}_{\lambda_1-1\,\lambda_2+1}
\cr
&& +{(\lambda_-{+1})(\lambda_-{+2})\over
(\lambda_-+\vep+1)(\lambda_-+\vep+2)}\,(\lambda_1+1)
\,(\lambda_2-1)\,{b}_{\lambda_1+1\,\lambda_2-1}\cr
&& -2\,{\lambda_-(\lambda_-+2\,\vep)+(2\,\vep+1)(
\vep-1)\over
(\lambda_-+\vep-1)(\lambda_-+\vep+1)}\,(\lambda_1+\vep)
\,\lambda_2\,{b}_{\lambda_1\,\lambda_2}=0\,.
 }} 
In fact we may use Jack polynomials as generating functions to solve
 (\ref{determb}) as follows,

\eqna{solsspec}{\eqalign{
&&\sum_{\lambda_1, \lambda_2}\Big({(\lambda_-\!+2\,\vep{-2})
(\lambda_-\!+2\,\vep{-1})\over
(\lambda_-{+\vep-2})(\lambda_-{+\vep-1})}\,(\lambda_1+\vep-1)
\,(\lambda_2+1)\,{b}_{\lambda_1-1\,\lambda_2+1}
\cr
&&+ {(\lambda_-{+1})(\lambda_-{+2})\over
(\lambda_-+\vep+1)(\lambda_-+\vep+2)}\,(\lambda_1+1)\,(\lambda_2-1)
\,{b}_{\lambda_1+1\,\lambda_2-1}\cr
&& - 2\,{\lambda_-(\lambda_-+2\,\vep)+(2\,\vep+1)(
\vep-1)\over
(\lambda_-+\vep-1)(\lambda_-+\vep+1)}\,(\lambda_1+\vep)\,\lambda_2
\,{b}_{\lambda_1\,\lambda_2}\Big)\,
P_{\lambda_1\,\lambda_2}^{(\vep)}(\xx,\zz)\cr
&&\,\,\, =
(\xx+\zz+2)(\xx+\zz-2)\,\sum_{\lambda_1,\lambda_2}
{b}_{\lambda_1\,\lambda_2}\,(\lambda_1+\vep)\,\lambda_2\,
P_{\lambda_1-1\,\lambda_2-1}^{(\vep)}(\xx,\zz)=0\,,}
}
which implies that $b_{\lambda_1\,\lambda_2}$ is non-zero only for
$\lambda_1=-\vep$ or $\lambda_2=0$.{\footnote{As
mentioned in Appendix \ref{AppB} we have that
$P_{\lambda_1\,\lambda_2}^{(\vep)}(\xx,\zz)=
P_{\lambda_2-\vep\,\lambda_1+\vep}^{(\vep)}(\xx,\zz)$
so that solutions of the form
$P^{(\vep)}_{-\vep\,\lambda+\vep}(\xx,\zz)=P^{(\vep)}_{\lambda\,0}(\xx,\zz)$.
However we may show that these are accounted for by the $\lambda_2=0$
possibility.}} 

It is not surprising that the restricted set of Jack polynomials
$P^{(\vep)}_{\lambda\,0}$ provide a basis for the $k=1$ solutions
as 
\eqn{integrability}{
D_{\vep}\,P^{(\vep)}_{\lambda\,0}(\xx,\zz)=0\,,
}
which is the form of the integrability constraints (\ref{ghope}).

Expanding the solution for $b_{\lambda_1\,\lambda_2}$ as
\eqn{solbloltwo}{
b_{\lambda_1\,\lambda_2}=\sum_{\lambda}
\de_{\lambda_1\lambda}\,\de_{\lambda_2
0}\,b_{\lambda}\,,
}
then we may find, in the way mentioned,
\eqna{acloltw}{\eqalign{
a_{\lambda_1\,\lambda_2}{}&=&\sum_{\lambda}\de_{\lambda_1\lambda}
\,\de_{\lambda_2
0}\,\Big(-b_\lambda+{\lambda+1\over \lambda+2\vep}\,b_{\lambda+1}\Big)
\,,
\cr
c_{\lambda_1\,\lambda_2}{}&=&\sum_{\lambda}\de_{\lambda_1\lambda}
\,\de_{\lambda_2
0}\,\Big({\lambda+2\vep-1\over
\lambda}b_{\lambda-1}-b_{\lambda}\Big)\,.}
}
We  may then reassemble the results to find 
that,
\eqn{fullsolone}{
G^{(\vep)}_1(\xx,\zz;\al,\bal)=
 u^\vep\, \H^{(\vep)}_1(\xx,\zz;\al,\bal)\,,
}
where,
\eqn{lastlyy}{
\H^{(\vep)}_1
(\xx,\zz;\al,\bal)=\sum_\lambda b_{1\,\lambda}\big(
{\tilde{P}}^{(\vep)}_{\lambda-1}(\xx,\zz)
-(\al+\bal){\tilde{P}}^{(\vep)}_{\lambda}(\xx,\zz)+\al\bal\,
{\tilde{P}}^{(\vep)}_{\lambda+1}(\xx,\zz)\big)\,,
}
for,
\eqn{alambda}{
b_{1\,\lambda}={\Gamma(\lambda+1)\over (2 \vep)_\lambda}\,b_\lambda\,.
}
Using (\ref{soline}) we may rewrite this for integer $\vep$ as
\eqn{fullsolg}{
G^{(\vep)}_1(\xx,\zz;\al,\bal)\big|_{\vep=1,2,\dots}=u^\vep
(D_\vep)^{\vep-1}\Big({(\xx\al-1)(\xx\bal-1)h(\xx)-(\zz\al-1)(\zz\bal-1)h(\zz)\over
\xx-\zz}\Big)\,,
}
for some arbitrary single variable function $h(\xx)$.  For $\bal=1$
in (\ref{fullsolg}), this agrees with the functional form of
(\ref{vepfokonet}). Similarly, for half integer $\vep$, we may find
$\tilde{P}_{\lambda}^{(\vep)}(\xx,\zz)\propto (D_\vep)^{\vep-{1\over
2}}(\xx \zz)^{{1\over 2}\lambda}P_{\lambda+2\vep-1}(s)$ 
for usual Legendre
polynomials and with $s=(\xx+\zz)/(2\,u^{1\over 2})$ - this
agrees with (\ref{identuseful}) for $h_j^{({1\over 2})}(\xx,\zz)
=\sum_\lambda a_\lambda \,\,(\xx \zz)^{{1\over
2}\lambda}P_{\lambda+2\vep-1}(s)$,
for some arbitrary $a_\lambda$.

\subsubsection{Absorption of $\H^{(\vep)}$ for $SO(3)$ solutions}

Here we prove a claim made in Section \ref{genksolo} that $\H^{(\vep)}$ of
(\ref{solgennet}) with (\ref{frumpet}) may be 
absorbed into $b_0$ through a redefinition for $\vep\neq 1$.  
In fact, this is a feature of the $SO(n)$ case as well whereby
the analysis here is also relevant - this is discussed in  more
detail for $k=2$ later.
This claim is easily seen
through an expansion in Jack polynomials whereby we may show
that, for $\vep\neq 1$,
\eqna{helo}{
\eqalign{
{\lim_{\mu\to 0}} \,u^{2\vep}\De_\vep
(\xx\al-1)(\zz\al-1)(\De_\vep)^{-1}\sum_{\lambda}a_\lambda\,
{(2\vep)_{\lambda-1}\over \Gamma(\lambda)}\,
P^{(\vep)}_{\lambda+\mu -\vep-1\,\mu-\vep}(\xx,\zz)= 
u^\vep\H^{(\vep)}(\xx,\zz;\al)\,,\cr}
}
with the  operator $\De_\vep$ defined in (\ref{delt}) and for which the
following holds, namely,
\eqn{actionfds}{
\De_f\,P_{\lambda_1\,\lambda_2}^{(\vep)}(\xx,\zz)=
\E_{\lambda_1\,\lambda_2}^{(f)}\,
P_{\lambda_1\,\lambda_2}^{(\vep)}(\xx,\zz)\,,\qquad
\E_{\lambda_1\,\lambda_2}^{(f)}=
(\lambda_1+1+\vep)_{f-1}(\lambda_2+1)_{{f}-1}\,,}
following from results in Appendix \ref{AppB}. 
Hence by making the redefinition,
\eqn{redfeern}{
b_0\rightarrow b_0-(\De_\vep)^{-1}\sum_{\lambda}a_\lambda\,
{(2\vep)_{\lambda-1}\over \Gamma(\lambda)}\,
P^{(\vep)}_{\lambda+\mu-\vep-1\,\mu-\vep}(\xx,\zz)\,,
}
and subsequently taking the limit $\mu\to 0$ we may absorb the
$\H^{(\vep)}$
contribution.  The limit is taken in this way because the modification
in (\ref{redfeern}) has an implicit factor coming from
$(\De_\vep)^{-1}$, whose action on the Jack polynomials gives $(\E^{(\vep)}_{\lambda+\mu-\vep-1\,\mu-\vep})^{-1}$,
which has poles for  $\mu=0$ and $\vep\neq 2,3,\dots$. 

For $\vep=2$ we may reproduce (\ref{goronet}), (\ref{helooop}) and  (\ref{whereo}),
whereby defining
\eqn{hollo}{
\sum_\lambda a_\lambda \tilde{P}^{(2)}_{\lambda-1}=-D_2\sum_\lambda
a_\lambda {\xx^{\lambda+2}-\zz^{\lambda+2}\over
\xx-\zz}=-D_2\Big({h(\xx)
-h(\zz)\over
\xx-\zz}\Big)\,,
}
so that
\eqn{pros}{
\H^{(2)}(\xx,\zz;\al)=D_2\Big({(\xx \al-1)h(\xx)-(\zz\al-1)h(\zz)\over \xx-\zz}\Big)\,,
}
then the modification from (\ref{redfeern}) is given by
\eqn{goro}{
{1\over u^2}\sum_\lambda a_\lambda
(\E^{(2)}_{\lambda-3\,\,-2})^{-1}\tilde{P}^{(2)}_{\lambda-1}
={1\over u^2}D_2\sum_\lambda {a_\lambda\over \lambda}{\xx^{\lambda+2}
-\zz^{\lambda+2}\over \xx-\zz}=-{1\over u^2}D_2\Big({g(\xx)-g(\zz)\over
\xx-\zz}\Big)\,,
}
for which $g$ satisfies (\ref{whereo}) in terms of $h$.

\subsubsection{Proof of (\ref{opidnet}) for general $\vep$}

Much use was made of the operator identity (\ref{opidnet})
in finding solutions to the Ward identities for $SO(3)$ R symmetry.  Indeed this
identity is relevant for the $SO(n)$ R symmetry case also.
Turning to the proof of (\ref{opidnet}), we wish to show that, for arbitrary 
$f(\xx,\zz)$,
\eqn{showthatkeo}{
{1\over \xx^2}u^{2 \vep} \pr_\xx f(\xx,\zz)={1\over x}u^\vep\pr_\xx u^\vep
\De_\vep
(\xx+\zz)u^{-1}(\De_\vep){}^{-1}f(\xx,\zz)-\pr_\xx u^{2\vep} \De_\vep u^{-1}
(\De_\vep){}^{-1} f(\xx,\zz)\,.
}
To prove this we consider $f(\xx,\zz),\,g(\xx,\zz),\,h(\xx,\zz)$ where
\eqn{expansions}{
f=\sum_{\lambda_1,\lambda_2}f_{\lambda_1\,\lambda_2}
P^{(\vep)}_{\lambda_1\,\lambda_2}\,,\quad
g=\sum_{\lambda_1,\lambda_2}g_{\lambda_1\,\lambda_2}
P^{(\vep)}_{\lambda_1\,\lambda_2}\,,\quad
h=\sum_{\lambda_1,\lambda_2}h_{\lambda_1\,\lambda_2}
P^{(\vep)}_{\lambda_1\,\lambda_2}\,,
} 
and attempt to find conditions on $g,\,h$ coming from
\eqn{condseqn}{
{1\over \xx^2}u^{2\vep}\pr_\xx f(\xx,\zz)={1\over \xx}u^\vep\pr_\xx u^\vep
g(\xx,\zz)+\pr_\xx u^{2\vep}h(\xx,\zz)\,.
}
{}From (\ref{condseqn}) with (\ref{expansions}) we may obtain two independent
recurrence relations determining $g_{\lambda_1\,\lambda_2},\,
h_{\lambda_1\,\lambda_2}$ in terms of $f_{\lambda_1\,\lambda_2}$.
Defining
\eqn{anotherdef}{
h_{\lambda_1\,\lambda_2}=-{(\lambda_1+\vep+1)(\lambda_2+1)\over 
(\lambda_1+2\vep)(\lambda_2+\vep)}\,\tilde{h}_{\lambda_1+1\,\lambda_2+1}\,,
}
we may obtain a recurrence relation from these of the form
(\ref{determb}) with
\eqn{determbbb}{
b_{\lambda_1\,\lambda_2}\rightarrow f_{\lambda_1\,\lambda_2}-
\tilde{h}_{\lambda_1\,\lambda_2}\,,
}
determining $h_{\lambda_1\,\lambda_2}$ in terms of
$f_{\lambda_1\,\lambda_2}$.
Similarly we may find that
\eqna{detgtermshf}{\eqalign{
(\lambda_1+\lambda_2+2\vep)g_{\lambda_1\,\lambda_2}={(\lambda_1+\vep+1)
(\lambda_-+1)\over \lambda_-+\vep+1}\Big(f_{\lambda_1+1\,\lambda_2}+
{\lambda_2\over
\lambda_1+2\vep}\tilde{h}_{\lambda_1+1\,\lambda_2}\Big)\cr
+{(\lambda_2+1)(\lambda_-+2\vep-1)\over \lambda_-+\vep-1}\Big(
f_{\lambda_1\,\lambda_2+1}+{\lambda_1+\vep\over \lambda_2+\vep}
\tilde{h}_{\lambda_1\,\lambda_2+1}\Big)\,.
}
}
We therefore obtain that $\tilde{h}_{\lambda_1\,\lambda_2}
=f_{\lambda_1\,\lambda_2}+\sum_\lambda\,h_\lambda\,\de_{\lambda_1,\lambda}
\de_{\lambda_2,0}$ so that,
\eqn{solsopidnet}{
h_{\lambda_1\,\lambda_2}=-{(\lambda_1+\vep+1)(\lambda_2+1)\over 
(\lambda_1+2\vep)(\lambda_2+\vep)}\,f_{\lambda_1+1\,\lambda_2+1}\,,
}
and 
\eqn{solsopidnetll}{
g_{\lambda_1\,\lambda_2}={(\lambda_1+\vep+1)
(\lambda_-+1)\over (\lambda_1+2\vep)(\lambda_-+\vep+1)}
f_{\lambda_1+1\,\lambda_2}
+{(\lambda_2+1)(\lambda_-+2\vep-1)\over
(\lambda_2+\vep)( \lambda_-+\vep-1)}f_{\lambda_1\,\lambda_2+1}\,.
}
With (\ref{expansions}) we therefore obtain using (\ref{actionfds})
\eqn{hgintermsofde}{
h(\xx,\zz)=-\De_\vep\,u^{-1}\,(\De_\vep)^{-1}f(\xx,\zz)\,,\quad 
g(\xx,\zz)=\De_\vep(\xx+\zz)u^{-1}(\De_\vep)^{-1}f(\xx,\zz)\,,
}
proving (\ref{showthatkeo}) and hence (\ref{opidnet}).

\subsection{The general $k$ case}\label{solgn4}

In this section we solve the Ward identities (\ref{condchvar}) for
general $k$, corresponding to the case when the four-point function
has operators belonging to the rank $k$ symmetric traceless
representation of the $SO(n)$ R symmetry group.  We start
with the simplest physical case - the $k=2$ one - whereby we make 
further use
of the operator identity
(\ref{opidnet}), along with certain extensions of it,
to solve the Ward identities in a manner not unlike the $k=2$
case considered in Section \ref{genksolo}.  The difference is that
in contrast to the $SO(3)$ case, where we could reduce the 
equations to $k=1$ ones, here the reduced equations are not
of the $k=1$ form for $SO(n)$.  Nevertheless, the analysis for
$k=1$ of $SO(n)$ is important here because it implies that the
solutions to the reduced equations are expressible in terms of
expansions in the restricted set of Jack polynomials
$\tilde{P}^{(\vep)}_\lambda(\xx,\zz)$.  This simplifies the analysis 
considerably.

\subsubsection{The $k=2$ case}\label{k2case}
We start with the $k=2$ case for which we have the following
three independent equations following from the Ward
identities (\ref{condchvar}), namely,
\eqn{ward}{
\sum_{i=0}^2\,{1\over \xx^i}u^{i \vep}\,\pr_\xx \, a_{i 0}=0\,,\quad
\sum_{{i,j\geq 0\atop
i+j=2}}(1-\xx)^j\,v^{-j\vep}\,\pr_\xx\,a_{ij}=0\,,\quad \sum_{i=0}^2
\Big({\xx-1\over
\xx}\Big)^{\!i}
\,\Big({u\over v}\Big)^{\!i\vep}\,\pr_\xx\,a_{0i}=0\,,
}
along with three more obtained from symmetry under $\xx\leftrightarrow \zz$.

To solve these we employ two further operator identities in addition to
(\ref{opidnet}), namely,
\eqna{moreident}{\eqalign{
{}&&{1\over (1-\xx)^j}\,v^{j\vep}\,\pr_\xx\,(D_\vep)^{\vep-1}\,v={1\over
(1-\xx)^{j-1}}
v^{(j-1)\vep}\,\pr_\xx\,v^{\vep}\,(D_\vep)^{\vep-1}(2-\xx-\zz)\cr
{}&& \qquad\qquad\qquad\qquad \qquad \qquad \qquad -{1\over
(1-\xx)^{j-2}}v^{(j-2)\vep}\,\pr_\xx\,v^{2\vep}(D_\vep)^{\vep-1}\,,\cr
{}&&\Big({\xx-1\over \xx}\Big)^j\Big({u\over v}\Big)^{j\vep}\pr_\xx\,v^{2\vep}
(D_\vep)^{\vep-1}\,u=\Big({\xx-1\over \xx}\Big)^{j-1}\Big({u\over
v}\Big)^{(j-1)\vep}\,\pr_\xx\,
(uv)^\vep\,(D_\vep)^{\vep-1}(2 u-\xx-\zz)
\cr
{}&&\qquad\qquad\qquad\qquad \qquad \qquad \qquad
-\Big({\xx-1\over \xx}\Big)^{j-2}
\Big({u\over
v}\Big)^{(j-2)\vep}\,\pr_\xx\,
u^{2\vep}\,(D_\vep)^{\vep-1}v \,,}
}
although these may be obtained from (\ref{opidnet}); the first from symmetry
under $\xx\to 1-\xx,\,\zz\to 1-\zz$ and the second from symmetry under
$\xx\to \xx/(\xx-1),\,\zz\to \zz/(\zz-1)$.  The second case may be seen
using
\eqn{use}{
D_{\vep}|_{u\to 1/u,\,v\to v/u}=u^{\vep+2}\,D_{\vep}\,u^{-\vep}\,,
}
along with{\foot{We may see this quite easily from Jack polynomials
with the operator $\De_\vep=(D_\vep)^{\vep-1}u^{\vep-1}$. 
We may show that 
\eqna{all}{
\De_\vep |_{u\to 1/u,\,v \to
v/u}P^{(\vep)}_{\lambda_1\,\lambda_2}(\xx,\zz)=
\De_\vep |_{u\to 1/u,\,v \to
v/u}P^{(\vep)}_{-\lambda_2\,-\lambda_1}(1/\xx,1/\zz)\cr
=u^{2\vep}\De_\vep\,u^{-2\vep}P^{(\vep)}_{\lambda_1\,\lambda_2}(\xx,\zz)\quad
{\rm for}\quad \vep=0,{\ts{1\over 2}},1,{\ts{3\over 2}},
\dots,\quad \lambda_-=0,1,2\dots,\nonumber
}
so that
\eqn{the}{
\De_\vep |_{u\to 1/u,\,v\to v/u}=u^{2\vep}\De_\vep\,u^{-2\vep}\quad
\Rightarrow \quad \De_\vep |_{u\to u/v,\,v\to 1/v}=v^{2\vep}\De_\vep\,
v^{-\vep-1}\,.\nonumber
}}}

\eqn{identcrosso}{
(D_{\vep}|_{u\to 1/u,\,v\to v/u})^{\vep-1} =
u^{2\vep}(D_{\vep})^{\vep-1}u^{-2}\quad \Rightarrow \quad 
(D_{\vep}|_{u\to u/v,\,v\to 1/v})^{\vep-1} =
v^{2\vep}(D_{\vep})^{\vep-1}v^{-2}\,.
} 
Examining the second two equations in (\ref{ward}) and using (\ref{moreident})
we may, by choosing,
\eqn{choiceket}{
a_{02}(\xx,\zz)=v^{2\vep}\,\De_\vep\,u\,a(\xx,\zz)\,,
}
(where $a(\xx,\zz)=a(\zz,\xx)$ is general)
reduce them to the following two equations,
namely,
\eqna{twoeqsket}{\eqalign{
{}&&\pr_\xx\big(a_{00}-u^{2\vep}\De_\vep v a\big)+{\xx-1\over \xx}\Big({u\over
v}\Big)^{\vep}
\pr_\xx\big(a_{01}+(u\,v)^{\vep}\De_\vep (2 u-\xx-\zz)a\big)=0\,,\cr
{}&&\pr_\xx
\big(a_{11}+v^{\vep}\De_\vep\,u(2-\xx-\zz)a\big)+{1\over 1-\xx}v^\vep 
\pr_\xx\big(a_{20}-\De_\vep u \,v\, a\big)=0\,,
}}
(along with two similar obtained by $\xx\leftrightarrow \zz$)
which is {\it{almost}} of the $k=1$ form (\ref{fset}).
In fact what the $k=1$ analysis of the last section implies
is that the solutions to (\ref{twoeqsket}) are expandable without
loss of generality in terms of the restricted set of Jack
polynomials $P^{(\vep)}_{\lambda\,0}(\xx,\zz)
\propto\tilde{P}^{(\vep)}_{\lambda}(\xx,\zz)$.  

For $\vep=2$ we may solve these equations directly using the
analysis of the last section.  We may find that
\eqna{partsolstice}{\eqalign{
a_{00}&\ea & u^{4}\De_2\,v\,a+u^{2}D_2\Big({g_1(\xx)-g_1(\zz)\over \xx-\zz}\Big)\,,\cr
a_{01}&\ea & (uv)^2\,\De_2(\xx+\zz-2 u)a+v^2D_2\Big({g_2(\xx)-g_2(\zz)\over \xx-\zz}\Big)\,,\cr
a_{11}&\ea & v^{2}\De_2 u(\xx+\zz-2 )a+v^{2}D_2\Big({g_3(\xx)-g_3(\zz)\over \xx-\zz}\Big)\,,\cr
a_{20}&\ea & \De_2\,u\,v\,a+D_2\Big({g_4(\xx)-g_4(\zz)\over \xx-\zz}\Big)\,,}
}
where 
\eqna{wherestice}{
g_1(\xx)&\ea&-{\xx-1\over \xx}g_2(\xx)+A_0{1\over \xx}+A_1+A_2\xx+A_3\xx^2\,,\cr
g_4(\xx)&\ea&(\xx-1)g_3(\xx)+B_1+B_2\xx+B_3\xx^2+B_4\xx^4\,,
}
with $A_i,\,B_i$ being arbitrary constants. Similarly solving the first equation in
(\ref{ward}) and using (\ref{opidnet}) we find that,
\eqn{lastpartstice}{
a_{10}=-u^{2}\De_2\,v(\xx+\zz)a+D_2\Big({g_5(\xx)-g_5(\zz)\over \xx-\zz}\Big)+u^2D_2\Big({g_6(\xx)-g_6(\zz)\over \xx-\zz}\Big)\,,
} 
where
\eqna{furtherexps}{
g_5(\xx)&\ea & (\xx-1)g_2(\xx)+B_1'+B_2'\xx+B_3'\xx^2+B_4'\xx^3\,,\cr
g_6(\xx)&\ea & -{\xx-1\over \xx}g_3(\xx)+A_0'{1\over \xx}+A_1'+A_2'\xx+A_3'\xx^2\,,
}
with $A_{\smash{i}}',\,B_{\smash{i}}'$ being arbitrary constants.
With these expressions for $a_{ij}$ we may write the full solution to
the Ward identities for
the $\vep=2$ case as
\eqna{fullsoltwoostice}{\eqalign{
G^{(2)}_{2}(\xx,\zz;\al,\bal)&\ea & u^{4}\De_2\,(\xx\al-1)
(\zz\al-1)(\xx\bal-1)
(\zz\bal-1)a(\xx,\zz)\cr
&{}& +u^{2}
\H^{(2)}_1(\xx,\zz;\al,\bal)+u^{4}\al\bal\,
\H^{(2)}_2(\xx,\zz;\al,\bal)\,,}
}
where
\eqn{whatnow}{
\H^{(2)}_{i}(\xx,\zz;\al,\bal)=D_{2}\Big({(\xx\al-1)(\xx\bal-1)h_i(\xx)-(\zz\al-1)(\zz\bal-1)h_i(\zz)\over
\xx-\zz}\Big)\,,
}
for
\eqn{exstice}{
h_1(\xx)={1\over \xx}g_2(\xx)+A_0{1\over \xx}+B_4'\xx\,,\qquad
h_2(\xx)={1\over \xx}g_3(\xx)+A_0'{1\over \xx}+B_4\xx\,,
}
being arbitrary single variable functions.

Using expansions in terms of $P^{(\vep)}_{\lambda\,0}(\xx,\zz)$, then
for general $\vep$ we  may  relatively
easily find that, for arbitrary $b_{1\,\lambda},\,b_{2\,\lambda}$,
\eqna{partsolo}{\eqalign{
a_{00}&\ea & u^{2\vep}\De_\vep\,v\,a+u^{\vep}\sum_\lambda\big(-b_{1\,\lambda}+
b_{1\,\lambda+1}\big)\tilde{P}^{(\vep)}_{\lambda}\,,\cr
a_{01}&\ea & (uv)^\vep\,\De_\vep(\xx+\zz-2 u)a+v^\vep\sum_\lambda b_{1\,\lambda}
\tilde{P}^{(\vep)}_\lambda\,,\cr
a_{11}&\ea & v^{\vep}\De_\vep u(\xx+\zz-2 )a+v^{\vep}\sum_\lambda b_{2\,\lambda}
\tilde{P}^{(\vep)}_\lambda\,,\cr
a_{20}&\ea & \De_\vep\,u\,v\,a+\sum_{\lambda}\big(b_{2\,\lambda-1}-b_{2\,\lambda}\big)
\tilde{P}^{(\vep)}_{\lambda}\,,}
}
then from the first equation in (\ref{ward}) and using (\ref{opidnet}) we
may find,
\eqn{lastpart}{
a_{10}=-u^{\vep}\De_\vep\,v(\xx+\zz)a+\sum_\lambda\big(b_{1\,\lambda-1}
-b_{1\,\lambda}\big)\tilde{P}^{(\vep)}_\lambda+u^\vep\sum_\lambda
\big(-b_{2\,\lambda}+b_{2\,\lambda+1}\big)\tilde{P}^{(\vep)}_\lambda\,.
}
Taking account of these solutions then we may
easily reassemble them to find that
\eqna{fullsoltwoo}{\eqalign{
G^{(\vep)}_{2}(\xx,\zz;\al,\bal)&\ea & u^{2\vep}\De_\vep\,(\xx\al-1)
(\zz\al-1)(\xx\bal-1)
(\zz\bal-1)a(\xx,\zz)\cr
&{}& +u^{\vep}
\H^{(\vep)}_1(\xx,\zz;\al,\bal)+u^{2\vep}\al\bal\,
\H^{(\vep)}_2(\xx,\zz;\al,\bal)\,,}
}
where $\H^{(\vep)}_i(\xx,\zz;\al,\bal)$ 
is given by (\ref{lastlyy}) with $b_{1\,\lambda}\to b_{i\,\lambda}$.

Due to 
the arbitrary freedom we have in $a(\xx,\zz)$, we may  re-define
\eqn{letting}{
a(\xx,\zz)\to a(\xx,\zz)- u^{-2}\,(\De_\vep)^{-1}\,\sum_\lambda
b_{2\,\lambda}
\,\,
{\tilde{P}}^{(\vep)}_{\lambda+1}(\xx,\zz)\,,
}
whereby using that for $p=0,1,2$,
\eqn{idents}{
\De_\vep \Big({\xx+\zz\over u}\Big)^p
\,{\tilde{P}}^{(\vep)}_{\lambda}(\xx,\zz)=
\E^{(\vep)}_{\lambda\,0}\,{\tilde{P}}^{(\vep)}_{\lambda-p}(\xx,\zz)\,,
} 
we may dispose of
$b_{2\,\lambda}$ contributions in (\ref{fullsoltwoo}) 
 for $\vep\neq 1,2$.  
For $\vep=2$, those left over may be easily computed to be given by
\eqn{bracing}{
G^{(2)}_2(\xx,\zz;\al,\bal)\sim u^2\,
\sum_{\lambda}\,b_{2\,\lambda-2}\,{\lambda\over
\lambda+2}\big(
{\tilde{P}}^{(\vep)}_{\lambda-1}(\xx,\zz)
-(\al+\bal){\tilde{P}}^{(\vep)}_{\lambda}(\xx,\zz)+\al\bal\,
{\tilde{P}}^{(\vep)}_{\lambda+1}(\xx,\zz)\big)\,,}
so that by re-defining
\eqn{redefoo}{
b_{1\,\lambda}\to b_{1\,\lambda}-{\lambda\over \lambda+2}b_{2\,\lambda-2}\,,
}
we may absorb $b_{2\,\lambda}$ into $b_{1\,\lambda}$ 
contributions in this case as well.{\foot{We 
see later, when performing the conformal
partial wave expansion,  that such re-absorption
of $b_{2\,\lambda}$ contributions is not possible for $\vep=1$.}}

In the case $\vep=2$, a more direct way of seeing absorbtion for one
of $\H_1^{(2)},\,\H_2^{(2)}$ is through use of the  following identity, namely,
\eqna{scrumpet}{
(\xx\zz)^4D_2\,(\xx\al-1)(\zz\al-1)(\xx\bal-1)(\zz\bal-1)
  \,{1\over \xx \zz}\,D_2\Big({h(\xx)-h(\zz)\over \xx-\zz}\Big)\cr
=(\xx\zz)^2\H^{(2)}_1(\xx,\zz;\al,\bal)+
\al\bal(\xx\zz)^4\H^{(2)}_2(\xx,\zz;\al,\bal)\,,
}
for
\eqn{gh}{
h_1(\xx)=-2 h(\xx)+\xx h'(\xx)\,,\qquad h_2(\xx)=-{1\over \xx}h'(\xx)\,.
}
Thus we may absorb $\H^{(2)}_2$ via modifications of the unrestricted
two-variable part and of $\H^{(2)}_1$, which is essentially the point
made above using expansions.

We thus have that,
\eqn{fullsoltw}{
G^{(\vep)}_{2}(\xx,\zz;\al,\bal)\big|_{\vep\neq 1}
= u^{2\vep}\De_\vep\,(\xx\al-1)
(\zz\al-1)(\xx\bal-1)
(\zz\bal-1)a(\xx,\zz)+u^\vep\,\H^{(\vep)}_1(\xx,\zz;\al,\bal)\,,
}
whereby for $\vep\neq 2$ we may absorb $\H^{(\vep)}_1$ through
redefinitions of $a(\xx,\zz)$.{\foot{We make the same
choice of modification of the two variable function as before in (\ref{helo}): 
we may find that
\eqna{soper}{
\eqalign{
{\lim_{\mu\to 0}} \,u^{2\vep}\De_\vep
(\xx\al-1)(\zz\al-1)(\xx\bal-1)(\zz\bal-1)(\De_\vep)^{-1}
\sum_{\lambda}b_{1\lambda}
\,
{(2\vep)_{\lambda-1}\over \Gamma(\lambda)}\,
P^{(\vep)}_{\lambda+\mu-\vep-1\,\mu-\vep}(\xx,\zz)= 
u^\vep\,\H_1(\xx,\zz;\al,\bal)
\,.\nonumber}
}
}}

Further absorbtion of the $\H^{(2)}_1$ contributions for $\vep= 2$ is not possible by the modification $a(\xx,\zz)\to a(\xx,\zz)+f(\xx,\zz)/u$. Indeed,  $\H^{(2)}_1$ contains three out of the six monomials in the polynomial in $\alpha, \alpha'$. Making a modification of $a(\xx,\zz)$ which might absorb these three terms, but which should not affect the remaining three, we must require that 
\eqn{satisfywhatmydear}{
 D_2\,u^2 f=0\,,\qquad D_2 u(\xx+\zz)f=0\,,\qquad D_2(\xx+\zz)^2
 f=0\,.
}
These equations are incompatible, as is easily seen by using the first of them in the other two and remembering that $f(\xx,\zz)$ must be a symmetric function. 

However, in the case $\vep= 2$ there is a different mechanism of absorption. Using the identity
\eqna{conversion}{
&&D_2\Big((\xx\al-1)(\zz\al-1)(\xx\bal-1)(\zz\bal-1)\Big({h(\xx)-h(\zz)\over
  (\xx-\zz)^3}\Big)\Big)\cr
&&-\al\bal
D_2\Big({(\xx\al-1)(\xx\bal-1)h(x)-(\zz\al-1)(\zz\bal-1)h(\zz)\over 
\xx-\zz}\Big)\cr
&&=
{1\over
  (\xx-\zz)^4}\,\big((\xx\al-1)^2(\xx\bal-1)^2h'(\xx)+(\zz\al-1)^2(\zz\bal-1)^2h'(\zz)\big)\,,
} 
we can rewrite (\ref{fullsoltw}) as follows:
\eqna{nameo}{
&&G^{(2)}_{2}(\xx,\zz;\al,\bal)\cr
&&=-{(\xx\zz)^4\over (\xx-\zz)^4}\,\De\Big((\xx\al-1)(\zz\al-1)(\xx\bal-1)(\zz\bal-1)F(\xx,\zz)\Big)\cr
&&-{1\over
  (\xx-\zz)^4}\,\big((\xx\al-1)^2(\xx\bal-1)^2F(\xx,\xx)+(\zz\al-1)^2(\zz\bal-1)^2F(\zz,\zz)\big)\,,
}   
where we have  used the differential operator $\Delta$ defined by the relation
\eqn{ourd}{
D_2\,{1\over (\xx-\zz)^2}f(\xx,\zz)=-{1\over (\xx-\zz)^4}\,\De\,f(\xx,\zz)\,.
}
The new function  $F(\xx,\zz)$ is obtained from $a(\xx,\zz)$ and $h(\xx)$ in the following way:
\eqn{setttt}{
F(\xx,\zz)={(\xx-\zz)^2}a(\xx,\zz)+{h(\xx)-h(\zz)\over \xx-\zz}\,.
}
Thus, assuming regularity of the function $a(\xx,\zz)$ as $\xx \to \zz$, the single variable function $h(\xx)$ is identified with   $F(\xx,\xx)$,
\eqn{have}{
F(\xx,\xx)=h'(\xx)\,.
}
We conclude that in this case the set of two functions $a(\xx,\zz)$ and $h(\xx)$ has effectively been reduced to the {\it single function of two variables} $F(\xx,\zz)$. The form (\ref{nameo}) first appeared in \cite{Heslop6d}.

We remark that both forms of the amplitude, (\ref{fullsoltw}) and (\ref{nameo}), are regular in the limit $\xx \to \zz$ under the assumption that the functions $a(\xx,\zz)$ and $h(\xx)$ are differentiable.

\subsubsection{The general $k$ case}\label{genkn4}
A similar analysis as that done for $k=2$ above becomes cumbersome, to
say that least, for higher $k$.  Indeed we can see this already
for the $k=3$ case which is considered in  Appendix \ref{AppC}. 
A way out of this difficulty is to find a sufficiently arbitrary
solution to the Ward identities (\ref{condchvar}) whose freedom
may then be used to reduce the constraining equations on 
the correlation function to a set of simpler ones.  This is what we do next.

As a partial solution to the Ward identities (\ref{condchvar}) it seems natural to consider 
\eqna{partialsol}{
\F_k^{(\vep)}(\xx,\zz;\al,\bal)=\!\!\!\!
\sum_{{0\leq n+m\leq k-2}}\!\!\!\!&& \!\!\!\!\!\!\!\!\!\!\!\,G_{nm}(\xx,\zz)(\al\bal)^n
\big((\al-1)(\bal-1)\big)^m\cr
&&\!\!\!\!\!\!\!\!\!\!\!\!\!\!\times \De_\vep (\xx\al-1)(\zz \al-1 )(\xx\bal-1)(\zz\bal-1) 
\F_{nm}(\xx,\zz)\,,
}
where  $G_{nm},\,\F_{nm}$ are symmetric in $\xx,\zz$ and the restriction
$m+n\leq k-2$ follows as the maximum degree of $G^{(\vep)}_k$ in $\al,\bal$
is $k$.   

The constraints on this, if it is to satisfy the Ward identities for
general $\F_{mn}$,
 are greatly simplified if we first
prove that,
\eqn{firstprove}{
\xx {\pr\over \pr \xx}\De_\vep (\xx\al-1)(\zz\al-1)f(\xx,\zz)\Big|_{\al\to
1/\xx}=\vep \De_\vep (\al (\xx+\zz)-2)f(\xx,\zz)\Big|_{\al\to 1/\xx}\,.
}
With the identity (\ref{firstprove}), then the Ward identities (\ref{condchvar}), with 
(\ref{partialsol}),
simply imply that,
\eqn{implication}{
\Big((\xx-1)\xx{\pr \over \pr \xx}+\vep(m-(n+2)(\xx-1))\Big)
G_{nm}(\xx,\zz)=0\,,
}  
which may be simply solved by, taking account of $\xx,\zz$ symmetry, 
\eqn{sol}{
G_{nm}(\xx,\zz)\propto u^{(2 +m+n)\vep}v^{-m\vep}\,.
}

Due to the simple action of $\De_\vep$ on Jack polynomials, perhaps the
simplest way of proving (\ref{firstprove}) is to first expand,
\eqn{expandf}{
f(\xx,\zz)=(\De_\vep)^{-1}\sum_{\lambda_1,\lambda_2}
\,a_{\lambda_1\,\lambda_2}\,P^{(\vep)}_{\lambda_1\,\lambda_2}(\xx,\zz)\,,
}
for arbitrary $a_{\lambda_1\,\lambda_2}$.
By using the identity (from results in Appendix \ref{AppB})
\eqna{id}{\eqalign{
&{}& \xx{\pr \over \pr \xx}P_{\lambda_1\,\lambda_2}^{(\vep)}(\xx,\zz)=\half
(\lambda_1+\lambda_2)P^{(\vep)}_{\lambda_1\,\lambda_2}(\xx,\zz)\cr
&& \qquad \qquad \qquad \qquad +{1\over
\xx-\zz}\,{\lambda_-(\lambda_-+2 \vep)\over
2(\lambda_-+\vep)}\,\big(P^{(\vep)}_{\lambda_1+1\,\lambda_2}(\xx,\zz)
-P^{(\vep)}_{\lambda_1\,\lambda_2+1}(\xx,\zz)\big)\,,}
}
along with repeated application of
$(\xx+\zz)P^{(\vep)}_{\lambda_1\,\lambda_2}(\xx,\zz)$ as in Appendix \ref{AppB}, we
may determine that, with (\ref{expandf}),
\eqn{determine}{
\xx {\pr \over \pr \xx} \De_\vep (\xx\al-1)
(\zz\al-1)f(\xx,\zz)\Big|_{\al\to 1/\xx}={1\over
\xx}\Big(f_+(\xx,\zz)+f_-(\xx,\zz)\Big)\,,
}
where,
\eqn{fs}{
f_+(\xx,\zz)=\vep(\vep-1)\sum_{\lambda_1,\lambda_2}
\Big(a_{\lambda_1-1\,\lambda_2}{1\over
\lambda_1+\vep}+a_{\lambda_1\,\lambda_2-1}\,{1\over
\lambda_2}\Big)P^{(\vep)}_{\lambda_1\,\lambda_2}(\xx,\zz)\,,
}
being symmetric in $\xx,\zz$, and
\eqna{fa}{\eqalign{
f_-(\xx,\zz)=-\vep{1\over \xx-\zz}
\sum_{\lambda_1,\lambda_2}\Big(\!\!\!\!\!\!\!\!\!&{}&
a_{\lambda_1-1\,\lambda_2}{(\lambda_-+2\vep-1)(\lambda_-+2\vep-2)\over
(\lambda_-+\vep-1)(\lambda_-+\vep-2)}\cr
&{}&- 2
a_{\lambda_1-1\,\lambda_2-1}\,{\lambda_-(\lambda_-+2\vep)
+(\vep-1)(2\vep+1)\over
(\lambda_-+\vep-1)(\lambda_-+\vep+1)}\cr
&{}&+ a_{\lambda_1\,\lambda_2-2}{(\lambda_-+1)(\lambda_-+2)
\over (\lambda_-+\vep+1)(\lambda_-+\vep+2)}\Big)
P^{(\vep)}_{\lambda_1\,\lambda_2}(\xx,\zz),}
}
being anti-symmetric in $\xx,\zz$.  It is not then difficult to show
that (\ref{determine}) with (\ref{fs}) and (\ref{fa})  
coincides with the right hand
side of (\ref{firstprove}), whereby we may determine that,
\eqna{determinemoreo}{\eqalign{
f_{+}(\xx,\zz)&\ea &\vep \De_\vep (\xx+\zz)f(\xx,\zz)- \vep {\xx+\zz\over u} 
\De_\vep f(\xx,\zz)\,,\cr
f_{-}(\xx,\zz)&\ea &\vep {\xx-\zz \over
u}\De_\vep f(\xx,\zz)\,.}
}

In summary the partial solution we have found to the Ward identities
(\ref{condchvar}) 
is given by,

\eqna{partialsoltw}{\eqalign{
&& \F_k^{(\vep)}(\xx,\zz;\al,\bal)=\!\!\!\!\!\sum_{{0\leq n+m\leq k-2}}\!\!\!\!\!
\,u^{(m+n+2)\vep}\,v^{-m\vep}\,(\al\bal)^n\,\big((\al-1)(\bal-1))^{m}\cr
&&\qquad\qquad\qquad \qquad \times \De_\vep (\xx\al-1)(\zz \al-1 ) 
(\xx\bal-1)(\zz\bal-1)\F_{nm}(\xx,\zz)\,,}
}
in terms of  $\half k(k-1)$ arbitrary two-variable functions 
$\F_{nm}(\xx,\zz)$.
For $k=2$, ({\ref{partialsoltw}) clearly coincides 
with (\ref{fullsoltw}).  When $\vep=1$, then
$\De_\vep\to 1$ and the solution coincides with two-variable part of the
four-dimensional solution obtained earlier.

We may use (\ref{partialsoltw}) to find the solution to
the Ward identities (\ref{condchvar}) more generally.  To show this we use the
arbitrariness in $\F_{nm}(\xx,\zz)$ to cancel all 
$a_{nm},\,m\geq 2,$
 in the
original four point function
$G_k^{(\vep)}(\xx,\zz;\al,\bal)$ so that{\foot{This is certainly possible as
there are $\half k(k-1)$ $a_{nm},\,m\geq 2,$ in 
$G_k^{(\vep)}(\xx,\zz;\al,\bal)$ - in fact there is some freedom left
over
in $\F_{mn}(\xx,\zz)$ due to $\De_\vep$ generally having a non-trivial
kernel, which has consequences for the restricted solutions we obtain.}}
\eqna{reducedequation}{
&& G_k^{(\vep)}(\xx,\zz;\al,\bal)-\F_k^{(\vep)}(\xx,\zz;\al,\bal)\cr
&& =\sum_{n=0}^k
u^{n\vep}
(\al\bal)^n\, {a}'_{n0}(\xx,\zz)+ \sum_{n=0}^{k-1}u^{n\vep}\Big({u\over
  v}\Big)^\vep
(\al\bal)^n(\al-1)(\bal-1)\,{a}'_{n1}(\xx,\zz)\,,
}
for some symmetric ${a}'_{n0}(\xx,\zz)\,,a'_{n1}(\xx,\zz)$.
The left-hand side of (\ref{reducedequation})
obviously satisfies the Ward identities ({\ref{condchvar}) and this
places constraints on ${a}'_{n0}\,,a'_{n1}$.
These constraint equations following from the Ward
identities are given by
\eqna{constraintw}{
&& {1\over 1-\xx}v^\vep\,\pr_\xx a'_{00}-{1\over \xx}u^\vep\,\pr_\xx
    a'_{01}=0\,,\cr
&& {1\over 1-\xx}v^\vep\,\pr_\xx a'_{10}+\pr_\xx\,a'_{01}-{1\over \xx}
u^\vep\,\pr_\xx
    a'_{11}=0\,,\cr
&& \qquad\qquad\qquad \vdots\cr
&& {1\over 1-\xx}v^\vep\,\pr_\xx a'_{n0}+\pr_\xx\,a'_{n-1\,1}-{1\over \xx}u^\vep\,\pr_\xx
    a'_{n1}=0\,,\cr
&& \qquad\qquad\qquad \vdots\cr
&& {1\over 1-\xx}v^\vep\,\pr_\xx a'_{k-1\,0}+\pr_\xx\,a'_{{k-2}\,1}-
{1\over \xx}u^\vep\,\pr_\xx
    a'_{k-1\,1}=0\,,\cr
&& {1\over 1-\xx}v^\vep\,\pr_\xx a'_{k0}+\pr_\xx\,a'_{k-1\,1}=0\,.}
Using the $k=1$ analysis we may immediately solve the first and last
of these equations to find that
\eqna{tofindforfirstandlast}{
&& a'_{00}=u^\vep\sum_{\lambda}\big(-b_{1\lambda}+b_{1\,\lambda+1}\big)
\tilde{P}^{(\vep)}_\lambda\,,\qquad
a_{01}'=v^{\vep}\sum_\lambda
b_{1\lambda}\,\tilde{P}^{(\vep)}_\lambda\,,\cr
&& a'_{k0}=\sum_{\lambda}\big(b_{k\,\lambda-1}-b_{k\lambda}\big)
\tilde{P}^{(\vep)}_\lambda\,,\quad \qquad
a_{k-1\,1}'=v^\vep\sum_\lambda
b_{k\lambda}\,\tilde{P}^{(\vep)}_\lambda\,,
}
for arbitrary $b_{1\lambda},\,b_{k\lambda}$.  Assuming
\eqn{assumpt}{
a'_{n-1\, 1}(\xx,\zz)=v^{\vep}\sum_\lambda\,b_{n \lambda}\,
\tilde{P}^{(\vep)}_\lambda(\xx,\zz)\,,\qquad n=1,\dots k\,,
}
 we may prove by induction that this is consistent with
(\ref{constraintw}) for
\eqn{forothergiveb}{
a'_{n0}(\xx,\zz)=\sum_\lambda\big(b_{n\,\lambda-1}-b_{n\,\lambda}+u^\vep(
-b_{n+1\,\lambda}+b_{n+1\,\lambda+1})\big)\tilde{P}^{(\vep)}_\lambda(\xx,\zz)\,.}
Substituting (\ref{assumpt}) and (\ref{forothergiveb}) into 
(\ref{reducedequation}), we may thus find that the full solution
to the Ward identities (\ref{condchvar}) is given by
\eqn{fullsol}{
G_k^{(\vep)}(\xx,\zz;\al,\bal)=\F_k^{(\vep)}(\xx,\zz;\al,\bal)
+\sum_{n=1}^{k}u^{n\vep}(\al\bal)^{n-1} 
\H_{n}^{(\vep)}(\xx,\zz;\al,\bal)\,,
}
where $\F_k^{(\vep)}(\xx,\zz;\al,\bal)$ is given by (\ref{partialsoltw})
and $\H^{(\vep)}_j(\xx,\zz;\al,\bal)$
is given by (\ref{lastlyy}) for $b_{1\lambda}\to b_{j\lambda}$.
 The $k$ different $b_{n\lambda}$ are
arbitrary and in four dimensions their contribution
corresponds to $k$ different single variable contributions.
In fact, for $\vep\neq 1$ we may further use the arbitrariness in
$\F_{mn}(\xx,\zz)$ to cancel most $\H^{(\vep)}_n$ contributions,
similarly to the $k=2$ case.{\foot{This is easily seen as follows:
 we may redefine $\F_{n-2\,0}$ as in
(\ref{letting}), for $b_{2\,\lambda}\to b_{n\,\lambda}$, to absorb 
$\H_{n}^{(\vep)},\,\,n\geq 2,$ for $\vep\neq 1,2$. 
After redefining $\F_{n-2\,0}$ in this way, the remaining contributions
from $\H_{n}^{(\vep)},\,\,n\geq 2,$ for $\vep=2$
may be absorbed into $\H_{n-1}^{(\vep)}$ as in (\ref{redefoo})
for $b_{1\,\lambda}\to b_{n-1\,\lambda}$.  In this way we
end up with only $\H_{1}^{(\vep)}$ contributions - these may be
absorbed,
as in the $k=2$ case, into $\F_{00}$ for $\vep\neq 1,2$.  We stress that for odd
dimensions the modification to $\F_{00}$ which cancels $\H_1^{(\vep)}$
is defined only in the limit $\mu\to 0$, with $\mu$ being some
regulating
parameter.}  In this way we find (\ref{mainsol2}).

We may show directly that the $\H_j(\xx,\zz,\al,\bal)$
contributions to the four-point function of the form 
\eqn{formofrest}{
G_{j}(\xx,\zz)\,(\al\bal)^j
\,\H^{(\vep)}_j(\xx,\zz;\al,\bal)\,,}
satisfy the Ward identities (\ref{condchvar}), where $G_{j}$ are some symmetric
functions,
whereby using
\eqn{identitty}{
\Big(\xx {\pr\over \pr \xx}-\vep\,\al {\pr\over \pr
\al}+\vep\Big)\H^{(\vep)}_j(\xx,\zz;\al,\bal)\Big|_{\al\to 1/\xx}=0\,,
}
the Ward identities simply imply that
\eqn{soltwooo}{
G_{j}(\xx,\zz)\propto u^{(j+1)\vep}\,.
}

The proof of (\ref{identitty})  is trivial once we show that,
\eqna{eqnanitent}{\eqalign{
&& \Big(\xx{\pr \over \pr \xx}-\vep \al{\pr \over
  \pr\al}+\vep\Big)f_{\lambda}(\xx,\zz;\al)\Big|_{\al\to 1/\xx}=
{\half}(\lambda+2\vep){\tilde{P}}_{\lambda}^{(\vep)}(\xx,\zz)\cr
&& +{1\over \xx-\zz}{1\over
2(\lambda+\vep)}
\big(\lambda(\lambda+1){\tilde{P}}_{\lambda+1}^{(\vep)}(\xx,\zz)-\xx
\zz (\lambda+2 \vep-1)(\lambda+2
\vep){\tilde{P}}_{\lambda-1}^{(\vep)}(\xx,\zz)\big)\,,
}}
for $f_{\lambda}(\xx,\zz;\al)={\tilde{P}}_{\lambda}^{(\vep)}(\xx,\zz)$ or
$f_{\lambda}(\xx,\zz;\al)=\al {\tilde{P}}_{\lambda+1}^{(\vep)}(\xx,\zz)$.

In analogy with the $SO(3)$ general $k$ solution (\ref{solgennetyeah}), we may write
(\ref{partialsoltw}) in a basis independent manner by employing
a new operator,
\eqna{newoperator}{
\DD'_\vep&\ea&\bigg(D_\vep-\vep{1\over
v}\Big(D_0^--D_1^++\vep{\pr\over \pr\si}\si\Big)\tau{\pr\over \pr \tau}\cr
&& +\vep{1\over uv}\Big(-v
D^+_1+\vep\Big(v{\pr\over \pr \si}\si+{\pr\over \pr \tau}\tau +1\Big)\Big)\Big({\pr \over \pr
\si}\si+{\pr\over \pr \tau}\tau\Big)\bigg)^{\vep-1}\,,}
for $\si=\al\bal$, $\tau=(\al-1)(\bal-1)$, where $D^+_0,\,D^+_1$ are 
given in Appendix \ref{AppB}, whereby we may write
\eqna{basisind}{
\F^{(\vep)}_k(\xx,\zz;\al,\bal)&\ea&\big(\si^2\,\DD'_{\vep}\,u v+\tau^2\, \DD_\vep'\,
u+\DD_\vep'\,v
-\si\,\DD_\vep'\,v(u+1-v)\cr
&&-\tau\,\DD_\vep'\,(u+v-1)-\si\tau\,\DD_\vep'\,
u(v+1-u)\big)\M^{(\vep)}_{k-2}(\xx,\zz;\al,\bal)\,,
}
for arbitrary $\M^{(\vep)}_{k-2}(\xx,\zz;\al,\bal)$ of degree $k-2$ in
$\al$ or
$\bal$. For $\vep=1$, when $\DD'_\vep\to 1$, then the prefactor in
(\ref{basisind}) factorizes into
$(\xx\al-1)(\zz\al-1)(\xx\bal-1)(\zz\bal-1)$
and we recover the unrestricted two variable part of the solution in
$d=4$, $\N=4$.

\section{The conformal partial wave expansion for the $k=2$ case}\label{CPWA}

Due to invariance of the  four point function
under R symmetry, we may expand
\eqn{proj}{\eqalign{
G_k^{(\vep)}(\xx,\zz;\al,\bal) = \sum_{a=0}^k\sum_{b=0}^{a} \,
Y_{a b}^{(n)}(\al,\bal)\,A_{a
 b}(\xx,\zz)\,,}
}
where $Y_{a b}^{(n)}(\al,\bal)$ 
(of degree at most $a$ in $\al$ or $\bal$) 
are eigenfunctions of the $SO(n)$ Casimir operator
 $L^2={1\over 2}\,L^{\mathbf{i}\mathbf{j}}\,L^{\mathbf{i}\mathbf{j}}$, where $L^{\mathbf{i}\mathbf{j}}=L^{\mathbf{i}\mathbf{j}}_1+L^{\mathbf{i}\mathbf{j}}_2$ with
\eqn{gens}{
L^{\mathbf{i}\mathbf{j}}=Y^{\mathbf{i}}\,{\pr\over \pr Y^{\mathbf{j}}}-Y^{\mathbf{j}}\,{\pr\over \pr Y^{\mathbf{i}}}\,,
}
in terms of the null variables in (\ref{lightR}), 
so that
\eqn{exp}{
L^2\,Y_{a b}^{(n)}(\al,\bal)
=- 2\,(a(a+n-3)+b(b+1))\,Y_{a b}^{(n)}(\al,\bal)\,.
}
An expansion for $Y_{ab}^{(n)}(\al,\bal)$ is given in \cite{fadho}, for
general $k$ and $n$, in terms of Jack polynomials.  Other expansions
may be found in \cite{NO} where these harmonic polynomials are given a
more complete treatment.

In terms of superconformal four-point functions, 
$Y_{a b}^{(n)}(\al,\bal)$ correspond to projections on to
irreducible representations in the product of two rank-$k$ symmetric
traceless representations of $SO(n)$ whereby for $d=3$ and
$SO(8)$ then $(a,b)\rightarrow [0,a{-b},2b,0]$, for $d=4$ and $SO(6)$
then $(a,b)\rightarrow [a{-b}, 2 b, a{-b}]$, 
for $d=5$ and $SO(3)$ the relevant cases
are $(a,a)\rightarrow [2 a]$,{\foot{This is related to the fact that
for $n=3$ then $\al=\bal$, and $Y_{ab}^{(3)}(\al,\bal)=0$
unless $b>a-2$.  Otherwise, in terms of Legendre polynomials,  
$Y_{aa}^{(3)}(\al,\bal)\propto
P_{2a}(2\,\al-1)$, $Y_{a\,a-1}^{(3)}(\al,\bal)\propto
P_{2a-1}(2\,\al-1)$ \cite{NO}.}}  for $d=6$ and $SO(5)$ then
$(a,b)\rightarrow [2(a{-b}),2b]$.

Also, the functions $A_{a b}(\xx,\zz)$ generally 
admit an expansion in terms of appropriate conformal partial
waves $\GG^{(\ell)}_\De(\xx,\zz)$ which then represents a rank-$\ell$
symmetric traceless  operator, of dimension $\De$ belonging to the  $(a, b)$
R symmetry representation, in the operator
product expansion of $ \O^k(x_1,y_1)\O^k(x_2,y_2)$.

{}From \cite{fadho} and in $d=2(\vep+1)$ dimensions, such conformal partial
waves may be expanded as
\eqn{confpartwave}{
\GG^{(\ell)}_\De(\xx,\zz)=\sum_{m,n\geq 0}\,r_{nm}(\De,\ell)\,P_{{1\over
2}(\De+\ell)+m\,{1\over 2}(\De-\ell)+n}^{(\vep)}(\xx,\zz)\,,
}
where the coefficients $r_{mn}$ are given by (we are assuming that
the conformal dimensions of the operators at $x_1,\,x_2$ are the same,
similarly for those at $x_3,\,x_4$)
\eqn{rrh}{\eqalign{
r_{mn}(\De,\ell)=(-{\ts{1\over 2}})^\ell(\half(\De+\ell))_m{}^2\,
(\half(\De-\ell)-\vep)_n{}^2\,\hat{r}_{mn}(\De,\ell) \, ,}
}
with $\hat{r}_{mn}$ being given recursively by,{\foot{This recurrence
relation has been solved explicitly in \cite{fadho} in terms of ${}_4F_3$
generalized hypergeometric functions.}}
\eqna{rrhhh}{\eqalign{
&& \big(m(m+\De+\ell-1)+n(n+\De-\ell-2\,\vep-1)\big)\hat{r}_{mn}\cr
&& ={\ell+m-n-1+2\,\vep\over \ell+m-n-1+\vep}\,\hat{r}_{m-1
n}+{\ell+m-n+1\over \ell+m-n+1+\vep}\,\hat{r}_{m n-1}\,,
}}
with $\hat{r}_{00}=(2\vep)_\ell/(\vep)_\ell$.

For the purposes of analysing the $k=2$ case we need only $Y_{a b}^{(n)}(\al,
\bal),\,a=0,1,2,$
\eqna{Yab}{\eqalign{
&& Y_{00}^{(n)}(\al,\bal) = 1  \, , \cr
&& Y_{10}^{(n)}(\al,\bal) =  \si - \tau \, , \cr
&& Y_{11}^{(n)}(\al,\bal) = \si + \tau - {\ts {2\over n}}\, , \cr
&& Y_{20}^{(n)}(\al,\bal) = \si^2 + \tau^2 - 2\si\,\tau
- {\ts {2\over n-2}}(\si + \tau) + {\ts {2\over (n-2)(n-1)}} \, , \cr
&& Y_{21}^{(n)}(\al,\bal) =  \si^2 - \tau^2 - {\ts {4\over n+2}}
(\si - \tau) \, , \cr
&& Y_{22}^{(n)}(\al,\bal) =  \si^2 + \tau^2 + 4\si\,\tau 
- {\ts {8\over n+4}}(\si + \tau) + {\ts {8\over (n+2)(n+4)}} \,
,}}
where $\si=\al\bal$ and $\tau=(\al-1)(\bal-1)$, 
which have been computed in \cite{NO}.  

In terms of components, for
\eqn{deftorro}{
G_2^{(\vep)}(\xx,\zz;\al,\bal)=\sum_{{0\leq m+n\leq
2}}(\al\bal)^n\,\big((\al-1)(\bal-1)\big)^m\,b_{n m}(u,v)\,,}
then
\eqna{aprojs}{\eqalign{
&& A_{00}=b_{00}+{\ts {1\over
n}}\,(b_{10}+b_{01})+{\ts {2\over
(n+2)(n-1)}}\,(b_{20}+b_{02})+{\ts{n-2\over (n-1)n
(n+2)}}\,b_{11}\,,\cr
&& A_{11}=\half(b_{10}+b_{01})+{\ts{2\,n\over
(n-2)(n+4)}}\,(b_{20}+b_{02})+{\ts{n-4\over
(n-2)(n+4)}}\,b_{11}\,,\cr
&& A_{10}=\half (b_{10}-b_{01})+{\ts{2\over n+2}}\,(b_{20}-b_{02})\,,\cr
&& A_{20}=\thir (b_{20}+b_{02})-{\ts{1\over 6}}\,b_{11}\,,\cr
&& A_{21}=\half (b_{20}-b_{02})\,,\cr
&& A_{22}={\ts{1\over 6}}(b_{20}+b_{11}+b_{02})\,,}
}
so that we may easily read off various contributions from (\ref{fullsoltw}).

\subsection{The analysis of the unrestricted two-variable part}

We start with contributions of long superconformal multiplets, where
the dimensions of  operators contributing in the operator product
expansion are unrestricted, save for unitarity constraints.
  The
corresponding contributions from the two-variable functions to the 
projections on to the various $R$ symmetry channels are given by, from
(\ref{fullsoltwoo}), with (\ref{aprojs}),
\eqna{longproj}{\eqalign{
&& A^{\rm long}_{22}={\ts{1\over 6}}\,u^{2\vep}\,\De_\vep\,u^2 \,a\,,\cr
&& A^{\rm long}_{21}=\half\, u^{2\vep}\,\De_\vep\,u (v-1)\,a\,,\cr
&& A^{\rm long}_{20}={\ts{1\over 6}}\,u^{2 \vep}\,
\De_\vep\,u(3(v+1)-u)\,a\,,\cr
&& A^{\rm long}_{10}=\half \, u^{2\vep}\,
\De_\vep\,(v-1)\big((v+1)-{\ts{n-2\over n+2}}\, u)\,a\,,\cr
&& A^{\rm long}_{11}=\half \,
u^{2\vep}\,\De_\vep\,\big((v-1)^2-{\ts{n-4\over
n-2}}\,u\,(v+1)+{\ts{2(n-4)\over (n-2)(n+4)}}\,u^2\big)\,a\,,\cr
&& A^{\rm long}_{00}={\ts{1\over
4}}\,u^{2\vep}\,\De_\vep\,\big((v+1)^2-{\ts{n-4\over n}}(v-1)^2\cr
{}&&\qquad\quad\qquad\qquad\qquad\qquad-{\ts{4(n-2)\over
n(n-1)}}\,u\,(v+1)+{\ts{4(n-2)\over n(n-1)(n+2)}}\,u^2\big)\,a\,.
}}
The simplest possibility is to expand
\eqn{usqg}{
u^2\,a=(\De_\vep)^{-1}\sum_{m,n\geq
0}\,r_{mn}(\De,\ell)
\,P_{{\alpha}_++m\,{\alpha}_-+n}^{(\vep)}(\xx,\zz)\,,\quad \alpha_{\pm}
=\half(\De\pm\ell)-2\,\vep\,,
}
so that
\eqn{solitary}{
A^{\rm long}_{22}(u,v)={\ts{1\over 6}}\,\GG^{(\ell)}_{\De}(\xx,\zz)\,,
}
representing just a single operator in the $(2, 2)$ $R$ symmetry channel.
Using results derived in Appendix \ref{AppD} we may find that, with the choice
of $a$ as in (\ref{usqg}),
\eqna{vmo}{\eqalign{
A^{\rm long}_{21}&&\!\!\!\!={\De-\ell-4\vep\over
\De-\ell-2 \vep-2}\,\GG^{(\ell+1)}_{\De-1}+{1\over 4}\,{\De+\ell-2\vep\over
\De+\ell-2}{\ell(\ell+2\vep-1)\over (\ell+\vep-1)(\ell+\vep)}\,
\GG^{(\ell-1)}_{\De-1}\cr
&&+{1\over 16}\,{(\De+\ell)(\De+\ell+2\vep-2)\over
(\De+\ell-1)(\De+\ell+1)}{(\De-1)(\De-2\vep)\over
(\De-\vep-1)(\De-\vep)}\,\GG^{(\ell+1)}_{\De+1}\cr
&&+{1\over 64}\,{(\De-\ell-2)(\De-\ell-2\vep)\over
(\De-\ell-2\vep+1)(\De-\ell-2\vep-1)}{\ell(\ell+2\vep-1)\over
(\ell+\vep-1)(\ell+\vep)}\cr
{}&&\qquad\qquad\qquad\times{(\De-1)(\De-2\vep)\over
(\De-\vep-1)(\De-\vep)}\,\GG^{(\ell-1)}_{\De+1}\,,}
}
and this satisfies positivity for the twist $\De-\ell\geq 4\vep$.
Other expansions may be found in Appendix \ref{AppD}.

The spectrum of operators appearing in the conformal partial wave
expansion of $A_{R}^{\rm{long}}$ is consistent with
that of a
long superconformal-multiplet with quasi-primary field in the $(0,0)$ (singlet)
representation of $SO(n)$.  From the conformal partial wave expansion,
the spectrum expected is, for $(\De')_{\ell'}$ representing an
operator of dimension $\De'$ and in the spin representation labelled
by $\ell'$,
\eqna{listo}{
\begin{array}{ccccccc}R&(0,0)&(1,0)&(1,1)&(2,0)&(2,1)&(2,2)\cr
&(\De)_{\ell,\ell\pm 2,\ell\pm 4}&(\De\pm 1)_{\ell\pm 1,\ell\pm
3}&(\De)_{\ell,\ell\pm 2}&(\De)_{\ell,\ell\pm 2}&(\De\pm 1)_{\ell\pm
1}&(\De)_{\ell}\cr
&(\De\pm 2)_{\ell,\ell\pm 2}&(\De\pm 3)_{\ell\pm 1}&
(\De\pm 2)_{\ell\pm 2,\ell}&
(\De\pm 2)_{\ell}&&\cr
&(\De\pm 4)_\ell&&&&&\end{array}
}
which is consistent with what we expect from the OPE \cite{fersok}
as shown from the relevant  representation theory in \ref{AppE}.
That the spectrum of operators (\ref{listo}) arises from the conformal
partial wave expansion of (\ref{longproj}) is essentially 
shown in Appendix \ref{AppD}.  These results are also consistent
with the corresponding case for $d=4,\N=4$ considered in \cite{DO}.

\subsection{The analysis of $\H_1^{(\vep)}$ for general $\vep$}

Much simpler to analyse are the contributions to the conformal partial
wave expansion coming from the $\H_1$ 
parts of (\ref{fullsoltwoo}) which
turn out to correspond to twist $2\vep$ operators.  
With the definition (\ref{confpartwave}),
we may write such conformal partial waves simply as,
\eqn{defgvep}{
\GG^{(\ell)}_{\ell+2\vep}(\xx,\zz)=u^\vep
\sum_{m=0}^{\infty}r_m(\ell)\,{\tilde{P}}^{(\vep)}_{\ell+m}(\xx,\zz)\,,
}
where,
\eqn{eqforrm}{
r_{m}(\ell)=\Gamma(\vep)\,(-\half)^\ell
{(\ell+\vep)_m(\ell+\vep)_{m+1-\vep}\over m! (2\ell+2\vep)_m}\,.
}
With the choice,
\eqn{alam}{
b_{1\,\lambda}=2 \sum_{m=0}^{\infty}r_m(\ell+2)\,\de_{\lambda,\ell+m+1}\,,
}
and using the following two relations, namely,
\eqna{tworels}{\eqalign{
&& u^\vep \,\sum_{m=0}^{\infty} \big(r_{m}(\ell)-2\,
r_{m+1}(\ell)\big){\tilde{P}}^{(\vep)}_{\ell+m}(\xx,\zz)\cr
&&={\ell\over
\ell+\vep-1}\,\GG^{(\ell-1)}_{\ell+2\vep-1}+{(\ell+\vep)(\ell+2\vep-1)\over
(2\ell+2\vep-1)(2\ell+2\vep+1)}\,\GG^{(\ell+1)}_{\ell+2\vep+1}\,,
}}
and
\eqna{secrelss}{\eqalign{
&&u^\vep \,\sum_{m=0}^{\infty} \big(r_{m+2}(\ell)-\,
r_{m+1}(\ell)\big){\tilde{P}}^{(\vep)}_{\ell+m}(\xx,\zz)\cr
&&\qquad\qquad \qquad  ={\ell(\ell-1)\over
4(\ell+\vep-2)(\ell+\vep-1)}\,\GG^{(\ell-2)}_{\ell+2\vep-2}\cr
&&\qquad\qquad \qquad  -{(\ell+\vep)(\ell+\vep-1)+\vep(\vep-2)\over
2
(2\ell+2\vep-3)(2\ell+2\vep+1)}\,\GG^{(\ell)}_{\ell+2\vep}\cr
&&\qquad\qquad \qquad 
+{(\ell+\vep)(\ell+\vep+1)(\ell+2\vep-1)(\ell+2\vep)\over
4(2\ell+2\vep-1)(2\ell+2\vep+1)^2(2\ell+2\vep+3)}
\GG^{(\ell+2)}_{\ell+2\vep+2}\,,
}}
we may obtain the following contributions to $A_R(\xx,\zz)$,
\eqna{contribtwo}{\eqalign{
A_{11}\sim \GG^{(\ell+2)}_{\ell+2\vep+2}\,,\quad A_{10}\sim {\ell+2\over
\ell+\vep+1}\,\GG^{(\ell+1)}_{\ell+2\vep+1}+{(\ell+\vep+2)(\ell+2\vep+1)\over
(2\ell+2\vep+3)(2\ell+2\vep+5)}\,\GG^{(\ell+3)}_{\ell+2\vep+3}\,, \cr
A_{00}\sim {(\ell+2)(\ell+1)\over
2(\ell+\vep)(\ell+\vep+1)}\,\GG^{(\ell)}_{\ell+2\vep}+
\Big({2\over n}-{(\ell+\vep+2)(\ell+\vep+1)+\vep(\vep-2)\over
(2\ell+2\vep+1)(2\ell+2\vep+5)}\Big)\,\GG^{(\ell+2)}_{\ell+2\vep+2}\cr
+{(\ell+\vep+2)(\ell+\vep+3)(\ell+2\vep+1)(\ell+2\vep+2)\over
2(2\ell+2\vep+3)(2\ell+2\vep+5)^2(2\ell+2\vep+7)}\GG^{(\ell+4)}_{\ell+2\vep+4}\,,}}
with other contributions vanishing.
For $(\vep,n)=(1,6)$ this agrees with similar results in \cite{DO}
where the relevant contribution was denoted by $\B_\ell$ and the
corresponding operators belong to a semi-short $\N=4$ superconformal multiplet
whereby the quasi-primary field has $(\De',\ell')=(\ell+2,\ell)$
and is an $SU(4)_R$ singlet.  The coefficients in (\ref{contribtwo}) are
positive for the range of $(\vep,n)$ values of relevance and the
lowest dimension operator contributing is again an R symmetry  
singlet with $(\De',\ell')=(\ell+2\vep,\ell)$.  For
$(\vep,n)=(\half,8)$ the contribution from the singlet operator with
$(\De',\ell')=(\ell+3,\ell+2)$ is absent.

\subsection{Modifications for $d=4$ and $\N=4$}

For $(\vep,n)=(1,6),(2,5)$ making the modification ({\ref{letting}) does not 
cancel all the
$\H_{2}$ contributions in (\ref{fullsoltwoo}).   
We have already dealt with those for $(\vep,n)=(2,5)$. 
For $(\vep,n)=(1,6)$, those left over may be easily computed to be given by
\eqn{bracinguu}{
G^{(1)}_2(\xx,\zz;\al,\bal)\sim
\sum_{\lambda}\,b_{2\,\lambda}\,\big((\al+\bal)\,u\,
P^{(1)}_{\lambda+1}(\xx,\zz;\al,\bal)-
P^{(1)}_{\lambda+2}(\xx,\zz;\al,\bal)\big)\,.}
We may make contact with known results \cite{DO} by
re-defining 
$b_{1\,\lambda}$ so that,
\eqn{lettingtw}{
b_{1\,\lambda} \to b_{1\,\lambda} - b_{2\,\lambda-1},
}
and choosing
\eqn{choice}{
b_{2\,\lambda}=2 \sum_{m=0}^{\infty}r_m(\ell+1)\, \de_{\lambda, \ell+m-1}\,.
}
We may then find that the conformal partial wave expansion of the remaining
$b_{2\,\lambda}$ contributions is given by,
\eqna{Aftwo}{\eqalign{
&& A_{21}\sim  \GG_{\ell+3}^{(\ell+1)}\,, \qquad  A_{20}\sim  
\GG_{\ell+2}^{(\ell)}
+ {(\ell+2)^2\over (2\ell+3)(2\ell+5)}\, \GG_{\ell+4}^{(\ell+2)} \, ,\cr
&& A_{11}\sim  2 \, \GG_{\ell+2}^{(\ell+2)}+{\ts{1\over 2}}\, 
\GG_{\ell+2}^{(\ell)}+{(\ell+2)^2\over 2(2\ell+3)(2\ell+5)}\,
\GG_{\ell+4}^{(\ell+2)}\,,\cr
&& A_{10} \sim  2 \, \GG_{\ell+1}^{(\ell+1)}
+ {2(\ell+2)^2\over (2\ell+3)(2\ell+5)}\,
\GG_{\ell+3}^{(\ell+3)}+{\ts{1\over 2}} \, \GG_{\ell+1}^{(\ell-1)} \, ,\cr
&& \quad +{(2\ell+3)^2-3\over
  4(2\ell+1)(2\ell+5)}\,\GG_{\ell+3}^{(\ell+1)}+
{(\ell+2)^2(\ell+3)^2\over 2(2\ell+3)(2\ell+5)^2(2\ell+7)}\,
\GG_{\ell+5}^{(\ell+3)}\cr
&& A_{00} \sim  \GG_{\ell}^{(\ell)} +
{2(\ell+1)(\ell+2)\over 3(2\ell+1)(2\ell+5)}\, 
\GG_{\ell+2}^{(\ell+2)}+ {(\ell+2)^2(\ell+3)^2\over 
(2\ell+3)(2\ell+5)^2(2\ell+7)}\,
\GG_{\ell+4}^{(\ell+4)}\cr
&&\quad {}+{\ts{1\over 15}}\, \GG_{\ell+2}^{(\ell)}
+{(\ell+2)^2\over 15(2\ell+3)(2\ell+5)}\, 
\GG_{\ell+4}^{(\ell+2)} \, , }
}
so that we have only twist zero and twist one contributions.  To
achieve this form we have used (\ref{tworels}), (\ref{secrelss}) and
\eqn{identyo}{
\sum_{m=0}^{\infty}r_{m-1}(\ell)\, {\tilde{P}}_{\ell+m}^{(1)}(\xx,\zz)=\half \, 
\GG_{\ell+2}^{(\ell)}(\xx,\zz)-2 \, \GG_{\ell+1}^{(\ell+1)}(\xx,\zz)\,,
}
and the following trivial identities, namely,
\eqn{alongwitho}{
r_{m-1}(\ell)-2 r_{m}(\ell)=r_{m}(\ell-1)+{(\ell+1)^2\over
(2\ell+1)(2\ell+3)}\, r_{m-2}(\ell+1)\,,
}
(which, when combined with (\ref{identyo}), is useful for determining $A_{10}$) and
\eqna{alongwithtw}{\eqalign{
&& r_{m+1}(\ell)-r_{m}(\ell)={1\over 4}\, r_{m+1}(\ell-2)
+{\ell(\ell+1)-1\over 6 (2\ell-1)(2\ell+3)}\,r_{m-1}(\ell)\cr
&&\quad +{(\ell+1)^2(\ell+2)^2\over 4(2\ell+1)(2\ell+3)^2(2\ell+5)}\,
r_{m-3}(\ell+2)\,,}
}
(which, again when combined with (\ref{identyo}), 
is useful for determining $A_{00}$).  

The expansion (\ref{Aftwo}) is significant for four dimensions because it in
fact represents contributions from a certain $\N=4$ non-unitary
multiplet, starting from an $SU(4)_R$ singlet.  This point and the
construction of such multiplets has been discussed in \cite{thesis}.  
It is a general feature of the $d=4$ $\N=4$ superconformal
four-point functions discussed here, for the single variable
parts of the Ward identity solutions \cite{NO}.

\vfil \eject

\section*{Acknowledgements}
We are grateful to H. Osborn for showing us his paper \cite{NO} before
publication and for numerous helpful discussions.
F.A.D. thanks G. Arutyunov for discussions and is grateful to LAPTH for their hospitality during a stay funded by an Alliance grant from the British Council. L.G. and E.S. acknowledge the participation of G. Arutyunov at the early stages of this work. E.S. profited from discussions with V. Dobrev, P. Heslop, P. Howe and I. Todorov.


\appendix{Superconformal invariants for the case $d=4$ ${\cal N}=4$}\label{AppA}

In the main text of this paper we needed just the linearized superconformal invariants $\hat z, \hat{z}',\hat{w}, \hat{w}'$. They could be easily obtained by examining the linearized supersymmetry transformations. In principle, the same procedure could give the full non-linear version of these invariants.  However, there exists an alternative and more efficient method which we are going to apply to the case $d=4$ ${\cal N}=4$.\footnote{E.S. acknowledges an enlightening discussion on this point with Paul Howe.}  In \cite{HH} it was pointed out that $\hat z, \hat{z}',\hat{w}, \hat{w}'$ can be obtained as the eigenvalues of the four-point covariant
\begin{equation}\label{XXX}
  X_{12}X^{-1}_{23}X_{34}X^{-1}_{41}\,,
\end{equation}
where the  supermatrices
\begin{equation}\label{d4sm4}
  X^{AB'} = 
  \begin{pmatrix}
    x^{\a\b'} & \q^{\a b'} \\
    \bq^{a\b'} & y^{ab'} 
  \end{pmatrix}\,,
\end{equation}     
combine the even and odd variables at each point of analytic superspace. 

Here we are going to this explicitly, with two additional important simplifications. Firstly, we  use the superconformal transformations to fix the superspace frame discussed in the main text:
\begin{equation}\label{26}
   X_2 =  \Sigma\,, \quad X_3 = X^{-1}_4 = 0\,.
\end{equation}
Here $\Sigma$ is a constant supermatrix obtained by setting $\q_2=0$ and $x^\mu_2=(1,0,\ldots,0)$, $y^i_2=(1,0,\ldots,0)$. Then the covariant (\ref{XXX}) is reduced to just $X\equiv X_{12}$. Secondly, instead of the explicit diagonalization of this matrix proposed in \cite{HH}, it is easier to solve the eigenvalue equation 
\begin{equation}\label{07}
  \mbox{Ber} (X- \lambda \mathbb{I}) = 0\,,
\end{equation}
where $\mbox{Ber}$ is the Berezinian (superdeterminant) and $\lambda$ are the eigenvalues. To zeroeth order in the $\q$ expansion these eigenvalues coincide with the even variables $z, z', w, \bw$. Their superconformal completions are then obtained as ``perturbations" around $z, z', w, \bw$. 

{}
{}

\subsection{The eigenvalues of the supermatrix}

For a general supermatrix $X$of $GL(n|m)$ given in block notation
\begin{equation}
X = 
\left( \begin{array}{cc}
 Z & \Theta \\
\bar{\Theta}& W\\
\end{array}\right)\,,
\end{equation} 
the Berezinian (superdeterminant) is defined as
\begin{equation}
{\rm Ber}(X) = {{\rm Det}(Z-\Theta W^{-1}\bar{\Theta}) \over {\rm Det}(W)} 
= {{\rm Det}(Z) \over {\rm Det}(W-\bar{\Theta} Z^{-1}\Theta)}\,,
\end{equation}
The Berezinian is a multiplicative function, invariant under conjugation, {e.g.},
\begin{equation}
{\rm Ber}(XY) = {\rm Ber}(X){\rm Ber}(Y) \,, \qquad
{\rm Ber}(MXM^{-1}) = {\rm Ber}(X)\,.
\end{equation}
As a consequence, the roots and poles of the rational function ${\rm Ber}(X-\lambda \mathbb{I})$ are 
invariant under conjugation.      
The eigenvalue equation for the supermatrix (\ref{d4sm4}) reads
\begin{eqnarray}
0 &\ea & {\rm Ber}(X_{12}-\lambda \mathbb{I})  \nonumber\\
&\ea &\left[\left( (z-\lambda)-\frac{\q^{11}\bar{\q}^{11}}{w-\lambda}- \frac{\q^{12}\bar{\q}^{21}}{w'-\lambda} \right)
\left((z' -\lambda)-\frac{\q^{21}\bar{\q}^{12}}{w-\lambda}- \frac{\q^{22}\bar{\q}^{22}}{\bw -\lambda} \right)\right.\nonumber\\
&&\left. -
\left(\frac{\q^{11}\bar{\q}^{12}}{w-\lambda}+ \frac{\q^{12}\bar{\q}^{22}}{\bw -\lambda}\right)
\left(\frac{\q^{21}\bar{\q}^{11}}{w-\lambda}+ \frac{\q^{22}\bar{\q}^{21}}{\bw -\lambda} \right)\right]/(w-\lambda)(\bw -\lambda)\,.
\end{eqnarray}
Thus, the invariants (eigenvalues) are the two roots and two poles of the rational fraction above.  The first eigenvalue is found as an expansion around $z$
\begin{equation}
\hat{z}= z- \Delta z = z- \Delta z_{1}- \Delta z_{2}- 
\Delta z_{3}- \Delta z_{4}\,.
\end{equation}
It is a root of the Berezinian or equivalently of the polynomial of degree six given by the numerator of the Berezinian:
\begin{eqnarray}
 &&\left((z-\lambda)(w-\lambda)\Big(\bw -\lambda\Big)-\q^{11}\bar{\q}^{11}(\bw -\lambda)- \q^{12}\bar{\q}^{21}(w-\lambda) \right)\nonumber\\
 &&\times \left((z' -\lambda)(w-\lambda)(\bw -\lambda)-\q^{21}\bar{\q}^{12}(\bw -\lambda)- \q^{22}\bar{\q}^{22}(w-\lambda) \right) \nonumber\\
&& - \left(\q^{11}\bar{\q}^{12}(\bw -\lambda)+ \q^{12}\bar{\q}^{22}(w-\lambda)\right)
\left(\q^{21}\bar{\q}^{11}(\bw -\lambda)+ \q^{22}\bar{\q}^{21}(w-\lambda) \right)  =0
\end{eqnarray}
This equation may be written in a more suggestive form using the variable $\Delta z$
\begin{eqnarray}
 &&\,\,\,\,\left( \Delta z\Big(1+ \frac{\Delta z}{ w-z}\Big)\Big(1+ \frac{\Delta z}{\bw -z}\Big)-\frac{\q^{11}\bar{\q}^{11}}{w-z } - \frac{ \q^{12}\bar{\q}^{21}}{\bw -z } \right) \nonumber\\
 &&\times \left( \Big(1+ \frac{\Delta z}{z' -z}\Big) 
\Big(1+ \frac{\Delta z}{ w-z}\Big)\Big(1+ \frac{\Delta z}{\bw -z}\Big)
- \frac{\q^{21}\bar{\q}^{12}}{(z' -z)(w-z) }- \frac{\q^{22}\bar{\q}^{22}}{(z' -z)(\bw -z)}  \right) \nonumber\\
&&- \left(  \frac{\q^{11}\bar{\q}^{12}}{ (w-z)(z' -z)^{{1\over 2}} }\Big(1+ \frac{\Delta z}{\bw -z}\Big) 
+ \frac{\q^{12}\bar{\q}^{22}}{(\bw -z)(z' -z)^{1\over 2}} \Big(1+ \frac{\Delta z}{w-z}\Big) \right) \nonumber\\
&& \times \left(\frac{\q^{21}\bar{\q}^{11}}{(w-z)(z' -z)^{1\over 2} }\Big(1+ \frac{\Delta z}{\bw -z}\Big)
+ \frac{\q^{22}\bar{\q}^{21}}{ (\bw -z)(z' -z)^{1\over 2}} \Big(1+ \frac{\Delta z }{ w-z}\Big) \right)=0 \label{sugg}
\end{eqnarray}
This form makes explicit the fact that each $\q$ will appear in $\Delta z$ dressed with some factor. As an  example,
$\q^{11}$ and $\bar{\q}^{11}$ are associated with the factor $(w-z)^{-1/2}$ while $\q^{22}$ and $\bar{\q}^{22}$ are associated with $\left( (\bw -z)(z' -z) \right)^{-1/2}$. The invariant $\hat{z}$ is then found iteratively as an expansion in the $\q$s and the net result is
\eqna{022}{
\Delta z_{1} &\ea & {\q^{11}\bar{\q}^{11} \over w-z}+{\q^{12}\bar{\q}^{21}\over \bw-z}\,, \cr
\Delta z_{2} &\ea & -\left({1 \over w-z} +
{1 \over \bw-z}\right){\q^{11}
\bar{\q}^{11}\q^{12}\bar{\q}^{21}\over(w-z)(\bw-z)}\cr
&&-{1 \over z'-z}\left({\q^{11}\bar{\q}^{11}\q^{21}\bar{\q}^{12}\over (w-z)^2} 
+{\q^{22}\bar{\q}^{22}\q^{12}\bar{\q}^{21}\over(\bw-z)^2} 
-{\q^{11}\bar{\q}^{12}\q^{22}\bar{\q}^{21}\over(w-z)(\bw-z)} 
-{\q^{12}\bar{\q}^{22}\q^{21}\bar{\q}^{11}\over(w-z)(\bw-z)} \right) \,,
\cr
 \Delta z_{3}
&\ea & 
-{\q^{22}\bar{\q}^{22}\q^{21}\bar{\q}^{12}\over(z'-z)^2(w-z)(\bw-z)}
\left({\q^{11}\bar{\q}^{11} \over w-z}+
{\q^{12}\bar{\q}^{21}\over \bw-z} \right)\cr
&& +{\q^{11}\bar{\q}^{11}\q^{12}\bar{\q}^{21}\over(w-z)(\bw-z)}
\left[\left({1 \over z'-z}+{2 \over w-z} +
{1 \over \bw-z} \right){\q^{21}\bar{\q}^{12}\over (z'-z) (w-z)} \right. \cr
&&\left. +\left({1 \over z'-z}+{1 \over w-z} +
{2 \over \bw-z} \right){\q^{22}\bar{\q}^{22}\over (z'-z) (\bw-z)} \right]\,,\cr\Delta z_{4}&\ea &0\,.}
Notice that the expansion is truncated and the top term $(\q)^8$ is absent. We shall come back to this result later on, when discussing the discrete symmetries of the problem.    

\subsection{Discrete symmetries}

\bigskip{\it Conjugation of the space-time coordinates}:
\bigskip

This is generated by conjugation with the matrix ${\cal M}_S$;
\begin{equation}
{\cal M}_S =  
\left( \begin{array}{cccc}
 0 &1 & 0 & 0 \\
1 & 0 &0 & 0 \\
0 &0 & 1 &0 \\
0&0 &0 &1  \\
\end{array}\right)\,, \qquad\, X_S = {\cal M}_S X {\cal M}^{-1}_S =
\left( \begin{array}{cccc}
 z'  &0 & \q^{21} & \q^{22} \\
0 &z &\q^{11} &\q^{12} \\
\bar{\q}^{12}&\bar{\q}^{11} & w &0 \\
\bar{\q}^{22}&\bar{\q}^{21} &0 &\bw   \\
\end{array}\right)\,,
\end{equation}  
This is a symmetry due to the identity
\begin{equation}
{\rm Ber}(X_S -\lambda \mathbb{I}) = {\rm Ber}(X -\lambda \mathbb{I})\,.
\end{equation}

\noindent
\bigskip{\it Conjugation of the internal coordinates}:\bigskip

This is generated by conjugation with the matrix ${\cal M}_H$;
\begin{equation}
{\cal M}_H = 
\left( \begin{array}{cccc}
1 &0 & 0 & 0 \\
0 & 1 &0 & 0 \\
0 &0 & 0 &1 \\
0&0 &1 &0  \\
\end{array}\right)\,, \qquad\, X_H = {\cal M}_H X {\cal M}^{-1}_H =
\left( \begin{array}{cccc}
 z &0 & \q^{12} & \q^{11} \\
0 &z'  &\q^{22} &\q^{21} \\
\bar{\q}^{21}&\bar{\q}^{22} & \bw  &0 \\
\bar{\q}^{11}&\bar{\q}^{12} &0 &w  \\
\end{array}\right)\,.
\end{equation}  
This is a symmetry due to the identity
\begin{equation}
{\rm Ber}(X_H -\lambda \mathbb{I}) = {\rm Ber}(X-\lambda \mathbb{I})\,.
\end{equation}

\noindent
\bigskip{\it Exchange of internal and space-time cross ratios}: \bigskip

This symmetry is specific to the ${\cal N}=4$ case. It is generated, externally to $gl(2|2)$, by conjugation with the matrix ${\cal J}$;
\begin{equation}
{\cal J} = 
\left( \begin{array}{cccc}
0 &0 & 1 & 0 \\
0 & 0 &0 & 1 \\
1 &0 & 0 &0 \\
0 & 1 & 0 &0  \\
\end{array}\right)
\,, \qquad\, X_J = {\cal J}X{\cal J}^{-1}=
\left( \begin{array}{cccc}
 w &0 &\bar{\q}^{11}&\bar{\q}^{12}   \\
0 &\bw  &\bar{\q}^{21}&\bar{\q}^{22} \\
\q^{11} & \q^{12}& z &0 \\
\q^{21} &\q^{22} &0 &z'   \\
\end{array}\right)\,,
\end{equation}
or, using $2\times2$ block notation,
\begin{equation}
X_J = 
\left( \begin{array}{cc}
 W & \bar{\Theta} \\
\Theta& Z\\
\end{array}\right)\,.
\end{equation} 

{}From the definition of the Berezinian one obtains 
\begin{equation}
{\rm Ber}(X_J-\lambda \mathbb{I})={\rm Ber}^{-1}(X-\lambda \mathbb{I})\,,
\end{equation}
so that this is also a discrete symmetry 
acting on the superconformal invariants. 
For example, one has $J(z)=w$, $J(\hat{z})= \hat{w}$, $J(\q^{ij})= \bar{\q}^{ij}$
and so on.

\bigskip
\noindent
{\it How these symmetries are used}:\bigskip

These symmetries can be used to obtain $\hat{z}'$, $\hat{w}$ and $\hat{w}'$ from $\hat{z}$. As an example, using the expression for $\Delta z_1$ and the conjugation for the space-time variables gives 
\begin{equation}
\Delta z' _{1}= \frac{\q^{21}\bar{\q}^{12}}{w-z' }+\frac{\q^{22}\bar{\q}^{22}}{\bw -z' }\,.
\end{equation}

The exchange symmetry may be used, in particular, to understand why the last term $\Delta z_{4}$ in (\ref{022}) vanishes while the corresponding top term is non-vanishing in the ${\cal N}=2$ case (see (\ref{015})). The most general form of the top term is
\begin{equation}
\Delta z_{4}= \q^{[8]} f_{z}(z, z' , w,\bw )\,,
\end{equation}
where $\q^{[8]}=\q^{11}\bar{\q}^{11}\q^{12}\bar{\q}^{21}\q^{21}\bar{\q}^{12}\q^{22}\bar{\q}^{22}$. Note that $\q^{[8]}$ is invariant under the discrete symmetries $S$, $H$, and $J$. 

{}From equation (\ref{sugg}) we know that the $\Delta z_{i}$ depend only on $z' -z$, $w-z$ and $\bw -z$, and analogously for $\Delta z' _{4}$, etc., so that one has
\begin{eqnarray}
&& \!\!\!\!\!\!\!\!\!\!\!\!\!\!\!\!\!\Delta z_{4}= \q^{[8]} F_{z}(z' -z, w-z,\bw -z) \, , \quad \quad
\Delta z' _{4}= \q^{[8]} F_{z}'(z-z' , w-z' ,\bw -z' ) , \nonumber\\
&& \!\!\!\!\!\!\!\!\!\!\!\!\!\!\!\!\!\Delta w_{4}= \q^{[8]} F_{w}(\bw -w, z-w,z' -w) \, , \quad \quad 
\!\!\!\!\!\Delta \bw _{4}= \q^{[8]} F_{\bw }(w-\bw , z-\bw ,z' -\bw ).
\end{eqnarray}

Let us now impose the discrete symmetries. First, the conjugation of the internal coordinates $H$ leaves $z$ and $z'$ invariant while exchanging $w$ and $\bw $, so we deduce that  both functions $F_{z}$ and $F_{z}'$ are symmetric under the exchange of the last two arguments. The same result for $F_{w}$ and $F_{\bw }$ is obtained by the use of the conjugation of spatial coordinates $S$. The conjugations $S$ and $H$ also imply the following identities between functions
\begin{equation}
F_{z}=F_{\smash{z}}' ,\qquad  F_{w}=F_{\bw } \,,
\end{equation}
respectively, so that we are left with only two functions. 

The difference between the ${\cal N}=2$ and ${\cal N}=4$ cases is that in the latter case these two functions have to be equal because of the exchange symmetry $J$. Since this symmetry exchanges spatial and internal coordinates, we have 
\begin{equation}
 {\cal F}=F_{z}=F_{\smash{z}}'= F_{w}=F_{\bw } \,,
\end{equation}
so that we are left with only one function ${\cal F}$. The function
${\cal F}$ has to satisfy various constraints:

\bigskip
\noindent
(A)  that the supertrace of the matrix $X$ has a vanishing $\q^{[8]}$ term so that $\Delta z_{4}+\Delta z' _{4}-\Delta w_{4}-\Delta \bw _{4}=0$,
hence the following linear equation is satisfied
\begin{eqnarray}
{\cal F}(z' -z, w-z,\bw -z)+{\cal F}(z-z' , w-z' ,\bw -z' )=\nonumber\\
{\cal F}(\bw -w, z-w,z' -w)+{\cal F}(w-\bw , z-\bw ,z' -\bw )\,; \label{conF}
\end{eqnarray}

\bigskip
\noindent
(B) (as already mentioned) that the function ${\cal F}$ is symmetric under the exchange of its last two arguments, ${\cal F}(a,b,c)={\cal F}(a,c,b)$\,;

\bigskip
\noindent
(C) from the perturbative approach, that the function ${\cal F}$ is a
rational function with monomials of global degree seven, in order for
$\Delta z_{4}$ to have the same dimension as $z$. 

\bigskip
Moreover, we have
already seen that with each $\q$ is associated a factor such that the
rational factor associated with  $\q^{[8]}$ is $(z'
-z)^{-2}(w-z)^{-2}(\bw -z)^{-2}$. Hence, we deduce from this result
and previous requirements that the function ${\cal F}$ must be of the
type
\begin{equation}
{\cal F}(z' -z, w-z,\bw -z)= {1\over {(z -z')^{2}(w-z)^{2}(\bw -z)^{2}}}\Big(-a   \frac{1}{z-z' }+ b \Big( \frac{1}{w-z }+\frac{1}{ \bw -z }\Big)\Big)\,,
\end{equation}
where $a$ and $b$ are complex parameters.

A glance at eq. (\ref{conF}) shows that such a function does not satisfy the constraint coming from the exchange symmetry unless it vanishes.

\subsection{Singularities}

 \bigskip{\it The variables $u$ and $v$}\bigskip

The singularity in $z' -z$ should not be present in physical quantities. In particular, it should be absent from  the supersymmetrized cross-ratios $\hat{u}$ and $\hat{v}$. Indeed, direct inspection shows that 
the $\hat{u}$ and $\hat{v}$ do not have this singularity. A simple argument explains this fact. It makes use of 
the properties of the matrix $X$ and of the symmetry by exchange of space-time and internal 
variables. 
Recall that
\begin{equation}
\hat{u}= \frac{\hat{z}\hat{z}'}{(1+ \hat{z})(1+\hat{z}') },\qquad \hat{v}= \frac{1}{ (1+ \hat{z})(1+\hat{z}')}\,,
\end{equation}
so that one has to check that the singularity is absent from $\hat{z}+\hat{z}'$ and $\hat{z}\hat{z}'$. These two combinations are related to the supertrace 
and the Berezinian of the matrix $X$, respectively:
\begin{eqnarray}
\hat{z}+\hat{z}' &\ea &\hat{w}+\hat{\bw } + \mathrm{Str}(X)\nonumber\\
\hat{z}\hat{z}' &\ea & \hat{w}\hat{\bw } \, \mathrm{{\rm Ber}}(X) \label{ste}\,.
\end{eqnarray}

Now, from the expression of $\hat{z}$ and the use of the various discrete symmetries, one gets that, apart from singularities of space-internal type (e.g., $w-z$, etc.), $\hat{z}$ and  $\hat{z}'$ have only the space-space singularity $z' -z$ while $\hat{w}$ and  $\hat{\bw }$ have only the internal-internal singularity $\bw -w$. The supertrace and Berezinian of the matrix $X$ are free from these two types of singularities. Hence, eqs. (\ref{ste}) tell us that the quantities $\hat{z}+\hat{z}'$, $\hat{z}\hat{z}'$, $\hat{w}+\hat{\bw }$ and $\hat{w}\hat{\bw }$   have neither $z' -z$, nor $\bw -w$ singularity but only singularities of the spatial-internal type. 

\bigskip
\noindent
{\it General functions of $\hat{z}$ and $\hat{z}'$}\bigskip

Let $F$ be a general function of $\hat{z}$ and $\hat{z}'$. In principle, this function will have some singularity in $z' -z$ unless it satisfies some requirements to be derived below. The quantity $F(\hat{z},\hat{z}')$ is an expansion in $\q$s 
\begin{equation}
F(\hat{z},\hat{z}')= F(z,z' ) + F^{[2]}+ F^{[4]}+ F^{[6]}+ F^{[8]}
\end{equation}
which can be obtained as a Taylor expansion around $F^{[0]}= F(z,z' )$. Due to its length, we shall not write the expansion explicitly but just the conditions for the absence of singularities that we have found.

The term with two $\q$s
\begin{equation}
F^{[2]}= - \left( \Delta z_1 F_{z} +\Delta z' _1 F_{z'}  \right)
\end{equation}
is non-singular in $z' -z$ because $\Delta z_1$ and $\Delta z' _1$ are not. The condition coming from the four-$\q$ term is 
 \begin{equation}
\left.\frac{F_{z'}-F_{z}}{z'-z}\right|_{z' -z \rightarrow 0} =
\left.\frac{2F_{z'-z}}{z'-z}\right|_{z' -z \rightarrow 0}\,,
\end{equation}
is non-singular. In the case ${\cal N}=2$ the expansion terminates at this level.
The ${\cal N}=4$ case is more complicated since higher-order singularities may be present in the six- and eight-$\q$ terms. In the six-$\q$ terms, a $(z' -z)^{-2}$ singularity could be present but is actually removed by the preceding requirement. The $(z' -z)^{-1}$ singularity is removed by requiring that 
\begin{equation}
\left.\partial_{z'-z}\left( \frac{F_{z'-z}}{z'-z}\right)\right|_{z'-z \rightarrow 0}
\end{equation}  
must be non-singular. In the eight-$\q$ term, the possible $(z' -z)^{-2}$ singularity is just not present and the  $(z' -z)^{-1}$ singularity is removed by requiring  that
\begin{equation}
\left.\frac{(F_{z'}-F_{z})_{zz' }}{z'  -z}\right|_{z' -z \rightarrow 0} 
\end{equation}
be non-singular. This is true if no $(z' -z)^3$ term is present in the expansion of $F$, which holds if $F$ is symmetric under the exchange of $z$ and $z' $.


\appendix{Jack polynomials}\label{AppB}

In two variables, Jack polynomials may be
expressed in terms  of  Gegenbauer polynomials as,
\eqn{Psol}{
P_{\lambda_1\lambda_2}^{(\vep)}(\xx,\zz) 
= {\lambda_- ! \over (2\, \vep)_{\lambda_-}}\,
(\xx\zz)^{{1\over 2}(\lambda_1+\lambda_2)}
C^{\,\vep}_{\lambda_-}(s) \, , \quad \lambda_- = \lambda_1 -
\lambda_2 \, ,\quad s={\xx+\zz\over 2\,(\xx\zz)^{1\over 2}}\,,  
}
where $\lambda_-=0,1,2,\dots,$ and the  normalisation here is such that
$P_{\lambda_1\lambda_2}^{(\vep)}(1,1)=1$.

We now summarize a few results  regarding Jack polynomials which are used
in the main text.
The first few cases, for integer $\vep$, are given by,
\eqna{Pss}{\eqalign{
P_{\lambda_1\lambda_2}^{(0)}(\xx,\zz)&\ea &  \half \big (
\xx^{\lambda_1} \zz^{\lambda_2} + 
\xx^{\lambda_2} \zz^{\lambda_1} \big )  \, , \cr
P_{\lambda_1\lambda_2}^{(1)}(\xx,\zz)&\ea & {1\over \lambda_- \! +1}
\bigg ({ \xx^{\lambda_1 +1 } \zz^{\lambda_2} 
- \xx^{\lambda_2} \zz^{\lambda_1 +1} \over
\xx-\zz}\bigg )  = {(\xx\zz)^{\lambda_2} \over \lambda_- \! +1}
\bigg ({ \xx^{\lambda_- +1 } - \zz^{\lambda_- +1} \over \xx-\zz}\bigg
) 
\, , \cr
P_{\lambda_1\lambda_2}^{(2)}(\xx,\zz)&\ea & {6\over \lambda_- \! + 2}
\bigg (  
{ \xx^{\lambda_1 + 3 } \zz^{\lambda_2} - \xx^{\lambda_2} \zz^{\lambda_1 +3} \over
( \lambda_- \! + 3)(\xx-\zz)^3 } - 
{ \xx^{\lambda_1 + 2 } \zz^{\lambda_2+1 } 
- \xx^{\lambda_2 +1 } \zz^{\lambda_1 +2} \over
(\lambda_- \! + 1)(\xx-\zz)^3}\bigg )\,,}
}
and others may be obtained through use of the recurrence relation,
\eqn{recurP}{
{(\xx-\zz)^2} P_{\lambda_1\lambda_2}^{(\vep{+1})}(\xx,\zz) =
{2(1+2\,\vep) \over \lambda_- \! + 1 + \vep} \Big (
P_{\lambda_1{+2}\,\lambda_2}^{(\vep)}(\xx,\zz)  - 
P_{\lambda_1{+1}\,\lambda_2{+1}}^{(\vep)}(\xx,\zz) \Big ) \, ,
}
following from standard identities for $C_n^{\vep}(s)$.

We also have that (at least for $\vep=0,\half,1,{\ts{3\over 2}},\dots$)
\eqn{integvep}{
C_n^{\vep}(s)=(-1)^{2\vep+1}C_{-n-2\vep}^{\vep}(s)\,,
}
which enables extension of Jack polynomials to some negative $\lambda_-$,
\eqn{neglambda}{
P^{(\vep)}_{\lambda_1\,\lambda_2}(\xx,\zz)=
P^{(\vep)}_{\lambda_2-\vep\,\lambda_1+\vep}(\xx,\zz)\,.
}

Among useful identities are
\eqn{Plc}{
(\xx\zz)^f  P_{\lambda_1\lambda_2}^{(\vep)}(\xx,\zz) =  
P_{\lambda_1{+f}\,\lambda_2{+f}}^{(\vep)}(\xx,\zz)\, ,
}
and 
\eqn{recurC}{\eqalign{
(\xx+\zz)P_{\lambda_1\lambda_2}^{(\vep)}(\xx,\zz) =  
{\lambda_-\! + 2 \,\vep \over \lambda_-{+ \vep}} \, 
P_{\lambda_1{+1}\,\lambda_2}^{(\vep)}(\xx,\zz) + 
{\lambda_- \over \lambda_- {+ \vep}}
\, P_{\lambda_1\, \lambda_2{+1}}^{(\vep)}(\xx,\zz) \, . }
}
Defining,
\eqn{defopera}{
D^{\pm}_{n}=\xx^n\,{\pr\over \pr \xx}\pm {\zz}^n\,{\pr \over \pr \zz}\,,
}
we have various derivative relations (which are used extensively in the
main text)
\eqna{derivrels}{\eqalign{
D^+_0
P_{\lambda_1\lambda_2}^{(\vep)}(\xx,\zz)&\ea &
 {(\lambda_1 + \vep) \lambda_- \over \lambda_-\! + \vep}\,
P_{\lambda_1{-1}\,\lambda_2}^{(\vep)}(\xx,\zz)+{ \lambda_2 
(\lambda_- +2\,\vep ) \over \lambda_- + \vep}
\, P_{\lambda_1\, \lambda_2{-1}}^{(\vep)}(\xx,\zz) \, ,\cr
D^+_2
P_{\lambda_1\lambda_2}^{(\vep)}(\xx,\zz)&\ea & 
{\lambda_1 (\lambda_- + 2\,\vep )\over \lambda_-\! + \vep}\,
P_{\lambda_1{+1}\,\lambda_2}^{(\vep)}(\xx,\zz) + { (\lambda_2 -  \vep ) 
\lambda_- \over \lambda_-\! +  \vep}
\, P_{\lambda_1\, \lambda_2{+1}}^{(\vep)}(\xx,\zz) \, ,\cr
D^+_3
P_{\lambda_1\lambda_2}^{(\vep)}(\xx,\zz)&\ea & 
{\lambda_1(\lambda_-+2\,\vep+1)(\lambda_-+2\,\vep)\over
(\lambda_-+\vep+1)(\lambda_-\!+\vep)}\,
P_{\lambda_1+2\,\lambda_2}^{(\vep)}(\xx,\zz)\cr
&{}&-\vep\,
{\lambda_-(\lambda_-+2\,\vep)+(\lambda_1+\lambda_2)(\vep-1)\over
(\lambda_-\!+\vep-1)(\lambda_-\!+
\vep+1)}\,P_{\lambda_1+1\,\lambda_2+1}^{(\vep)}(\xx,\zz)\cr
&{}&+{(\lambda_2- \vep)(\lambda_--1)\lambda_-\over
(\lambda_-+ \vep-1)(\lambda_-+ \vep)}\,
P_{\lambda_1\,\lambda_2+2}^{(\vep)}(\xx,\zz)\,,}
}
with other such being obtained recursively.{\foot{We may obtain
other such derivative relations recursively using the trivial identity
\eqn{tall}{
D^+_n=(\xx+\zz)D^+_{n-1}-\xx\zz\,D^+_{n-2}\,,\nonumber
}
along with (\ref{Plc}) and (\ref{recurC}). We may similarly find the action
of,
\eqn{fall}{
D^-_{n}={1\over
\xx-{\zz}}\,(2\,D^+_{n+1}-(\xx+\zz)D^+_{n})\,.
}
}}
Also, $P_{\lambda_1\,\lambda_2}^{(\vep)}(\xx,\zz)$ is an 
eigenfunction of the Euler
operator, $D^+_1$,
with
\eqn{actiondo}{
D^+_1\,P_{\lambda_1\,\lambda_2}^{(\vep)}(\xx,\zz)=(\lambda_1+\lambda_2)
\,P_{\lambda_1\,\lambda_2}^{(\vep)}(\xx,\zz)\,,
}
and the following second order
symmetric operator,
\eqn{defDv}{
D^J_\vep =  \xx^2 {\pr^2 \over \pr \xx^2} +  \zz^2  {\pr^2 \over \pr \zz^2}
+ 2\,\vep \, {1 \over \xx-\zz} D_2^- \, , 
}
with 
\eqn{Deig}{
D^J_\vep P_{\lambda_1\lambda_2}^{(\vep)}(\xx,\zz) = 
\big ( \lambda_1 (\lambda_1-1+2\,\vep)
+ \lambda_2 ( \lambda_2 -1) \big ) 
P_{\lambda_1\lambda_2}^{(\vep)}(\xx,\zz) \, , \quad
\lambda_1 \ge \lambda_2 \, .
}
We may rewrite the operator in (\ref{opanyd}) as 
\eqn{rewrite}{
D_{\vep}={\pr^2\over \pr \xx \pr {\zz}}-\vep {1\over \xx-\zz}\,
D_{0}^- ={1\over 2 \, \xx \,\zz}\,\big(-D^J_\vep+D^+_1{}^2
+(2\,\vep+1)\,D^+_1\big)\,,
}
so that
\eqn{eigsd}{
D_{\vep}\,P_{\lambda_1\,\lambda_2}^{(\vep)}(\xx,\zz)=\lambda_2(\lambda_1+
\vep)\,P_{\lambda_1-1\,\lambda_2-1}^{(\vep)}(\xx,\zz)\,,
}
which enables one to show (\ref{actionfds}).
We also have the special identities, following from (\ref{recurP}),
\eqn{specid}{
{1\over \xx-\zz}D^-_0\,P_{\lambda\,0}^{(\vep)}(\xx,\zz)
={\lambda\,(\lambda-1)\over
2\,(2\,\vep+1)}\,P_{\lambda{-2}\,0}^{(\vep+1)}(\xx,\zz)\,,
}
thus we have quite simply that,
\eqn{spect}{
D_{\vep}\,P_{\lambda\,0}^{(\vep')}(\xx,\zz)
={\lambda\,(\lambda-1)
\,(\vep'-\vep)\over
2\,(2\,\vep'+1)}\,P_{\lambda-2\,0}^{(\vep'+1)}(\xx,\zz)
}
which is used to prove (\ref{soline}).

\appendix{The $k=3$ case for $SO(n)$ R symmetry}\label{AppC} 

We here attempt to solve the Ward identities (\ref{condchvar})
 for the $k=3$ case, 
showing agreement with the general 
$k$ solutions.  The Ward identities involve eight independent
equations which we may conveniently take to be,
\eqn{ketward}{\eqalign{
\sum_{i=0}^3\, {1\over \xx^i} u^{i \vep}\,\pr_\xx \, a_{i 0}=0\,,\quad
\sum_{{i,j\geq 0\atop
i+j=3}}\,(1-\xx)^j\,v^{-j\vep}\,\pr_\xx\,a_{ij}=0\,,
\quad \sum_{i=0}^3\,\Big({\xx-1\over
\xx}\Big)^{i}\,\Big({u\over v}\Big)^{i\vep}\pr_\xx\,a_{0i}=0\,,}
}
along with three other such with $\xx\leftrightarrow \zz$ and,
\eqna{ketwardo}{\eqalign{
\pr_\xx a_{11}+2 {\xx^2\over
(\xx-1)}\Big({v\over u^2}\Big)^\vep \pr_\xx a_{00}+
{\xx\over \xx-1}\Big({v\over u}\Big)^\vep\pr_\xx a_{10}+\xx{1\over
u^\vep}\pr_\xx
a_{01}\cr
+{\xx-1\over \xx}\Big({u\over v}\Big)^{\vep}\pr_\xx a_{12}-{(\xx-1)^2\over \xx}
\Big({u\over v^2}\Big)^{\vep}\pr_\xx a_{03}=0\,,}
}
involving $a_{11}$,
along with another such with $\xx\leftrightarrow \zz$.
It is worth mentioning that any
equation involving $a_{11}$ necessarily involves at least six other of
the $a_{ij}$.

It proves convenient to define
\eqna{ketdefo}{
&& a_{03}=v^{2\vep}\De_\vep\,u c\,,\cr
&& a_{12}=v^{2\vep}\De_\vep\,u b+v^{\vep}\De_\vep (\xx+\zz-2)u c\,,\cr
&& a_{02}=v^{2\vep}\De_\vep\,u a+(u v)^{\vep}\De_\vep (\xx+\zz-2 u) c\,,
}
where $a,\,b,\,c$ are 
three arbitrary symmetric functions of
$\xx,\,\zz$.  With these definitions we may use the operator
identities (\ref{opidnet}) and (\ref{moreident}) to find directly for
the 
third equation in (\ref{ketward}) that,
\eqna{ketwardsimplifyf}{\eqalign{
&&\pr_\xx\big(a_{00}-u^{2\vep}\De_\vep
v\,a\big)\cr
&&\qquad\qquad +{\xx-1\over \xx}\Big({u\over v}\Big)^\vep \pr_\xx\big(
a_{01}-(uv)^\vep\De_\vep(\xx+\zz-2 u)a-u^{2\vep}\De_\vep v c\big)=0\,,}
}
so that we may use the $k=1$ analysis to show that
\eqna{partsolth}{\eqalign{
&& a_{00}=u^{2\vep}\De_\vep v a+u^{\vep}\sum_\lambda
    (-b_{1\,\lambda}+b_{1\,\lambda+1})
\tilde{P}_{\lambda}^{(\vep)}\,,\cr
&& a_{01}=(uv)^\vep\De_\vep(\xx+\zz-2u)a+u^{2\vep}\De_\vep
v\,c+v^{\vep}\sum_{\lambda}b_{1\,\lambda}\,
\tilde{P}_{\lambda}^{(\vep)}\,,}
}
for arbitrary $b_{1\,\lambda}$.
Similarly, we may find for the second equation in (\ref{ketward}) that,
\eqn{ketwardsimplifyt}{
\pr_\xx\big(a_{30}-\De_\vep \,u v b\big)
+{1\over
1-\xx}v^{\vep}\pr_\xx\big(a_{21}
-v^{\vep}\De_\vep(\xx+\zz-2)u\,b-\De_\vep uv\,c\big)=0\,,
}
so that
\eqna{partsolt}{\eqalign{
&& a_{21}=v^{\vep}\De_\vep(\xx+\zz-2)u\,b+\De_\vep uv\,c+v^{\vep}
\sum_{\lambda}b_{3\,\lambda}\,\tilde{P}_{\lambda}^{(\vep)}\,,\cr
&& a_{30}=\De_\vep\,u v
b+\sum_\lambda\big(b_{3\,\lambda-1}-b_{3\,\lambda}\big)
\tilde{P}_{\lambda}^{(\vep)}\,,
}
}
for arbitrary $b_{3\,\lambda}$.
Using (\ref{opidnet}) and (\ref{moreident})
and also the identity
\eqn{extral}{
\pr_\xx u^\vep
\,\tilde{P}_{\lambda}^{(\vep)}(\xx,\zz)
={1\over \xx}u^\vep \pr_\xx 
\,\tilde{P}_{\lambda+1}^{(\vep)}(\xx,\zz)\,,
} 
with (\ref{partsolt}) and (\ref{partsolth}), we may easily 
deal with the first equation in (\ref{ketward}) to find
that
\eqna{ketwardsimlify}{
&& \pr_\xx\big(a_{10}+u^{\vep}\De_\vep (\xx+\zz) v\,a-u^{2\vep}\De_\vep v
\,b-
{\ts{\sum_\lambda}}(b_{1\,\lambda-1}-b_{1\,\lambda})
\tilde{P}_{\lambda}^{(\vep)}\big)\cr
&&+{1\over \xx}u^{\vep}\pr_\xx\big(a_{20}-\De_\vep u v\, a
+u^\vep\De_\vep(\xx+\zz)v\, b+u^\vep {\ts{\sum_\lambda}}
(b_{3\,\lambda}-b_{3\,\lambda+1})
\tilde{P}_{\lambda}^{(\vep)}\big)=0\,. 
}
This may be solved similarly to the $k=1$ case and we find that
\eqna{partsol}{\eqalign{
&& a_{10} = -u^{\vep}\De_\vep (\xx+\zz) v\,a+u^{2\vep}\De_\vep v \,b
+\sum_\lambda \big(b_{1\,\lambda-1}-b_{1\,\lambda}+u^\vep a_\lambda\big)
\tilde{P}_{\lambda}^{(\vep)}\,,\cr
&& a_{20} = \De_\vep u v\, a- u^\vep\De_\vep(\xx+\zz)v\, b+
\sum_\lambda \big(u^\vep(-b_{3\,\lambda}+b_{3\,\lambda+1})-a_{\lambda-1}\big)
\tilde{P}_{\lambda}^{(\vep)}\,,
}}
with arbitrary $a_\lambda$.

More difficult to deal with is (\ref{ketwardo}).
We first deal with each of the two variable contributions
in turn.
Using (\ref{opidnet}) and (\ref{moreident}) we may find
that
\eqna{anotherket}{\eqalign{
{}&&  2 {\xx^2\over \xx-1}\Big({v\over u^2}\Big)^\vep\pr_\xx u^{2\vep}\De_\vep v\, a
-{\xx\over \xx-1}\Big({v\over u}\Big)^\vep\pr_\xx
\,u^{\vep}\De_\vep(\xx+\zz)v\,a\cr
{}&& \qquad \qquad\qquad + \xx {1\over u^\vep}\pr_\xx(uv)^\vep\De_\vep(\xx+\zz-2 u)a=-
\pr_x v^{\vep}\De_\vep(\xx+\zz-2)u\,a\,,}
}
for the $a$ contributions coming from $a_{00},\,a_{10},\,a_{01}$.  
Similarly we may find that
\eqn{anotherkett}{
{\xx\over \xx-1}\Big({v\over u}\Big)^\vep\pr_\xx u^{2\vep}\De_{\vep}v\,b
+{\xx-1\over \xx}\Big({u\over v}\Big)^{\vep}\pr_\xx v^{2\vep} \De_\vep
u\,b=-\pr_\xx (uv)^{\vep}\De_\vep\,(\xx+\zz-2 u)b\,,
} 
for the $b$ contributions coming from $a_{10},\,a_{12}$.
Similarly we may also find that
\eqna{anotherketf}{\eqalign{
&{}& \xx {1\over u^\vep}\pr_\xx u^{2\vep}\De_\vep v\,c+{\xx-1\over \xx}
\Big({u\over v}\Big)^\vep\pr_\xx v^{\vep}\De_\vep(\xx+\zz-2)u\,c\cr
&{}& \qquad \qquad \qquad -{(\xx-1)^2\over \xx}\Big({u\over v^2}\Big)^\vep
\pr_\xx v^{2\vep}\De_\vep u\,c=\pr_\xx u^{\vep}\De_\vep (\xx+\zz)v\,c\,,
}
}
for the $c$ contributions from $a_{01},\,a_{12},\,a_{03}$.
The only other contributions to (\ref{ketwardo})
come from $b_{1\,\lambda}$ and $a_\lambda$, however, due to
\eqna{fromforal}{
&& 2 {\xx^2\over
(\xx-1)}\Big({v\over u^2}\Big)^\vep \pr_\xx
\,u^\vep\big(\tilde{P}^{(\vep)}_{\lambda-1}(\xx,\zz)-\tilde{P}^{(\vep)}_{\lambda}(\xx,\zz)\big)+
{\xx\over \xx-1}\Big({v\over u}\Big)^\vep\pr_\xx 
\big(\tilde{P}_{\lambda+1}^{(\vep)}(\xx,\zz)-\tilde{P}_{\lambda}^{(\vep)}(\xx,\zz)\big)\cr
&& \qquad\qquad\qquad\qquad\qquad\qquad =-\xx{1\over
u^\vep}\,\pr_\xx \,v^\vep 
\tilde{P}^{(\vep)}_{\lambda}(\xx,\zz)\,,
}
those from $b_{1\,\lambda}$ vanish.
We thus find from (\ref{ketwardo}) that
\eqna{ketfindthat}{\eqalign{
&&\pr_\xx{\ts{\sum_\lambda}}a_\lambda\,\tilde{P}_{\lambda}^{(\vep)}
+{\xx-1\over \xx}\Big({u\over v}\Big)^\vep\pr_\xx
(a_{11}-v^{\vep}\,\De_\vep\,(\xx+\zz-2)\,u\,a\cr
&&\quad\qquad\qquad\qquad-(u\,v)^{\vep}\,\De_\vep\,
(x+\zz-2
u)\,b+u^{\vep}\,\De_\vep\,v\,(\xx+\zz)c\big)
=0\,,} 
}
so that the solution is given by
\eqna{expansionket}{\eqalign{
&&
a_{11}=v^{\vep}\,\De_\vep\,(\xx+\zz-2)\,u\,a+(u\,v)^{\vep}\,\De_\vep\,
(\xx+\zz-2
u)\,b\cr
&& \qquad\qquad\qquad\qquad -u^{\vep}\,\De_\vep\,v\,(\xx+\zz)c
+v^\vep\sum_\lambda b_{2\,\lambda}\,
\tilde{P}_{\lambda}^{(\vep)}\,,
}}
where 
\eqn{restdef}{
a_\lambda=-b_{2\,\lambda}+b_{2\,\lambda+1}\,.
}

With the solutions  (\ref{partsolt}), 
(\ref{partsolth}), (\ref{partsol}) and (\ref{expansionket})
we may then find that
\eqna{lastket}{\eqalign{
 G_3^{(\vep)}(\xx,\zz;\al,\bal)=&{}& \!\!\!\!\!\!\!\! 
u^{2\vep}\,\De_\vep\,(\xx\al-1)(\zz\al-1)(\xx\bal-1)(\zz\bal-1)a\cr
&{}& \!\!\!\!\!\!\!\!\!\!\!\!
+u^{3\vep}\,\al\bal\,\De_\vep\,(\xx\al-1)(\zz\al-1)(\xx\bal-1)(\zz\bal-1)b\cr
&{}&\!\!\!\!\!\!\!\!\!\!\!\! +\Big({u^{3}\over
v}\Big)^\vep\,(\al-1)(\bal-1)\,\De_\vep\,(\xx\al-1)(\zz\al-1)(\xx\bal-1)(\zz\bal-1)c\cr
&{}&\!\!\!\!\!\!\!\!\!\!\!\! +\sum_{n=1}^3 u^{n\vep}(\al\bal)^{n-1}\,
\H_{n}^{(\vep)}(\xx,\zz;\al,\bal)\,,}
}
which agrees with the general $k$ analysis.
Given the comments on the $k=2$ case, for $\vep\neq 1$ we may absorb 
the $\H^{(\vep)}_n,\,n=2,3$ contributions to (\ref{lastket}).  

\appendix{Recurrence relations for conformal partial waves}\label{AppD}
To perform the conformal partial wave expansion of (\ref{longproj})
requires other results than those mentioned in \cite{fadho}
essentially
because, whereas recurrence relations for $(v\pm 1)\GG^{(\ell)}_\De/u$
were found there (which are of direct relevance to four dimensions, see 
\cite{DO}), here matters are very much more complicated by
the presence of the operator $\De_\vep$.  Here we need results
of the form e.g. $u^{2\vep}\De_\vep (v\pm 1)u a$, for $a$ given by
(\ref{usqg}).  

In order to simplify the potentially very long winded recurrence relations for
conformal partial waves, we define
\eqn{defbF}{
F_{r,s}(\xx,\zz)=D(\De,\ell,r,s)\GG^{(\ell+r-s)}_{\De+r+s}(\xx,\zz)\,,
}
where $r,s$ are integers, and $D(\ell,\De,r,s)$ is a
constant so that
\eqna{factors}{\eqalign{
&& D(\De,\ell,r, s)=2^{r-s}\,\,{B^{{\rm{sgn}}\,
r}_{{1\over 2}(\De+\ell),r}}\,\,{B^{{\rm{sgn}}\,
s}_{{1\over 2}(\De-\ell)-\vep,s}}\cr
&&\quad \quad \qquad \times {A_{\ell+1,-{1\over 2}
({\rm sgn}(r-s)-1)(r-s)}}\,\,{A_{2-\De,-{1\over 2}
({\rm sgn}(r+s)+1)(r+s)}}\,,}
}
where $D(\De, \ell, 0,0)\equiv 1$ and
\eqn{defbcons}{
A_{\lambda,t }={(\lambda+\vep)_t(\lambda+\vep-1)_t\over 
(\lambda)_t(\lambda+2\vep-1)_t}\,,\quad
B_{\lambda,t}^{+}=16^{-t}{(\lambda)_t(\lambda+\vep-1)_t\over 
(\lambda-\half)_t(\lambda+\half)_t}\,,\quad 
B_{ \lambda,t}^{-}={(\lambda)_t\over (\lambda+1-\vep)_t}\,.}

We have trivially that,  recalling that $\al_{\pm}=\half(\De\pm\ell)-2\vep$,
\eqn{trivconf}{
u^{2\vep}\De_\vep \sum_{m,n\geq0}r_{m
n}(\De,\ell)(\E^{(\vep)}_{\al_++m\,\al_-+n})^{-1}\,\,
P_{\al_++m\,\al_-+n}^{(\vep)}(\xx,\zz)=F_{0,0}\,.
}
We may also find that,
\eqna{vminuso}{\eqalign{
&& u^{2\vep}\De_\vep\,\sum_{m,n\geq 0}r_{m
n}(\De,\ell)(\E^{(\vep)}_{\al_++m\,\al_-+n})^{-1}\,{(v-1)\over u}\,
P_{\al_++m\,\al_-+n}^{(\vep)}(\xx,\zz)\cr
&& \qquad\qquad =F_{-1,0}+F_{0,-1}+ F_{0,1}+ F_{1,0}\,,
}}
which is useful for expanding $\De_\vep u\,(v-1)a$, from which
(\ref{Aftwo})
 follows.  We may also determine
that,
\eqna{vminusoik}{\eqalign{
&& \half u^{2\vep}\De_\vep\,\sum_{m,n\geq 0}r_{m
n}(\De,\ell)(\E^{(\vep)}_{\al_++m\,\al_-+n})^{-1}\,{(v+1)\over u}\,
P_{\al_++m\,\al_-+n}^{(\vep)}(\xx,\zz)\cr
&&\qquad =F_{-1,-1}+F_{-1,1}+F_{1,-1}+F_{1,1}\cr
&&\qquad\,\,  +\quar\big(1-\half\vep(\vep-1)
\big(\A_{\ell+1}+\A_{2-\De}-(2\vep-1)(2\vep-3)
\A_{\ell+1}\A_{2-\De}\big)\big)\,F_{0,0}\,,
}}
 where
\eqn{vpluso}{
\A_{\lambda}={1\over (\lambda+\vep)(\lambda+\vep-2)}\,,
}
which is useful for expanding $\De_\vep u(v+1) a$.
Useful for expanding $\De_\vep (v-1)^2 a$ is the following, namely,
\eqna{vminusquared}{\eqalign{
&& u^{2\vep}\De_\vep\,\sum_{m,n\geq 0}r_{m
n}(\De,\ell)(\E^{(\vep)}_{\al_++m\,\al_-+n})^{-1}\,{(v-1)^2\over
u^2}\,P_{\al_++m\,\al_-+n}^{(\vep)}(\xx,\zz)\cr
&& \qquad = F_{-2,0}+F_{0,-2}+ F_{0,2}+F_{2,0}\cr
&& \qquad \quad +2(1-\vep(\vep-1)\A_{\ell+1})\big(F_{-1,-1}+F_{1,1}\big)
+2(1-\vep(\vep-1)\A_{2-\De})\big(F_{-1,1}+F_{1,-1}\big)\cr
&& \qquad\quad + C_{\De,\ell}F_{0,0}\,,}
}
where
\eqna{eqsforvmm}{\eqalign{
\!\!\!\!\!\!\!\!\!\!\!4 C_{\De,\ell}&&\!\!\!\!\!\!\!\!\! =1-\half
(2\vep-1)(2\vep-3)\big(\B_{\ell+\vep+1-\De}+\B_{\De+\ell-\vep-1}\big)\cr
&&+\vep(\vep-1)\big(\vep(\vep-1)\A_{\ell+1}\,
\A_{2-\De}-\A_{\ell+1}-\A_{2-\De}\big)\cr
&&\times \left(\half -(2\vep-1)(2\vep-3)\big(\quar \B_{\ell+\vep+1-\De}+\quar
\B_{\De+\ell-\vep-1}-3 \B_{\ell+\vep+1-\De}
\B_{\De+\ell-\vep-1}\big)\right),
}} 
where
\eqn{eqDconf}{
\B_\lambda={1\over (\lambda+\vep+2)(\lambda+\vep-2)}\,.
}
Useful for expanding $\De_\vep (v-1)(v+1) a$ is the following,
\eqna{vminusplus}{\eqalign{
&& \half u^{2\vep}\De_\vep\,\sum_{m,n\geq 0}r_{m
n}(\De,\ell)(\E^{(\vep)}_{\al_++m\,\al_-+n})^{-1}\,{v^2-1\over
u^2}\,P_{\al_++m\,\al_-+n}^{(\vep)}(\xx,\zz)\cr
&& \qquad = F_{-2,-1}+ F_{-1,-2}+ F_{-2,1}+F_{1,-2}+ F_{-1,2}+ 
F_{2,-1}+F_{1,2}+F_{2,1}\cr
&& \qquad \quad  +a_{\De,\ell}F_{0,-1}+b_{\De,\ell}F_{-1,0}+c_{\De,\ell}
F_{1,0}+d_{\De,\ell}F_{0,1}\,,}
}
where
\eqn{eqsforvpm}{
b_{\De,\ell}=a_{\De,-\ell-2\vep}\,,\qquad
c_{\De,\ell}=a_{2\vep+2-\De,\ell}\,,\qquad 
d_{\De,\ell}=a_{2\vep+2-\De,-\ell-2\vep}\,,
}
with
\eqna{eqsforvpmo}{\eqalign{
4 a_{\De,\ell}{}&&\!\!\!\!\!\!\!\!\! =3-{\ts{3\over
2}}\vep(\vep-1)(\C_{\ell+1}+\C_{2-\De})-(2\vep-1)(2\vep-3)\B_{\De+\ell-\vep-1}\cr
&{}& +\vep(\vep-1)(2\vep-1)(2\vep-3)\cr
&{}&\qquad \times\big(\C_{\ell+1}\C_{2-\De}+
\half \B_{\De+\ell-\vep-1}(\C_{\ell+1}+\C_{2-\De}
-10 \C_{\ell+1}\C_{2-\De})\big)\,,}
}
where
\eqna{eqsforvpmoo}{
\C_{\lambda}={1\over (\lambda+\vep+1)(\lambda+\vep-2)}\,.
}
Useful for expanding $\De_\vep(v+1)^2a$ is the following
\eqna{vplussq}{\eqalign{
&& \quar u^{2\vep}\De_\vep\,\sum_{m,n\geq 0}r_{m
n}(\De,\ell)(\E^{(\vep)}_{\al_++m\,\al_-+n})^{-1}\,{(v+1)^2\over
u^2}\,
P_{\al_++m\,\al_-+n}^{(\vep)}(\xx,\zz)\cr
&&\qquad\qquad = F_{-2,-2}+F_{-2,2}+ F_{2,-2}+ F_{2,2}\cr
&& \qquad\qquad\quad +{\ts{1\over
8}}(1-(2\vep-1)(2\vep-3)\B_{\ell+\vep+1-\De})(F_{-2,0}+F_{2,0})\cr
&& \qquad\qquad\quad +{\ts{1\over
8}}(1-(2\vep-1)(2\vep-3)\B_{\De+\ell-\vep-1})(F_{0,-2}+F_{0,2})\cr
&& \qquad\qquad\quad + e_{\De,\ell}F_{-1,-1}+f_{\De,\ell}F_{-1,1}
+g_{\De,\ell}F_{1,-1}
+h_{\De,\ell}F_{1,1}+D_{\De,\ell}F_{0,0}\,,
}}
where
\eqn{vpluspluso}{
f_{\De,\ell}=e_{-\ell-3,1-\De}\,,\quad
g_{\De,\ell}=e_{-\ell-1,1-\De}\,,\quad h_{\De,\ell}=e_{2+\De,\ell}\,,
}
with
\eqn{vplusplustw}{
2\,e_{\De,\ell}=1-\half
\vep(\vep-1)\big(\A_{\ell+1}+\B_{2-\De}
-(2\vep-1)(2\vep-3)\A_{\ell+1}\B_{2-\De}\big)\,.
}

The important point here is that from these results we may show 
compatibility of (\ref{longproj}) with (\ref{listo}) due to the
correspondence
\eqn{duetow}{
F_{r,s}\leftrightarrow (\De+r+s)_{\ell+r-s}\,,
}
between the conformal partial wave and an operator in the OPE.


\appendix{Representation theory}\label{AppE}
Following an approach like that for construction of superconformal
multiplets
in \cite{fhrep} based on using the Racah-Speiser algorithm (described
in \cite{fuchs}),
we show how we may verify that the conformal partial wave expansion results for
long multiplets are consistent with representation theory.  Generically
we have that the states appearing have $SO(d-1,1)\otimes SO(n)$
eigenvalues given by unconstrained tensor products of the form
\eqn{tensorprods}{
\prod_{I,A}(Q_{IA})^{n_{IA}}|{\rm{h}};{\rm{k}}\rangle^{\rm{hw}}\,,\quad
n_{IA}\in\{0,1\}\,,
}
where $|{\rm{h}};{\rm{k}}\rangle^{\rm{hw}}$ is a highest weight state, 
annihilated by
all generators of the appropriate superconformal algebra, apart from
the dilatation operator, for which it has eigenvalue $i\De$, the
momentum operators, the
super-charges $Q_{IA}$, where $I,A$ are, respectively,  $SO(d-1,1)$,
$SO(n)$ indices, the
lowering operators in the Chevalley-Serre basis of the algebra for
$SO(d-1,1)\otimes SO(n)$ and the generators of the Cartan sub-algebra
for which it has respective eigenvalues ${\rm{h}},\,{\rm{k}}$.

The $SO(d-1,1)\otimes SO(n)$ Cartan sub-algebra  eigenvalues of 
$Q_{IA}$ are given by
those in the weight system for some fundamental representation of the
group, which we denote generically by $[{\rm{i}}_I;{\rm{j}}_A]$.  
Thus, the states in
(\ref{tensorprods}) have ${\rm{h}}',\,{\rm{k}}'$ 
eigenvalues and conformal dimensions
$\De'$ given by,
\eqn{eigstate}{
[{\rm{h}}';{\rm{k}}']=\sum_{I,A}n_{IA}[{\rm{i}}_I;{\rm{j}}_A]
+[{\rm{h}};{\rm{k}}]\,,\qquad
\De'=\De+\half \sum_{I,A}n_{IA}\,.
}
Specific eigenvalues ${\rm{h}},\,{\rm{k}}$ requires use of the Racah-Speisser
algorithm when we may have cancellations among representations for
fixed $\De'$.

For example, for $d=3$ and $n=8$ we have that $Q_{IA}$ belongs to the $[\half]$
representation of $SO(2,1)$ and the $[1,0,0,0]$
representation of $SO(8)$, and thus that 
\eqn{adiiib}{
{\rm{i}}_I\in\{[\half],\,[-\half]\}\,,
}
and
\eqna{iiia}{\eqalign{
{\rm{j}}_A\!\!\!\!\!\!\!\!\!\!{}&&
\in\left\{[1,0,0,0],\,[-1,1,0,0],\,[0,-1,1,1],
\,[0,0,-1,1],\right.\cr
{}&&\quad\,\, \,\left.[0,0,1,-1],\,[0,1,-1,-1],\,[1,-1,0,0],
\,[-1,0,0,0]\right\}\,,}
}
whereby we may easily work out all possibilities for
${\rm{h}}',\,{\rm{k}}',\,\De'$ in (\ref{eigstate}).  For application to the
Racah-Speiser algorithm we may note the action of the generators of
the appropriate Weyl groups on generic weights; $[h]^\pi=[-h-1]$ for
$SO(2,1)$ and 
\eqna{actWeight}{\eqalign{
[k_{1},k_2,k_3,k_4]^{\pi_1}\!\!\!\!\!\!\!\!{}&& 
=[-k_1-2, k_1+k_2+1, k_3, k_4]\,,\cr
[k_{1},k_2,k_3,k_4]^{\pi_2}\!\!\!\!\!\!\!\!{}&& 
=[k_1, k_2+k_3+1,- k_3-2, k_4]\,,\cr
[k_{1},k_2,k_3,k_4]^{\pi_3}\!\!\!\!\!\!\!\!{}&& 
=[k_1,k_2+k_4+1,k_3,-k_4-2]\,,\cr
[k_{1},k_2,k_3,k_4]^{\pi_4}\!\!\!\!\!\!\!\!{}&&
=[k_1+k_2+1,-k_2-2,k_2+k_3+1,k_2+k_4+1]\,
}}
for $SO(8)$.  We may thus verify that (\ref{listo}) is compatible with
states appearing in a long superconformal multiplet starting from a
state with conformal dimension $\De'=\De-4$, with spin ${\rm{h}}=[\ell]$ and
being an
$SO(8)$ singlet, ${\rm{k}}=[0,0,0,0]$.

For $d=6$ and $n=5$ then
\eqn{adanoth}{
{\rm{i}}_I\in\{[1,0,0],\,[-1,1,0],\,[0,-1,1],\,[0,0,-1]\}\,,
}
and 
\eqn{adanothero}{
{\rm{j}}_A\in\{[1,0],\,[-1,1],\,[1,-1],\,[-1,0]\}\,.
}
The action of the generators of the appropriate Weyl groups is given
by
\eqna{adanothy}{\eqalign{
&&[h_1,h_2,h_3]^{\pi_1}=[-h_1-2,h_1+h_2+1,h_3]\,,\cr
&&[h_1,h_2,h_3]^{\pi_2}=[h_1+h_2+1,-h_2-2,h_2+h_3+1]\,,\cr
&&[h_1,h_2,h_3]^{\pi_3}=[h_1,h_2+h_3+1,-h_3-2]\,,
}}
for $SO(5,1)$ and
\eqn{adanothu}{
[k_1,k_2]^{\rho_1}=[-k_1-2,k_1+k_2+1]\,,\qquad
[k_1,k_2]^{\rho_2}=[k_1+2k_2+2,-k_2-2]\,,
}
for $SO(5)$.
Again we may verify that (\ref{listo}) is compatible with
states appearing in a long superconformal multiplet starting from a
state with conformal dimension $\De'=\De-4$, with 
spin ${\rm{h}}=[0,\ell,0]$ and
being an
$SO(5)$ singlet, ${\rm{k}}=[0,0]$.

\vfil \eject

 \end{document}